\documentclass{aa}
\usepackage{rotating}
\usepackage{natbib}
\usepackage{txfonts}
\usepackage{graphicx}
\def\la{\le}

\def\mld{\left (M/L\right )_{\rm d}}
\def\mlb{\left (M/L\right )_{\rm b}}

\begin{document}

\title{MACHOs in M31?\thanks{Based on observations made with the Isaac
Newton Telescope operated on the island of La Palma by the Isaac
Newton Group in the Spanish Observatorio del Roque de los Muchachos of
the Instituto de Astrofisica de Canarias} Absence of evidence but not
evidence of absence}

\author{Jelte~T.A.~de~Jong\inst{1}
        \and Lawrence~M.~Widrow\inst{2}
        \and Patrick Cseresnjes\inst{3}
        \and Konrad~Kuijken\inst{4,1}
        \and Arlin~P.S.~Crotts\inst{3}
        \and Alexander Bergier\inst{3}
        \and Edward~A.~Baltz\inst{5}
        \and Geza Gyuk\inst{6}
        \and Penny~D.~Sackett\inst{7}
        \and Robert~R.~Uglesich\inst{8}
        \and Will~J.~Sutherland\inst{9}\\
({\bf The MEGA collaboration})}

\institute{Kapteyn Astronomical Institute, University of Groningen, PO
Box 800, 9700 AV, Groningen, The Netherlands 
\and 
Department of Physics, Engineering Physics, and Astronomy, Queen's
University, Kingston, ON K7L 3N6, Canada
\and
Columbia Astrophysics Laboratory, 550 W 120th St., Mail Code 5247, New
York, NY 10027, United States 
\and 
Sterrewacht Leiden, University of Leiden, PO Box 9513, 2300 RA, Leiden, 
The Netherlands 
\and 
Kavli Institute for Particle Astrophysics and Cosmology, Stanford
University, PO Box 20450, MS 29, Stanford, CA 94309, United States
\and 
Department of Astronomy and Astrophysics, University of Chicago,
5640 South Ellis Avenue, Chicago, IL 60637, United States 
\and
Research School of Astronomy and Astrophysics, Australian National
University, Mt. Stromlo Observatory, Cotter Road, Weston ACT 2611,
Australia 
\and 
Laboratory of Applied Mathematics, Box 1012, Mount
Sinai School of Medicine, One Gustave L. Levy Place, New York, NY
10029, United States 
\and 
Institute of Astronomy, Madingley Rd, Cambridge CB3 0HA, United Kingdom }

\offprints{Jelte~T.A.~de~Jong, \email{jdejong@astro.rug.nl}}

\date{Received ? / Accepted ???}

\abstract{We present results of a microlensing survey toward
the Andromeda Galaxy (\object{M31}) carried out during four observing
seasons at the 
Isaac Newton Telescope (INT). This survey is part of the larger
microlensing survey toward M31 performed by the Microlensing
Exploration of the Galaxy and Andromeda (MEGA) collaboration.
Using a fully automated search algorithm, we identify 14 candidate
microlensing events, three of which are reported here for the first
time.  Observations obtained at the Mayall telescope are combined with
the INT data to produce composite lightcurves for these candidates.
The results from the survey are compared with theoretical predictions
for the number and distribution of events.  These predictions are
based on a Monte Carlo calculation of the detection efficiency and
disk-bulge-halo models for M31.  The models provide the full
phase-space distribution functions (DFs) for the lens and source
populations and are motivated by dynamical and observational
considerations.  They include differential extinction and span a wide
range of parameter space characterised primarily by the mass-to-light
ratios for the disk and bulge.  For most models, the observed event
rate is consistent with the rate predicted for self-lensing --- a
MACHO halo fraction of 30\% or higher can be ruled at the 95\%
confidence level.  The event distribution does show a large near-far
asymmetry hinting at a halo contribution to the microlensing signal.
Two candidate events are located at particularly large projected radii
on the far side of the disk.  These events are difficult to explain by
self lensing and only somewhat easier to explain by MACHO lensing.
A possibility is that one of these is due to a lens in a giant stellar
stream.

 }

\maketitle

\section{Introduction}
\label{sec:mulresult:intro}
Compact objects that emit little or no radiation form a class of
plausible candidates for the composition of dark matter halos.
Examples include black holes, brown dwarfs, and stellar remnants such
as white dwarfs and neutron stars.  These objects, collectively known
as Massive Astrophysical Compact Halo Objects or MACHOs, can be
detected indirectly through gravitational microlensing wherein light
from a background star is amplified by the space-time curvature
associated with the object \citep{paczynski86}.

The first microlensing surveys were performed by the MACHO
\citep{macho5.7} and EROS \citep{lasserre00,afonso03} collaborations
and probed the Milky Way halo by monitoring stars in the Large and
Small Magellanic Clouds.  While both collaborations detected
microlensing events they reached different conclusions.  The MACHO
collaboration reported results that favour a MACHO halo fraction of
20\%. On the other hand, the results from EROS are consistent with no MACHOs
and imply an upper bound of 20\% on the MACHO halo fraction.  The
two surveys are not inconsistent with each other since they probe
different ranges in MACHO masses.  They do leave open the question of
whether MACHOs make up a substantial fraction of halo dark matter and
illustrate an inherent difficulty with microlensing searches for
MACHOs, namely that they must contend with a background of
self-lensing events (i.e., both lens and source stars in the Milky Way or
Magellanic clouds), variable stars, and supernovae.  The Magellanic
Cloud surveys are also hampered by having only two lines of sight
through the Milky Way halo.

Microlensing surveys towards M31 have important advantages over the
Magellanic Cloud surveys \citep{crotts92}.  The microlensing
event rate for M31 is greatly enhanced by the high density of
background stars and the availability of lines-of-sight through dense
parts of the M31 halo.  Furthermore, since lines of sight toward the
far side of the disk pass through more of the halo than those toward
the near side, the event distribution due to a MACHO population should
exhibit a near-far asymmetry \citep{gyukcrotts00,kerins01,bgc}.

Unlike stars in the Magellanic Clouds, those in M31 are largely
unresolved, a situation that presents a challenge for the surveys but
one that can be overcome by a variety of techniques.  To date
microlensing events toward M31 have been reported by four different
collaborations, VATT-Columbia \citep{colvatt}, MEGA \citep{dejong04},
POINT-AGAPE \citep{paulin03,calchi03,calchi05} and WeCAPP \citep{riffeser03}.

Recently, the POINT-AGAPE collaboration presented an analysis of data
from three seasons of INT observations in which they concluded that
``at least 20\% of the halo mass in the direction of M31 must be in
the form of MACHOs'' \citep{calchi05}.  Their analysis is significant
because it is the first for M31 to include a model for the detection
efficiency.

The MEGA collaboration is conducting a microlensing survey in order to
quantify the amount of MACHO dark matter in the M31 halo.
Observations are carried out at a number of telescopes including the
2.5m Isaac Newton Telescope (INT) on La Palma, and, on Kitt Peak, the
1.3m McGraw-Hill, 2.4m Hiltner, and 4m Mayall telescopes.  The
observations span more than 4 seasons.  The first three seasons of INT
data were acquired jointly with the POINT-AGAPE collaboration though
the data reduction and analysis have been performed independently.

In \cite{dejong04} (hereafter Paper I) we presented 14 candidate
microlensing events from the first two seasons of INT data.  The
angular distribution of these events hinted at a near-far asymmetry
albeit with low statistical significance.  Recently \citet{an04}
pointed out that the distribution of variable stars also shows a
near-far asymmetry raising questions about the feasibility of the M31
microlensing program.  However, the asymmetry in the variable stars is
likely caused by extinction which can be modelled.

In this paper, we present our analysis of the 4-year INT data set.
This extension of the data by two observing seasons compared to Paper
I is a significant advance, but this data set is still only a subset
of the MEGA survey. The forthcoming analysis of the complete data set
will feature a further increase in time-sampling and baseline coverage
and length.
But there are more significant advances from Paper I.
We improve upon the photometry and data reduction in order to reduce
the number of spurious variable-source detections.  We fully automate
the selection of microlensing events and model the detection
efficiency through extensive Monte Carlo simulations.  Armed with
these efficiencies, we compare the sample of candidate microlensing
events with theoretical predictions for the rate of events and their
angular and timescale distributions.  These predictions are based on
new self-consistent disk-bulge-halo models \citep{m31models} and a
model for differential extinction across the M31 disk.  The models are
motivated by photometric and kinematic data for M31 as well as a
theoretical understanding of galactic dynamics.

Our analysis shows that the observed number of events can be explained
by self-lensing due to stars in the disk and bulge of M31, contrary to
the findings of \cite{calchi05}. Our results are consistent with a no
MACHO hypothesis, although we cannot rule out a MACHO fraction of 30\%.

Data acquisition and reduction methods are discussed in Sect.
\ref{sec:mulresult:data}.  The construction of a catalogue of
artificial microlensing events is described in Sect.
\ref{sec:mulresult:simulations}.  This catalogue provides the basis
for a Monte Carlo simulation of the survey and is used, in Sect.
\ref{sec:mulresult:selection}, to set the selection criteria for
microlensing events.  Our candidate microlensing events are presented
in Sect. \ref{sec:mulresult:sample}.  The artificial event catalogue
is then used in Sect. \ref{sec:mulresult:efficiency} to calculate the
detection efficiency.  Our extinction model is presented in
Sect. \ref{sec:mulresult:extinction}.  In
Sect. \ref{sec:mulresult:models} the theoretical models are described
and the predictions for event rate and distribution are presented.  A
discussion of the results and our conclusions are presented in
Sects. \ref{sec:mulresult:discussion} and
\ref{sec:mulresult:conclusions}.

\section{Data acquisition and reduction}
\label{sec:mulresult:data}

\begin{figure}
\centering
\vspace{0.27\hsize}{\tt 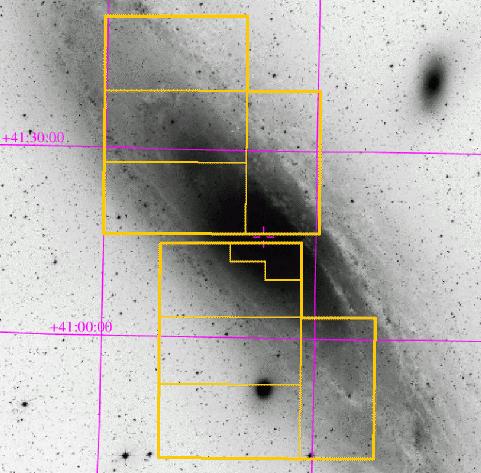}\vspace{0.27\hsize}
\caption{the layout of the two INT Wide Field Camera (WFC) fields in
  M31. A small part of the south field close to the bulge is not used
  since the image subtraction is not of high quality due to the high
  surface brightness.}
\label{fig:mulresult:intfields}
\end{figure}

Observations of M31 were carried out using the INT Wide Field Camera
(WFC) and spread equally over the two fields of view shown in Fig.
\ref{fig:mulresult:intfields}.  The WFC field of view is approximately
0.25$\Box^o$ and consists of four 2048x4100 CCDs with a pixel scale of
0.333\arcsec.  The chosen fields cover a large part of the far side
(SE) of the M31 disk and part of the near side.  Observations span
four observing seasons each lasting from August to January.  Since the
WFC is not always mounted on the INT, observations tend to cluster in
blocks of two to three weeks with comparable-sized gaps during which
there are no observations.

Exposures during the first (1999/2000) observing season were taken in
three filters, r$^{\prime}$, g$^{\prime}$ and i$^{\prime}$, which
correspond closely to Sloan filters.  For the remaining seasons
(2000/01, 2001/02, 2002/03), only the r$^{\prime}$ and i$^{\prime}$
filters were used.  Nightly exposure times for the first season were
typically 10 minutes in duration but ranged from 5 to 30 minutes.  For
the remaining seasons the default exposure time was 10 minutes per
field and filter.  Standard data reduction procedures, including bias
subtraction, trimming and flatfielding were performed in IRAF.

\subsection{Astrometric registration and image subtraction}

We use Difference Image Photometry (DIP) \citep{tc96} to detect
variable objects in the highly crowded fields of M31.  Individual
images are subtracted from a high quality reference image to yield
difference images in which variable objects show up as residuals.
Most operations are carried out with the IRAF package DIFIMPHOT.

Images are transformed to a common astrometric reference frame.  A
high signal-to-noise (S/N) reference image is made by stacking
high-quality images from the first season.  Exposures from a given
night are combined to produce a single ``epoch'' with Julian date
taken to be the weighted average of the Julian dates of the individual
exposures.

Average point spread functions (PSFs) for each epoch and for the
reference image are determined from bright unsaturated stars.  A
convolution kernel is calculated by dividing the Fourier transform of
the PSF from an epoch by the PSF transform from the reference image.
This kernel is used to degrade the image with better seeing (usually
the reference image) before image subtraction is performed
\citep{tc96}.

Image subtraction does not work well in regions with very high surface
brightness because of a lack of suitable, unsaturated stars.  For this
reason we exclude a small part of the south field located in a
high-surface brightness region of the bulge (see
Fig. \ref{fig:mulresult:intfields}).

\subsection{Variable source detection}

Variable sources show up in the difference images as residuals which
can be positive or negative depending on the flux of the source in a
given epoch relative to the average flux of the source as measured in
the reference image.  However, difference images tend to be dominated
by shot noise.  The task at hand is to differentiate true variable
sources from residuals that are due to noise.

The program SExtractor \citep{sextractor} is used to detect
``significant residuals'' in r$^{\prime}$ epochs, defined as groups of
4 or more connected pixels that are all at least 3$\sigma$ above or
below the background.  Residuals from different epochs are
cross-correlated and those that appear in two or more consecutive
epochs are catalogued as variable sources.  (Because of fringing, the
i$^{\prime}$ difference images are of poorer quality than the
r$^{\prime}$ ones and we therefore use r$^\prime$ data to make the
initial identification.)

\subsection{Lightcurves and Epoch quality}

The difference images for a number of epochs are discarded for a
variety of reasons.  Epochs with poor seeing do not give clean
difference images.  We require better than 2\arcsec~ seeing and
discard 7 epochs and parts of 12 epochs where this condition is not
met.  PSF-determination fails if an image is over-exposed.  We discard
7 epochs and parts of another 7 epochs for this reason.  Finally 2
epochs from the second and third seasons are discarded because of
guiding errors.

Lightcurves for the variable sources are obtained by performing
PSF-fitting photometry on the residuals in the difference images.
For every pixel the Poisson-noise is evaluated as well as the
fractional flux error due to photometric inaccuracies in the matching
and subtraction steps for the difference image in question. Fluxes and
their error bars are derived by optimal weighting of the individual
pixel values.  Lightcurves are also produced at positions where no
variability is identified and fit to a flat line.  These
lightcurves serve as a check on the contribution to the flux error
bars derived from the photometric accuracy of each difference image.
For each epoch, we examine by eye the distribution of the
deviations from the flat-line fits normalised by the photometric error
bar.  Epochs where this distribution shows broad non-Gaussian wings
are discarded since wings in the distribution are likely caused by
guiding errors or highly variable seeing between individual exposures.
For epochs where this normalised error distribution is approximately
Gaussian but with a dispersion greater than one, the error bars are
renormalised.

Approximately 19\% of the 209 r$^{\prime}$ epochs and 22\% of the 183
i$^{\prime}$ epochs are discarded. The number of epochs that remain for
each season, filter, and field are tabulated in Table
\ref{tab:mulresult:epochs}.  Though variable objects are detected in
r$^{\prime}$, lightcurves are constructed in both r$^{\prime}$ and
i$^{\prime}$.  In total, 105,447 variable source lightcurves are
generated.

\begin{table}
\centering
\caption{Overview of the number of epochs used for each  
field and filter.}
\begin{tabular}{l|cccc}
\hline
\hline
~ & r$^{\prime}$ & ~ & i$^{\prime}$ & ~\\
~ & North & South & North & South\\
\hline
99/00 & 48 & 50 & 21 & 18\\
00/01 & 58 & 57 & 66 & 62\\
01/02 & 28 & 30 & 27 & 28\\
02/03 & 35 & 32 & 33 & 30\\
\hline
Total & 169 & 169 & 147 & 138\\
\hline
\end{tabular}
\label{tab:mulresult:epochs}
\end{table}

\section{Artificial microlensing events}
\label{sec:mulresult:simulations}

This section describes the construction of a catalogue of artificial
microlensing lightcurves which forms the basis of our Monte Carlo
simulations.  We add artificial events to the difference images and
generate lightcurves in the same manner as is done with the actual
data.  The details of this procedure follow a review of microlensing
basics and terminology.

\subsection{Microlensing lightcurves}
\label{subsec:mulresult:mullc}

The lightcurve for a single-lens microlensing event is described by the
time-dependent flux \citep{paczynski86}:
\begin{equation}
F(t)~=~F_{\rm 0} \frac{u^2 + 2}{u \sqrt{u^2 + 4}}~\equiv~ F_0A(t)
\label{eq:mulresult:paczynski}
\end{equation}
where $F_{\rm 0}$ is the unlensed source flux and $A$ is the amplification.
$u=u(t)$ is the projected separation of the lens and the source
in units of the Einstein radius,
\begin{equation}
R_\textsc{E} = \sqrt{\frac{4Gm}{c^2} \frac{D_{\rm OL}D_{\rm LS}}{D_{\rm OS}}}~,
\label{eq:mulresult:r_e}
\end{equation}
where $m$ is the lens mass and the $D$'s are the distances between
observer, lens and source.  If the motions of lens, source, and
observer are uniform for the duration of the lensing event we can
write
\begin{equation}
u(t) = \sqrt{\beta^2 + \left( \frac{t-t_{\rm max}}{t_{\rm E}}\right)^2 }
\end{equation}
where $\beta$ is the impact parameter in units of $R_{\rm E}$, that is, the
mimimum value attained by $u$.  $t_{\rm max}$ is the time of maximum
amplification and $t_{\rm E}$ is the Einstein time, defined as the
time it takes the source to cross the Einstein radius.

In classical microlensing the measured lightcurves contain
contributions from unlensed sources. Blending, as this effect is
known, changes the shape of the lightcurve and can also spoil the
achromaticity implicit in equation \ref{eq:mulresult:paczynski}.  
In our survey, we measure flux differences that are created by
subtracting a reference image.  Since the flux from unlensed
sources is subtracted from an image to form the difference image,
blending is not a problem unless the unlensed sources are variable.
Blending by variable sources does introduce variations in the
baseline flux and adversely affects the fit.

For a difference image the microlensing lightcurve takes the form
\begin{equation}
\Delta F(t)\equiv F(t) - F_{\rm ref} ~=~ 
\Delta F_{\rm bl} + F_{\rm 0} (A(t)-1)
\label{eq:mulresult:difflux}
\end{equation}
where $F_{\rm ref}$ is the reference image flux and $\Delta F_{\rm
bl}\equiv F_{\rm 0}-F_{\rm ref}$. Thus, if in the reference image
the source is not lensed, $F_{\rm ref}=F_{\rm 0}$ and therefore $\Delta
F_{\rm bl}\equiv0$. Only if the source is amplified in the reference image
will $\Delta F_{\rm bl}$ be non-zero and negative.

For unresolved sources, a situation known as pixel lensing (and the one
most applicable to stars in M31), those microlensing events that can
be detected typically have high amplification.  In the high
amplification limit, $t_{\rm E}$ and $\beta$ are highly degenerate
\citep{gould96,baltzsilk00} and difficult to extract from the
lightcurve.  It is therefore advantageous to parameterise the event
duration in terms of the half-maximum width of the peak,
\begin{equation}
\label{eq:mulresult:tfwhm}
t_{\rm FWHM} = t_{\rm E} w(\beta)~,
\end{equation}
\noindent where
\begin{equation}
\label{eq:mulresult:tfwhma}
w(\beta) = 2 \sqrt{2 f( f(\beta^2)) - \beta^2 }
\end{equation}
\noindent and
\begin{equation}
\label{eq:mulresult:tfwhmb}
f(x) = \frac{x + 2}{\sqrt{x(x + 4)}} - 1
\end{equation}
\citep{gondolo99}.  $w(\beta)$ has the limiting forms $w(\beta\ll
1)\simeq\beta\sqrt{3}$ and $w(\beta\gg1)\simeq\beta(\sqrt{2}-1)^{1/2}$.

\subsection{Simulation parameters}
\label{subsec:mulresult:simulations:parameters}

The parameters that characterise microlensing events fall into two
categories: ``microlensing parameters'' such as $\beta$, $t_{\rm
max}$, and $t_{\rm E}$, and parameters that describe the source such
as its brightness $F_{\rm 0,r}$, its r$^{\prime}$-i$^{\prime}$ colour
$\mathcal{C}$, and its position.  We survey many lines-of-sight across
the face of M31.  Furthermore, all types of stars can serve as a
source for microlensing.  Therefore, our artificial event catalogue
must span a rather large parameter space.  This parameter space is
summarised in Table \ref{tab:mulresult:simulations} and motivated by
the following arguments:\\

\noindent
$\bullet$~{\it Peak times and baseline fluxes}

We demand that the portion of the lightcurve near peak amplitude is
well-sampled and therefore restrict $t_{\rm max}$ to one of the four
INT observing seasons.  The reference images are constructed from
exposures obtained during the first season. If a microlensing event
occurs during the first season and if the source is amplified in one
or more exposures during this season, the baseline in the difference
image will be below the true baseline.  For an actual event in
season one, this off-set is absorbed in one of the fit parameters
for the lightcurve.  For artificial events, the baseline is corrected
by hand.

\noindent
$\bullet$~{\it Event durations}

Limits on the duration of detectable events follow naturally from the
setup of the survey and the requirement that events are sampled
through their peaks.  Since the INT exposures are combined nightly,
events with $t_{\rm FWHM}<1\,{\rm day}$ are practically undetectable
except for very high amplifications.  On the other hand, events with
$t_{\rm FWHM}$ approaching the six-month length of the observing
season are also difficult to detect with the selection probability
decreasing linearly with $t_{\rm FWHM}$.
Because gaps in the time coverage of our survey will affect our
sensitivity to short events more strongly than to long events,
sampling should be denser at shorter timescales.  To limit computing
time and ensure statistically significant results spread over a wide
range of event durations, we simulate events at six discrete values
of $t_{\rm FWHM}$: 1, 3, 5, 10, 20 and 50 days.

\noindent
$\bullet$~{\it Source fluxes and colours}

Faint stars are more abundant than bright ones.  On the other hand,
microlensing events are more difficult to detect when the source is a
faint star.  The competition between these two effects means that
there is a specific range of the source luminosity function that is
responsible for most of the detectable microlensing events.

\begin{figure}
\centering
\includegraphics[width=8cm]{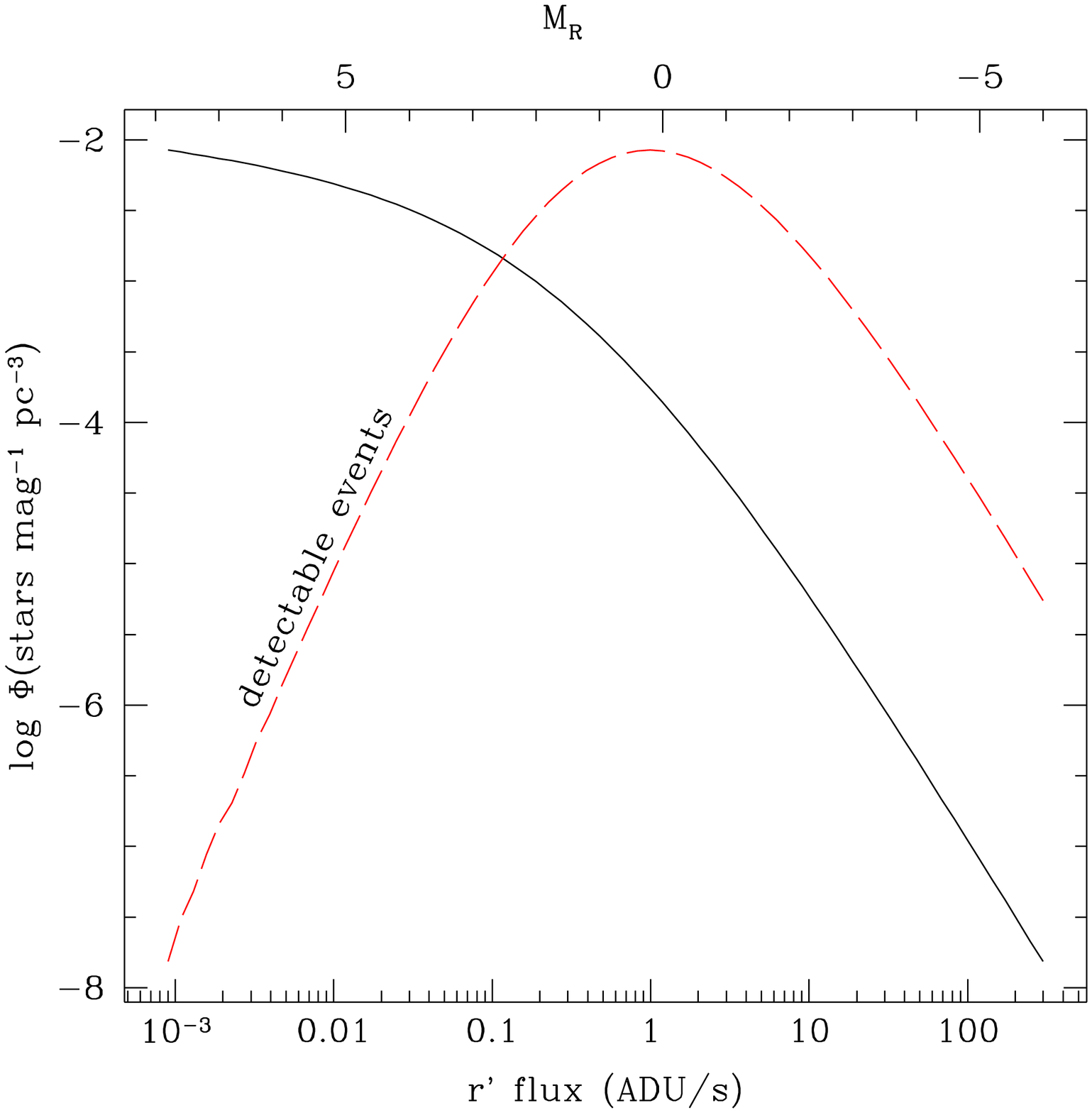}
\caption{The solid line in this figure shows the R-band luminosity
function from \cite{mamonsoneira}. Multiplying this function with the
square of the maximum impact parameter $\beta_{\rm max}$ needed to detect
a microlensing event gives the dashed line. The line shown is for a
detection threshold of 1 ADU\,s$^{-1}$ in r$^{\prime}$. The upper
horizontal axis shows absolute R-band magnitude, the lower axis the
corresponding r$^{\prime}$ flux.}
\label{fig:mulresult:detprob}
\end{figure}

The maximum flux difference during a microlensing event is
\begin{equation}
\Delta F_{\rm max} ~=~ F_{\rm 0} \left( \frac{\beta^2 + 2}{\beta
\sqrt{\beta^2 + 4}} ~-~ 1 \right)
\label{eq:mulresult:maxdf}
\end{equation}
where we are ignoring the $\Delta F_{\rm bl}$ term in equation
\ref{eq:mulresult:difflux}.  Let $\Delta F_{\rm det}$ be the detection
threshold for $\Delta F_{\rm max}$.  A lower bound on $\Delta F_{\rm
max}$ implies an upper bound on $\beta$ which, through equation
\ref{eq:mulresult:maxdf}, is a function of the ratio $F_{\rm 0}/F_{det}$:
$\beta_{\rm u} = \beta_{\rm u}\left (F_{\rm 0}/F_{det}\right )$.  The
probability that a given source is amplified to a detectable level
scales as $\beta_{\rm u}^2$.  In Fig. \ref{fig:mulresult:detprob} we
show both the R-band luminosity function, $N_*$, from
\cite{mamonsoneira} and the product of this luminosity function with
$\beta_{\rm u}^2$ assuming a detection threshold of $F_{\rm det}=
1\,{\rm ADU \,\rm s^{-1}}$.  The latter provides a qualitative picture
of the distribution of detectable microlensing events.  This
distribution peaks at an absolute R-band magnitude of approximately 0
indicating that most of the sources for detectable microlensing events
are Red Giant Branch (RGB) stars. 

Since there is no point in simulating events we cannot detect we let
the impact parameter $\beta$ vary randomly between 0 and $\beta_{\rm
u}$.  Table \ref{tab:mulresult:simulations} summarises the fluxes and
values for $\beta_{\rm u}$ used in the simulations.

For the artificial event catalogue, we use source stars with a
r$^{\prime}$ fluxes at several discrete values between 0.01 and 10
ADU\,s$^{-1}$. Typically the r$^{\prime}$$-$i$^{\prime}$ colours of RGB stars range between
$\mathcal{C} = 0.5$ and $2.0$.  We assume $\mathcal{C}=0.75$ for our
artificial events.  As a check of the dependence of the detection
efficiency with colour, we also simulate events with $\mathcal{C}=1.25$.

\noindent
$\bullet$~{\it Position in M31}

Lightcurve quality and detection efficiency vary with position in M31
for several reasons.  The photometric sensitivity and therefore the
detection efficiency depend on the amount of background light from M31
and are lowest in the the bright central areas of the bulge.
Difference images from these areas are also highly crowded with
variable-star residuals which influence the photometry and add noise
to the microlensing lightcurves.  To account for the
position-dependence of the detection efficiency, artificial events are
generated across the INT fields.  To be precise, the artificial event
catalogue is constructed in a series of runs.  For each run,
artificial events are placed on a regular grid with spacing of a 45
pixels ($\simeq 15\,\arcsec$) so that there are 3916 artificial events
per chip.  The grid is shifted randomly between runs by a maximum of
10 pixels.\\

\begin{table}
\centering
\caption{Fluxes and maximum impact parameters probed in the
simulations of microlensing events.
}
\begin{tabular}{lllll}
\hline
\hline
$F_{\rm 0,r}$ & $m_{\rm r}$ & $F_{\rm 0,i}$ & $m_{\rm i}$ & $\beta_{\rm u}$ \\
(ADU\,s$^{-1}$) & ~ & (ADU\,s$^{-1}$) & ~ & ~\\
\hline
0.01 & 29.5 & 0.011 & 28.75 & 0.01 \\
0.1 & 27.0 & 0.11 & 26.25 & 0.09 \\
0.5 & 25.2 & 0.55 & 24.45 & 0.35 \\
1.0 & 24.5 & 1.11 & 23.75 & 0.56 \\
10.0 & 22.0 & 11.1 & 21.25 & 1.67 \\
\hline
\end{tabular}
\begin{small}
\label{tab:mulresult:simulations}
\end{small}
\end{table}

To summarise, artificial events are characterised by the parameters
$t_{\rm FWHM}$, $F_{\rm 0}$, $\mathcal{C}$, $t_{\rm max}$, $\beta$,
and their angular position.  These events are added as residuals to the
difference images using the PSF in the subregion of the event.  The
residuals also include photon noise.  The new difference images are
analysed as in Sect. \ref{sec:mulresult:data} and lightcurves are
built for all artificial events detected as variable objects.

\section{Microlensing event selection}
\label{sec:mulresult:selection}

\begin{figure*}
\centering
\includegraphics[width=12cm,clip=]{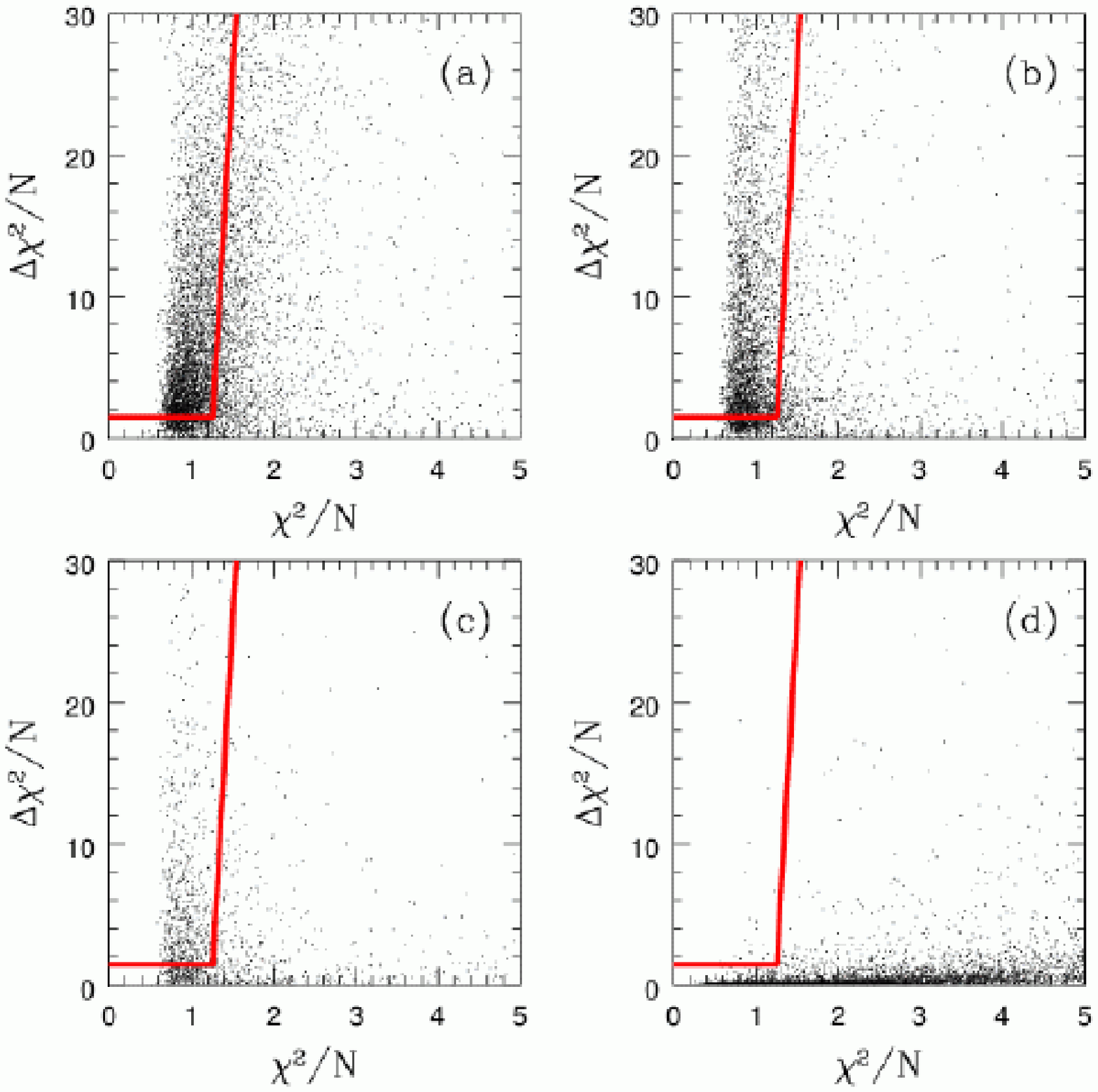}
\caption{
Scatter plots of $\Delta \chi2$ vs. $\chi2$
for simulated events with $t_{\rm FWHM}$=50 days (a), 10 days (b), 1 day (c),
and for the actual data for 1 CCD.  The solid lines correspond
to equations 9 and 10.
}
\label{fig:mulresult:chi2plots}
\end{figure*}

The vast majority of variable sources in our data set are variable
stars.  In this section we describe an automated algorithm that
selects candidate microlensing lightcurves from this rather formidable
background.  Our selection criteria pick out lightcurves that have a
flat baseline and a single peak with the ``correct'' shape.  The
criteria take the form of conditions on the $\chi^2$ statistic that
measures the goodness-of-fit of an observed lightcurve to equation
\ref{eq:mulresult:difflux}.  The fit involves seven free parameters:
$t_{\rm max}$, $\beta$, $t_{\rm E}$~, $F_{\rm 0,r}$, $F_{\rm 0,i}$,
$\Delta F_{\rm bl,r}$, and $\Delta F_{\rm bl,i}$.  To increase
computation speed we first obtain rough estimates for $t_{\max}$ and
$t_{\rm E}$ from the r$^{\prime}$ lightcurve and then perform the full
7-parameter fit using both r$^{\prime}$ and i$^{\prime}$ lightcurves.

Gravitational lensing is achromatic and therefore the observed colour
of a star undergoing microlensing remains constant in contrast with
the colour of certain variables.  While we do not impose an explicit
achromaticity condition, changes in the colour of a variable source
show up as a poor simultaneous r$^{\prime}$ and i$^{\prime}$ fit.
Because many red variable stars vary little in colour, as defined by
measurable differences in flux ratios, the lightcurve shape and
baseline flatness are better suited for distinguishing microlensing
events from long period variable stars (LPVs) than a condition
on achromaticity.

Lightcurves must contain enough information to fit adequately both the
peak of the microlensing event and the baseline.  We therefore impose
the following conditions: (1) The r$^{\prime}$ and i$^{\prime}$
lightcurves must contain at least 100 data points.  (2) The peak must
be sampled by several points well-above the baseline.  (3) The upper
half of the peak, as defined in the difference-image lightcurve, must
lie completely within a well-sampled observing period.  The second
condition can be made more precise.  We allow for one of the following
two possibilities: (a) 4 or more data points in the
r$^{\prime}$-lightcurve are $3\sigma$ above the baseline or (b) 2 or
more points in r$^{\prime}$ and 1 or more points in i$^{\prime}$ are $3\sigma$
above the baseline.  (The r$^{\prime}$ data is weighted more heavily
than the i$^{\prime}$ data because it is generally of higher quality
and because i$^{\prime}$ was not sampled as well during the first season.)
The third condition insures that we sample both rising and falling
sides of the peak.  We note that there are periods during the last two
seasons where we do not have data due to bad weather. The periods we
use are the following: 01/08/1999-13/12/1999, 04/08/2000-23/01/2001,
13/08/2001-16/10/2001, 01/08/2002-10/10/2002, and
23/12/2002-31/12/2002.

\begin{figure}
\centering
\includegraphics[width=8cm,clip=]{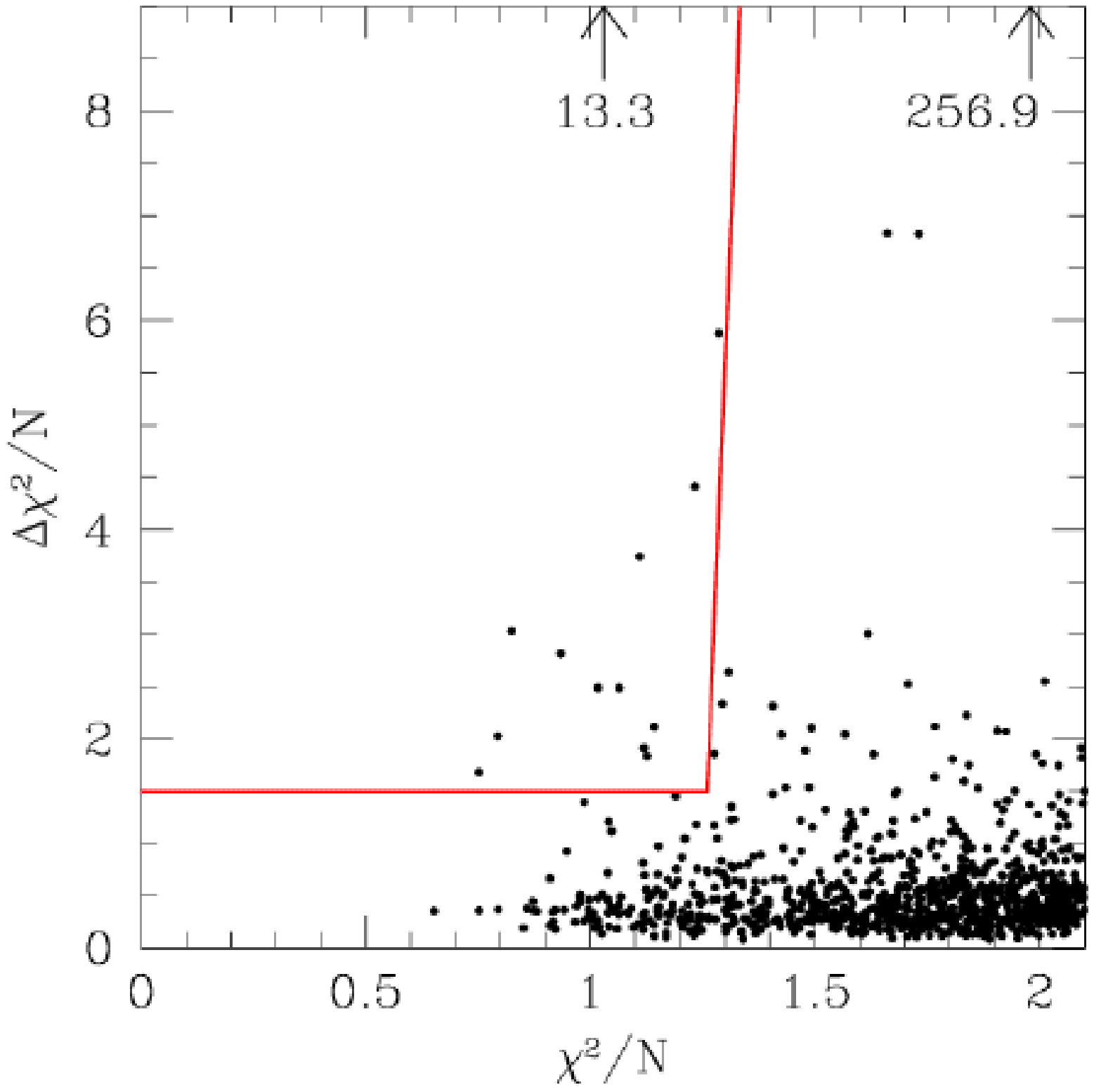}
\caption{$\Delta\chi^2/N$ versus $\chi^2/N$ for variable sources that
  satisfy selection criteria (1), (2) 
  and (3) for peak and lightcurve sampling. The solid line indicates
  criteria (4) and (5) for peak significance and goodness of
  fit. Criterium (5) depends on the number of points in the
  lightcurves, and the line drawn here is for $N$=309, the typical
  number of available data points per source. Two candidate events
  with higher $\Delta\chi^2/N$ are indicated with arrows, labelled with
  their $\Delta\chi^2/N$ value.}
\label{fig:mulresult:chi2select}
\end{figure}

The selection of candidate microlensing events is based on the
$\chi^2$-statistic for the fit of the observed lightcurve to equation
\ref{eq:mulresult:difflux} as well as $\Delta\chi^2\equiv \chi^2_{\rm
flat} - \chi^2$ where $\chi^2_{\rm flat}$ is the $\chi^2$-statistic
for the fit of the observed lightcurve to a flat line.  Our
$\chi^2$-cuts are motivated by simulations of artificial
microlensing events.  In Fig. \ref{fig:mulresult:chi2plots} we show
the distribution of artificial events with $t_{\rm FWHM} = 50,\,10,\,$ and
1 days (panels a, b, and c respectively) and for all variable sources
in one of the CCDs (panel d).  In Fig.
\ref{fig:mulresult:chi2select}, we show the variable sources from all
CCDs that satisfy conditions 1-3.  The plots are presented in terms
of $\chi^2/N$ and $\Delta\chi^2/N$ where $N$ is the number of data
points in an event.  We choose the following cuts:

\begin{equation}
\label{eq:mulresult:chi2cutsa}
\Delta\chi^2 > 1.5 N
\end{equation}

\noindent and

\begin{equation}
\label{eq:mulresult:chi2cutsb}
\chi^2 < \left (N-7\right )f\left (\Delta\chi^2\right )
+ 3\left (2\left (N-7\right )\right )^{1/2}
\end{equation}

\noindent where $f\left (\Delta\chi^2\right ) = \Delta\chi^2/100 + 1$.
The first criterion is meant to filter out peaks due to noise or
variable stars.  The second criterion corresponds to a $3\sigma$-cut
in $\chi^2$ for low signal-to-noise events.  The $\chi^2$ threshold
increases with increasing $\Delta\chi^2$.  Panels a-c of
Fig. \ref{fig:mulresult:chi2plots} show a trend where $\chi^2$
increases systematically with $\Delta\chi^2$.  This effect is due to
the photometry routine in DIFIMPHOT which underestimates the error in
flux measurements for high flux values.  The function $f$ is meant to
compensate for this effect.

The selection criteria appear as lines in
Figs. \ref{fig:mulresult:chi2plots} and
\ref{fig:mulresult:chi2select}.  (To draw these lines, we take $N=309$
though in practice $N$ is different for individual lightcurves.)

\section{Candidate events}
\label{sec:mulresult:sample}

Of the 105\,477 variable sources 28\,667 satisfy conditions 1-3.  Of
these, 14 meet the criteria set by equations
\ref{eq:mulresult:chi2cutsa} and \ref{eq:mulresult:chi2cutsb}.  The
positions of 12 of these events in the $\chi^2/N-\Delta\chi^2/N$ plane
are shown in Fig. \ref{fig:mulresult:chi2select}.

\subsection{Sample description}
\label{subsec:mulresult:sample}

In Table \ref{tab:mulresult:eventpars} we summarise the properties and
fit parameters of the 14 candidate microlensing events.  The first
column gives the assigned names of the events using the nomenclature
from Paper I.  The numbering reflects the fact that candidates 4, 5,
6, and 12 from Paper I are evidently variable stars since they peaked
a second time in the fourth season.  The other 10 events from Paper I
are ``rediscovered'' in the current more robust analysis.  Four
additional candidates, events 15, 16, 17, and 18, are presented.
Event 16 is the same as PA-99-N1 from \cite{paulin03} and was not
selected in our previous analysis because the baseline was too noisy
due to a nearby bright variable star.  It now passes our selection
criteria thanks to the smaller aperture used for the photometry (see
discussion below).  The three other events all peaked in the fourth
observing season and are reported here for the first time.

\begin{figure}
\centering
\vspace{0.27\hsize}{\tt 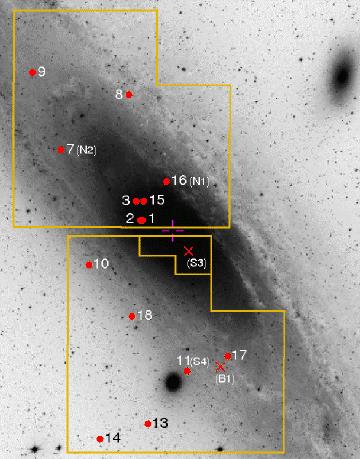}\vspace{0.27\hsize}
\caption{The locations of the 14 microlensing events within the INT
fields are shown here with the dots. Events 7 and 16 correspond with
events N2 and N1 from \cite{paulin03}. Their
event S3 is indicated with a cross and lies in the high surface
brightness region that we exclude from our analysis. Also marked with
a cross (B1) is the position of level 1 candidate 1 of \cite{belokurov04}.}
\label{fig:mulresult:eventpos}
\end{figure}

The coordinates of the events are given in columns 2 and 3 of Table
\ref{tab:mulresult:eventpars}; their positions within the INT fields
are shown in Fig. \ref{fig:mulresult:eventpos}.  The fit parameters,
$\chi^2$, and $\Delta\chi^2$ are given in the remaining columns.  In
Appendix \ref{ap:mulresult:lightcurves} we show the r$^{\prime}$ and
i$^{\prime}$ lightcurves, thumbnails from the difference images for a
number of epochs, and a comparison of $\Delta$r$^{\prime}$ and
$\Delta$i$^{\prime}$ for points near the peak.  The latter provides an
indication of the achromaticity of the event.  The lightcurves include
data points from observations at the 4m Mayall telescope on Kitt Peak
(KP4m) though the fits use only INT data.

\begin{table*}
\centering
\caption{Coordinates, highest measured difference flux, and some fit
  parameters for the 14 candidate microlensing events. The peak time
  $t_{\rm max}$ is in days after August 1st 1999 (JD 2451393).
}
\begin{tabular}{l|llcrrrrrr}
\hline
\hline
Candidate & RA & DEC & $\Delta$r$^{\prime}$ & $t_{\rm max}$ & $t_{\rm FWHM}$~ &
$\chi^2/N$ & $\Delta\chi^2/N$ & $F_{\rm 0,r}$ & r$^{\prime}$-i$^{\prime}$\\
event & (J2000) & (J2000) & (mag) & (days) & (days) & & & (ADU\,s$^{-1}$) & (mag) \\
\hline
MEGA-ML 1 & 0:43:10.54 & 41:17:47.8 & 21.8$\pm$0.4 & 60.1 $\pm$ 0.1 &
5.4 $\pm$ 7.0 & 1.12 & 1.91 & 0.1$\pm$0.3 & 0.6\\
MEGA-ML 2 & 0:43:11.95 & 41:17:43.6 & 21.51$\pm$0.06 & 34.0 $\pm$ 0.1
& 4.2 $\pm$ 0.7 & 1.06 & 2.48 & 3.4$\pm$1.7 & 0.3\\
MEGA-ML 3 & 0:43:15.76 & 41:20:52.2 & 21.6$\pm$0.1 & 420.03 $\pm$ 0.03
& 2.3 $\pm$ 2.9 & 1.14 & 2.11 & 0.08$\pm$0.21 & 0.4\\
MEGA-ML 7 & 0:44:20.89 & 41:28:44.6 & 19.37$\pm$0.02 & 71.8 $\pm$ 0.1
& 17.8 $\pm$ 0.4 & 1.98 & 256.9 & 6.8$\pm$0.4 & 1.5\\
MEGA-ML 8 & 0:43:24.53 & 41:37:50.4 & 22.3$\pm$0.2 & 63.3 $\pm$ 0.3 &
27.5 $\pm$ 1.2 & 0.82 & 3.03 & 20.4$\pm$22.9 & 0.6\\
MEGA-ML 9 & 0:44:46.80 & 41:41:06.7 & 21.97$\pm$0.08 & 391.9 $\pm$ 0.1
& 2.3 $\pm$ 0.4 & 1.02 & 2.49 & 0.9$\pm$0.4 & 0.2\\
MEGA-ML 10 & 0:43:54.87 & 41:10:33.3 & 22.2$\pm$0.1 & 75.9 $\pm$ 0.4 &
44.7 $\pm$ 5.6 & 1.28 & 5.88 & 1.4$\pm$0.5 & 1.1\\
MEGA-ML 11 & 0:42:29.90 & 40:53:45.6 & 20.72$\pm$0.03 & 488.43 $\pm$
0.04 & 2.3 $\pm$ 0.3 & 1.03 & 13.27 & 1.5$\pm$0.4 & 0.2\\
MEGA-ML 13 & 0:43:02.49 & 40:45:09.2 & 23.3$\pm$0.1 & 41.0 $\pm$ 0.3 &
26.8 $\pm$ 1.5 & 0.75 & 1.68 & 9.2$\pm$10.8 & 0.8\\
MEGA-ML 14 & 0:43:42.53 & 40:42:33.9 & 22.5$\pm$0.1 & 455.9 $\pm$ 0.1
& 25.4 $\pm$ 0.4 & 1.11 & 3.74 & 146$\pm$182 & 0.4\\
MEGA-ML 15 & 0:43:09.28 & 41:20:53.4 & 21.63$\pm$0.08 & 1145.5 $\pm$
0.1 & 16.1 $\pm$ 1.1 & 1.23 & 4.41 & 7.0$\pm$2.2 & 0.5\\
MEGA-ML 16 & 0:42:51.22 & 41:23:55.3 & 21.16$\pm$0.06 & 13.38 $\pm$
0.02 & 1.4 $\pm$ 0.1 & 0.93 & 2.81 & 2.6$\pm$0.7 & ~\\
MEGA-ML 17 & 0:41:55.60 & 40:56:20.0 & 22.2$\pm$0.1 & 1160.7 $\pm$ 0.2
& 10.1 $\pm$ 2.6 & 0.79 & 2.02 & 0.5$\pm$0.3 & 0.4\\
MEGA-ML 18 & 0:43:17.27 & 41:02:13.7 & 22.7$\pm$0.1 & 1143.9 $\pm$ 0.4
& 33.4 $\pm$ 2.3 & 1.13 & 1.83 & 13.7$\pm$16.3 & 0.5\\
\hline
\end{tabular}
\label{tab:mulresult:eventpars}
\end{table*}

\begin{figure}
\centering
\includegraphics[width=8cm,clip=]{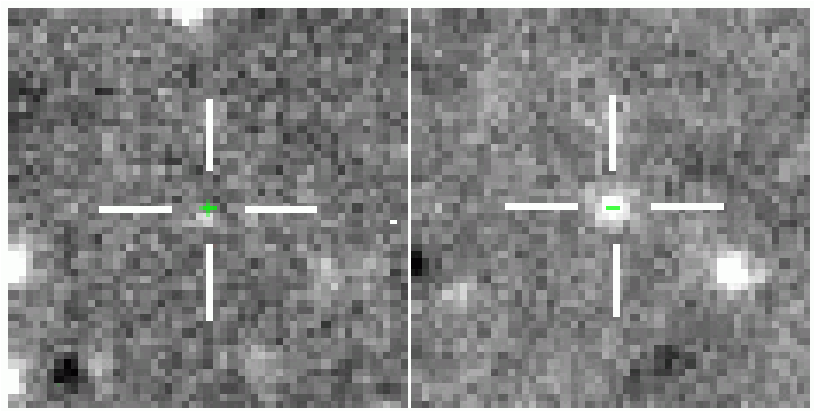}
\caption{Detail of two KP4m difference images centred on the position
of event 3. {\it Left:} October 27th 1997, almost 3 years before the
event peaks, a very faint residual is seen centred just 1 pixel
(0.21\arcsec) away from the event. {\it Right:} September 26th 2000,
during the peak of the event that is displaced from the position of
the faint variable.}
\label{fig:mulresult:event3_kp97}
\end{figure}

We have already seen that variable stars can mimic microlensing
events.  Blending of variable stars is also a problem since it leads
to noisy baselines.  This problem was rather severe in Paper I causing
us to miss event PA-99-N1 found by the POINT-AGAPE collaboration
\citep{paulin03}.  In an effort to reduce the effects of blending by
variable stars, we use a smaller aperture when fitting the PSF to
residuals in the difference images.  Nevertheless, some variable star
blending is unavoidable, especially in the crowded regions close to
the centre of M31.  Event 3 provides an example of this effect.  A
faint positive residual is visible in the 1997 KP4m difference image
as shown in Fig. \ref{fig:mulresult:event3_kp97}.  The residual is
located one pixel ($0.21''$) from the event and is likely due to a
variable star.  It corresponds to the data point in the lightcurve
$\sim$1000 days before the event and well-above the baseline (see
Fig.\,\ref{fig:mulresult:event3}).  The KP4m data point from 2004 is
also above the baseline but in this and other difference images, no
residual is visible.  The implication is that variable stars can
influence the photometry even when they are too faint to be detected
directly from the difference images.

Good simultaneous fits are obtained in both r$^{\prime}$ and
i$^{\prime}$ for all candidate events.  Event 7 has a high $\chi^2/N$
of 1.98, but since $\Delta\chi^2/N$ is very high, the event easily
satisfies our selection criteria.  In high S/N events, secondary
effects from parallax or close caustic approaches can cause measurable
deviations from the standard microlensing fit.  In addition, as
discussed above, we tend to underestimate the photometric errors at
high flux levels.  \cite{pa99n2} studied this event in detail and
found that the deviations from the standard microlensing shape of the
POINT-AGAPE lightcurve are best explained by a binary lens.  The
somewhat high $\chi^2$ for events 10 and 15 are probably because they
are located in regions of high surface brightness.

All of the candidate events are consistent with achromaticity, though
for events with low S/N, it is difficult to draw firm conclusions
directly from the lightcurves or $\Delta r'$ vs. $\Delta i'$ plots.
The values for $F_{\rm 0,r}$ and $\mathcal{C}$ for the events give some
indication of the properties of the source stars.  The unlensed fluxes
are consistent with the expected range of $0.1-10\,{\rm ADU}\,{\rm
s}^{-1}$ and the colours for most of the events are typical of RGB
stars.  Note however that for many of the events, the uncertainties
for $F_{\rm 0}$, $\beta$, and $t_{\rm FWHM}$~are quite large.  These
uncertainties reflect degeneracies among the lightcurve fit
parameters.

The number of candidate events varies considerably from season to
season.  We find 7 events in the first season, 4 in the second season,
none in the third season and 3 in the fourth season.  The paucity of
events during the third and fourth seasons is not surprising given
that we have fewer epochs for those seasons (see table
\ref{tab:mulresult:epochs}).  In particular, the gaps in time coverage
during those seasons conspired against the detection of short duration
events.

\subsection{Comparison with other surveys}
\label{subsec:mulresult:othersurveys}

\begin{figure}
\includegraphics[width=8cm,clip=]{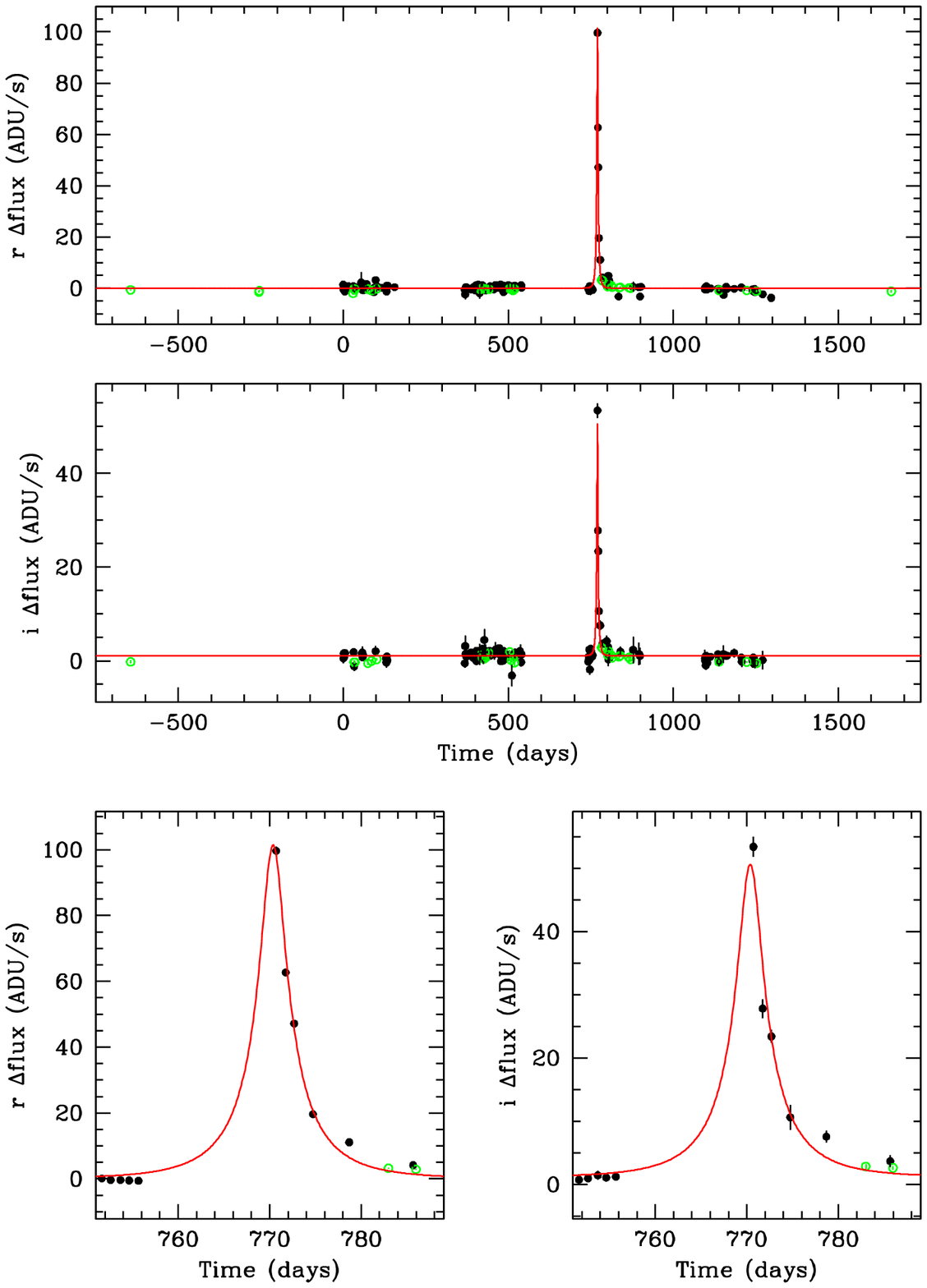}
\caption{Our photometry for microlensing event candidate 1 from
  \cite{belokurov04}.}
\label{fig:mulresult:eventb1}
\end{figure}

\begin{figure}
\centering
\includegraphics[width=7cm]{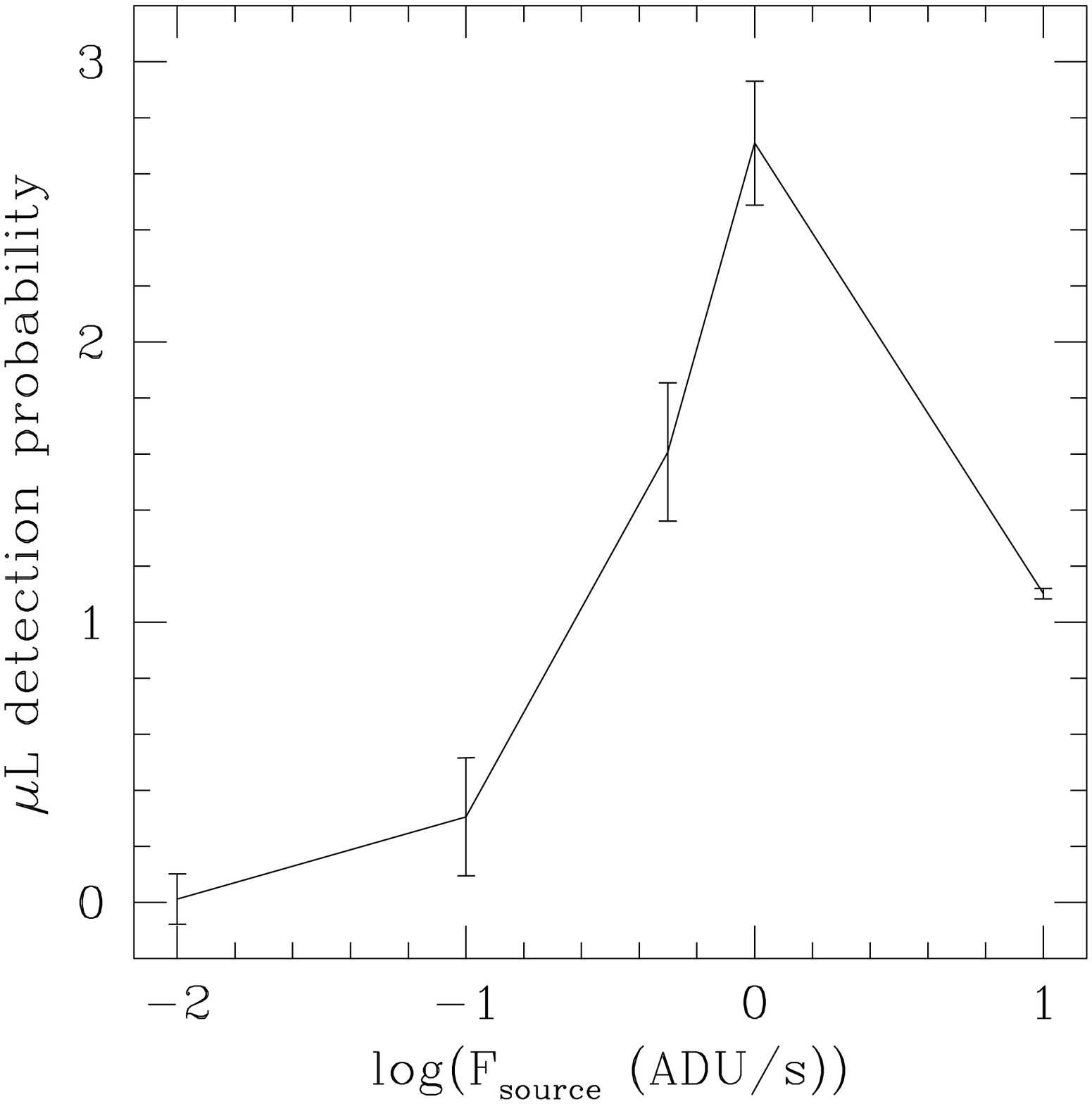}
\caption{ Relative probability of detecting a microlensing event of a
source star with a certain intrinsic flux. This
probability is the product of the number of available stars (taken
from the luminosity function), the square of the maximum impact
parameter for which an event can be detected, and the detection
efficiency for each source population, averaged over all
$t_{\rm FWHM}$.}
\label{fig:mulresult:de1}
\end{figure}

The POINT-AGAPE collaboration published several analyses of the INT
observations.  In \cite{paulin03} they presented four convincing
microlensing events from the first two observing seasons using
stringent selection criteria.  In particular, they restricted their
search to events with high S/N and $t_{\rm FWHM}<25\,{\rm days}$.
They argued that one of these events (PA-00-S3) is probably due to a
stellar lens in the M31 bulge.  This event lies in the region of the
bulge excluded from our analysis (see Fig.
\ref{fig:mulresult:intfields}).  The other three events, PA-99-N1,
PA-99-N2, and PA-00-S4, correspond respectively to our events 15, 7,
and 11.  Evidently, the remaining eight events from our analysis of
the first two INT seasons did not satisfy their rather severe
selection criteria.

In \citet{belokurov04}, the POINT-AGAPE 
collaboration analysed data from the first three
INT observing seasons without any restrictions on the event duration.
Using different selection criteria from their previous analyses, they
found three high quality candidates.  Two of these events were already
known (PA-00-S4 or MEGA-ML-11 and PA-00-S3).  The one new event is
present in our survey but does not pass our selection criteria because
of a high $\chi^2$.  The lightcurve for this event, along with our
best-fit model, is shown Fig. \ref{fig:mulresult:eventb1}.  The model
does not do a good job of reproducing the observed lightcurve
behaviour.  In particular, the observed lightcurve appears to be
asymmetric about the peak time $t_{\rm max}$.  The observed
r$^{\prime}$-lightcurve is systematically below the model 15-20
days prior to $t_{\rm max}$.  Both r$^{\prime}$ and i$^{\prime}$
lightcurves are above the model 10-15 days after $t_{\rm max}$.  Since
there are no data available on the rising part of the peak, $t_{\rm
max}$ is poorly constrained and may in fact be less than the $770\,{\rm
days}$ used in the fit.  The shape of our r$^{\prime}$ lightcurve is
similar to the one presented in \cite{belokurov04} (NB. They removed
one epoch close to the peak centre that is present in our lightcurve.)
In i$^{\prime}$ the peak shapes are somewhat different.

Peak asymmetries can be caused by secondary effects such as parallax.
In our opinion, a more likely explanation for this case is that the
event is a nova-like eruptive variable.  Granted, the event appears to
be achromatic.  But classical novae can be achromatic on the declining
part of the lightcurve (see, for example, \cite{darnley04}), precisely
where there is data.  If this is a classical nova, it would be a very
fast one, with a decline rate corresponding to $\sim$0.6 mag per day.

\begin{figure*}
\centering
\includegraphics[width=12cm,clip=]{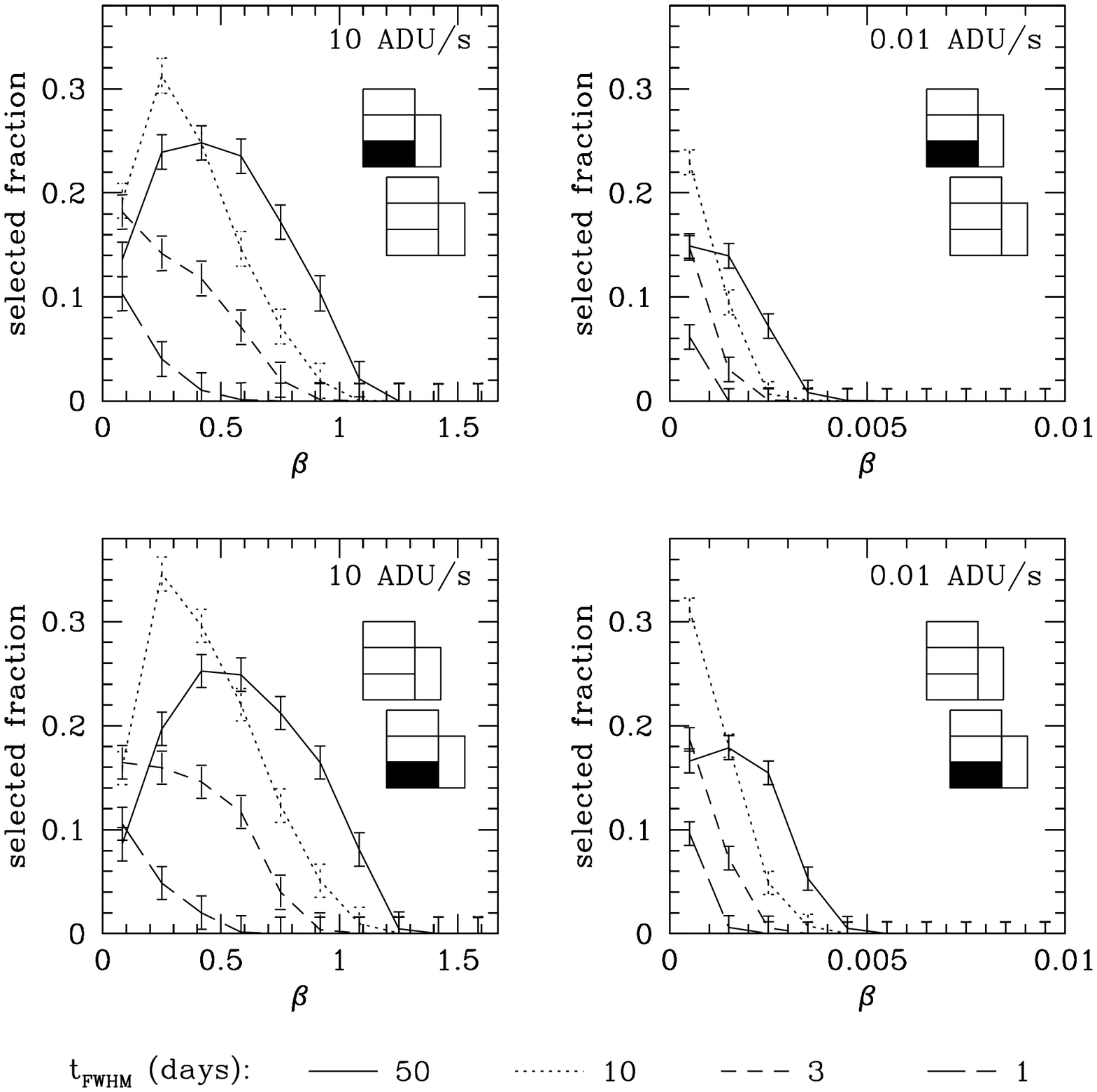}
\caption{Detection efficiencies as function of impact parameter
$\beta$ for different values of $t_{\rm FWHM}$~ (50, 10, 3 and 1 days).
The two upper panels show the fraction of simulated events that pass
the microlensing selection criteria for 2 source fluxes, 10 and 0.01
ADU\,s$^{-1}$, in the south-east chip of the north field. The lower panels
show the same for the south-east chip of the south field.}
\label{fig:mulresult:de2}
\end{figure*}

\cite{calchi05} found six candidate microlensing events in an analysis
of the three-year INT data set. Of these events, four are the same as
reported by \cite{paulin03} and two are new events: PA-00-N6 and
PA-99-S7. The latter of these is located in the bright part of the
southern field excluded in our analysis
(Fig. \ref{fig:mulresult:intfields}). Candidate event PA-00-N6 is
present in our data, but was only detected in one epoch in our
automatic SExtractor residual detection step and therefore did not
make it into the catalogue of variable sources.  \cite{calchi05} do
not detect our events 1, 2, 3, 8, 9, 10, 13, and 14, which all peak in the
first two observing seasons.  Evidently, these events do not satisfy
their S/N constraints.

\section{Detection efficiency}
\label{sec:mulresult:efficiency}

We determine the detection efficiency for microlensing events by
applying the selection criteria from Sect.
\ref{sec:mulresult:selection} to the catalogue of artificial events
from Sect. \ref{sec:mulresult:simulations}.  As discussed above,
simulated lightcurves are generated by adding artificial events to the
difference images and then passing the images through the photometry
analysis routine designed for the actual data.  Those lightcurves that
satisfy the selection criteria for microlensing form a catalogue of
simulated {\it detectable} microlensing events.  The detection
efficiency is the ratio of the number of these events to the original
number of artificial events.

We first check that our artificial event catalogue includes the
portion of the source luminosity function responsible for most of the
detectable events.  The function $N_*\beta_{\rm u}^2$ in Fig.
\ref{fig:mulresult:detprob} is meant to give a qualitative picture of
the detectability of microlensing as a function of source luminosity.
Here we consider the function $P_{\rm det}\equiv N_* \beta_{\rm u}^2
\epsilon$ where $\epsilon$ is detection efficiency as a function of
$F_{0,r}$ integrated over $\beta$, $t_{\rm FWHM}$ and position.
$P_{\rm det}$ gives the relative probability for detection of a
microlensing event as a function of the source luminosity.  As shown
in Fig.  \ref{fig:mulresult:de1}, the range $0.01$ to $10\,{\rm
ADU\,s^{-1}}$ adequately covers the peak of this probability
distribution.

Our goal is to represent the detection efficiency in terms of a simple
portable function of a few key parameters.  We adopt a strategy
whereby the detection efficiency is modelled as functions of $t_{\rm
FWHM}$ and $\Delta F_{\rm max}$ for individual subregions of the two
fields.  The parameters $\beta$ and $t_{\rm max}$ are ``integrated
out'' and $\mathcal{C}$ is fixed to the value $0.75$.  This strategy
is motivated by the following considerations.

In Fig. \ref{fig:mulresult:de2} we plot the detection efficiencies
as a function of $\beta$ for four different values of $t_{\rm
FWHM}$.  In each of the panels, the efficiencies are integrated over
position within a single chip of the INT fields.  The top (bottom)
panels are for the south-east chip of the north (south) field.  The
right (left) panels are for bright (faint) source stars.  The general
trend is for the detection efficiency to increase with increasing
$t_{\rm FWHM}$ and decreasing $\beta$.  This trend is expected
since longer duration events are more likely to be observed near the
peak and smaller values of $\beta$ imply larger amplification factors.
For $F(r)=10\,{\rm ADU\,s^{-1}}$, $t_{\rm FWHM}\ge 10\,{\rm days}$ and
small $\beta\la 0.7$, the detection efficiencies decrease with
decreasing $\beta$.  The decrease is more severe for the $t_{\rm
FWHM}=50$ day events where the detection efficiency actually drops
below that for the $t_{\rm FWHM}=10$ day events.  The problem may be
that we underestimate the photometric error at high fluxes therefore
causing $\chi^2$ to be systematically high.  Moreover, $50$ days is a
substantial fraction of the observing season and therefore some long
duration events may not meet the requirement that the peak be entirely
within a single season.

\begin{figure*}
\centering
\includegraphics[width=12cm,clip=]{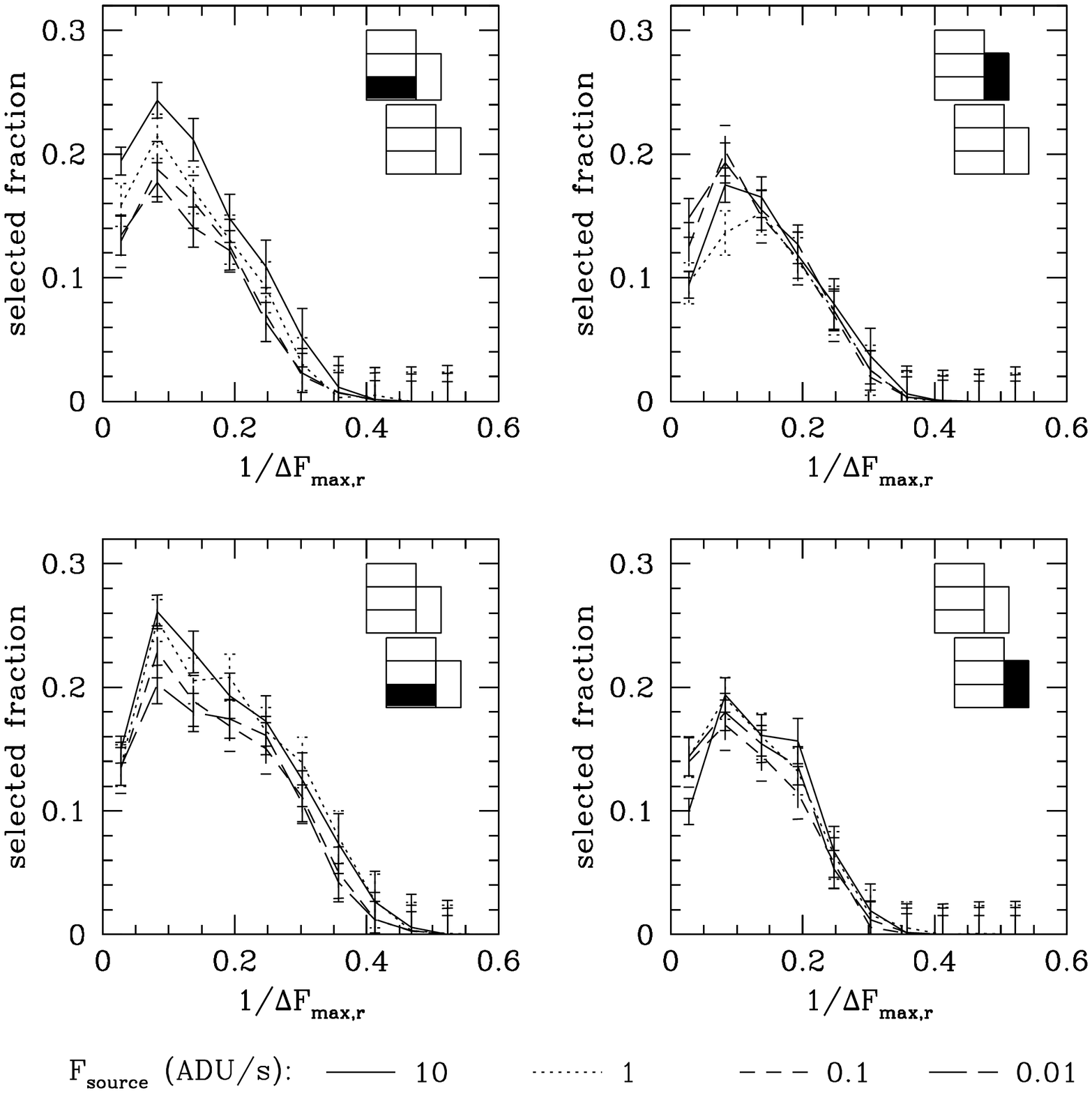}
\caption{
Detection efficiencies as a function of $1/\Delta F_{\rm max}$ for
$t_{\rm FWHM}$ = 50 days and $F_{\rm 0,r}$ = 10 ADU\,s$^{-1}$ (solid
line), 1 ADU\,s$^{-1}$ (dotted line), 0.1 ADU\,s$^{-1}$ (long-dashed),
and 0.01 ADU\,s$^{-1}$ (short-dashed line). In general the
lines overlap within the errors.
}
\label{fig:mulresult:de3}
\end{figure*}

Since the shape of the microlensing lightcurve does not depend
strongly on $\beta$ we expect no significant dependence of the
detection efficiency on the intrinsic source brightness.  This point
is illustrated in Fig. \ref{fig:mulresult:de3} where we plot the
detection efficiencies as a function of $1/\Delta F_{\rm max}$ for
events with $t_{\rm FWHM}=50$ days.  We integrate the efficiencies
over positions within single CCDs and show the results for four of
the eight CCDs in our fields.  The curves vary by at most 30\% over
three orders of magnitude in $F(r)$.  The implication is that an
explicit $F(r)$ dependence in the detection efficiency will not change
the results significantly.

\begin{figure}
\centering
\includegraphics[width=8cm]{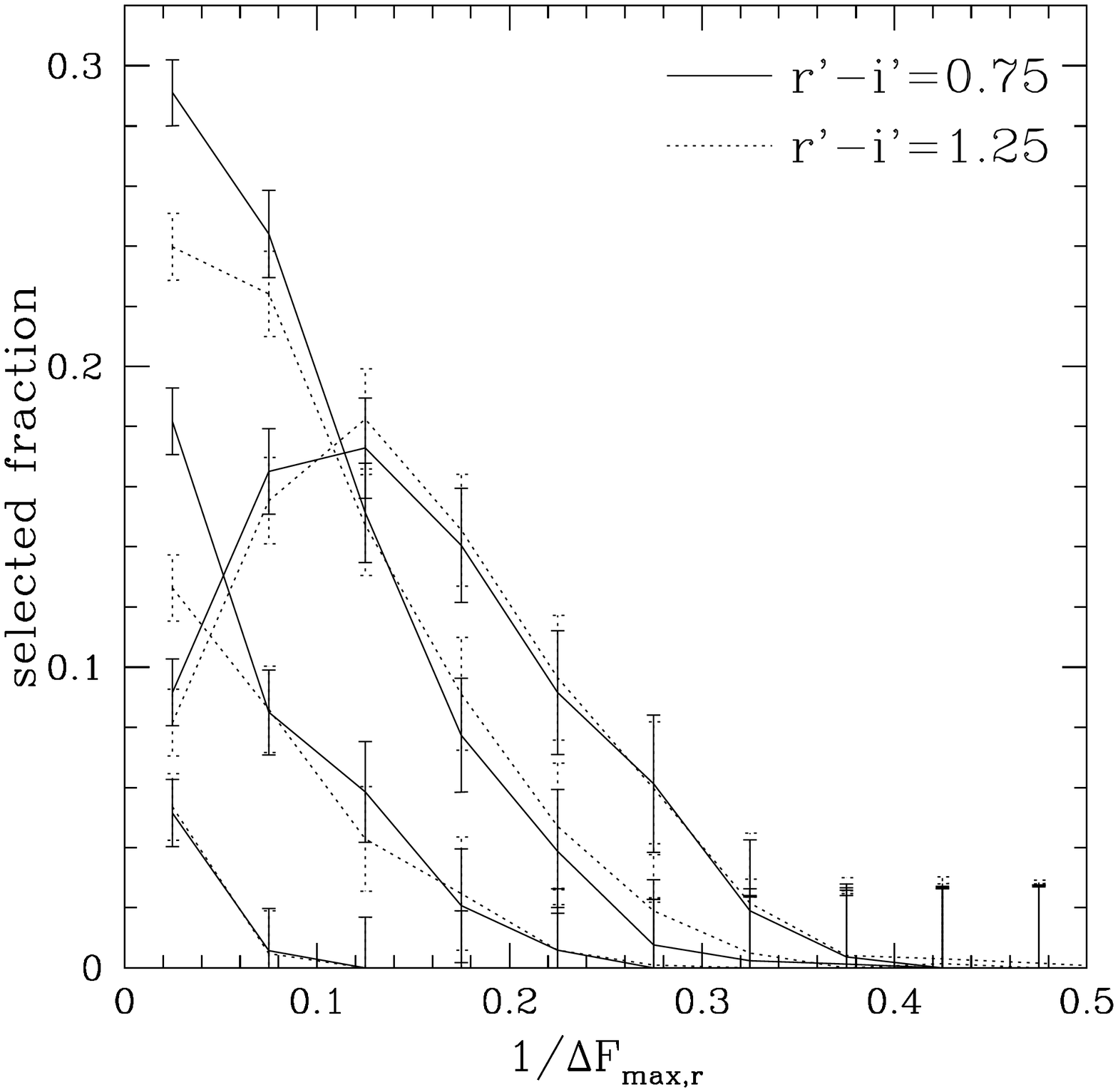}
\caption{Colour dependence of the detection efficiency. For
$t_{\rm FWHM}$~'s of 1, 3, 10 and 50 days the detection efficiencies are
shown for the 2 different source colours simulated. The colour has no
noticeable effect, except for the highest signal-to-noise events.}
\label{fig:mulresult:de4}
\end{figure}

We next test whether the detection efficiency depends on the colour
$\mathcal{C}$ of the source.  In addition to the main artificial event
catalogue, we generate artificial events with $\mathcal{C}=1.25$ and
r$^{\prime}$ unchanged for a part of the north field.  Fig.
\ref{fig:mulresult:de4} compares the detection efficiencies for the
two colours and shows that there is no significant difference, except
for the very highest signal to noise events.  The discrepancy at high
S/N reflects the problem discussed above with our estimates of the
photometric errors at high flux.  This problem is worse for redder
sources which have a higher i$^{\prime}$-band flux.

\begin{figure}
\centering
\includegraphics[width=8cm]{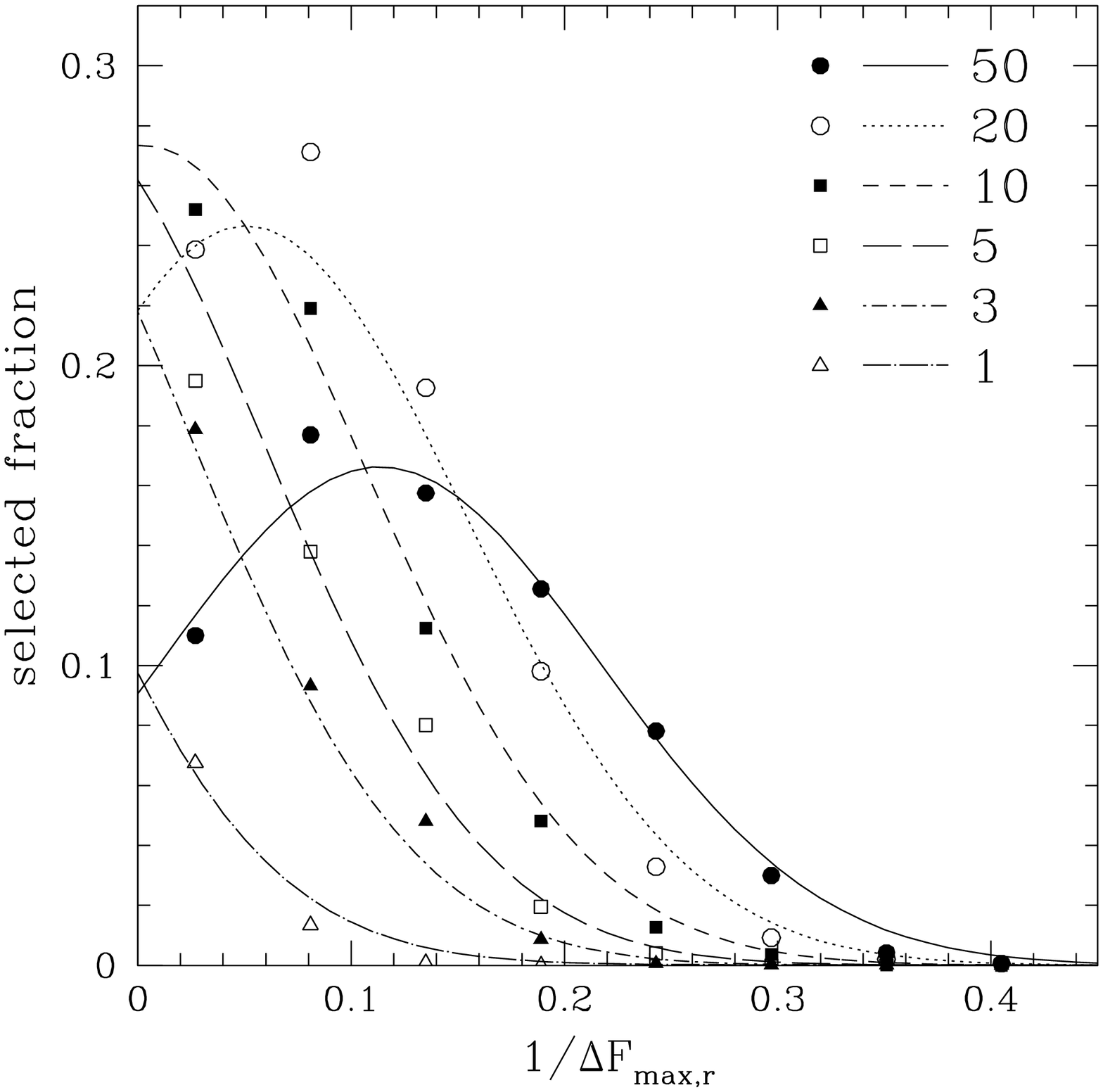}
\caption{
Detection efficiencies as a function of $1/\Delta F_{\rm max}$ for
different values of $t_{\rm FWHM}$.  The symbols give the results
of the Monte Carlo calculation for one chip.  The lines correspond
to the fitting formula, equation \ref{eq:mulresult:fitfunc}.}
\label{fig:mulresult:de5}
\end{figure}

Motivated by the shapes of the curves in Fig.
\ref{fig:mulresult:de4}, we choose a Gaussian in $1/\Delta F_{\rm max}$
where the position of the peak depends on $t_{\rm FWHM}$.  The
explicit functional form is taken to be:

\begin{equation}
\label{eq:mulresult:fitfunc}
\epsilon ~=~ c_1 \left (1-t_{\rm FWHM}/112\right )
e^{-c_2(1/\Delta F_{\rm max}-c_3)^2}
\end{equation}
\noindent where
\begin{equation}
\label{eq:mulresult:fitfuncb}
c_3 ~=~ d_1 \cdot \ln(t_{\rm FWHM}) ~+~ d_2~.
\end{equation}

\noindent The factor multiplying the Gaussian takes into account the
sharp decrease in detection efficiency for events with duration
comparable to or longer than the observing season.  The parameters
$c_1$, $c_2$, $d_1$ and $d_2$ are determined by fitting simultaneously
the detection efficiencies for all values of $t_{\rm FWHM}$~ to
equation \ref{eq:mulresult:fitfunc}.  Fig. \ref{fig:mulresult:de5}
shows an example of these fitting formulae to the detection
efficiencies.

Fig. \ref{fig:mulresult:de3} illustrates the dependence of the
detection efficiencies on location in the INT fields.  This dependence
is due mainly to variations in galaxy surface brightness but also to
the presence of bad pixels and saturated-star defects.  As discussed
above, we account for the spatial dependence by fitting the detection
efficiency separately for subregions of the fields.  To be precise, we
divide each chip into 32 subregions, $\sim$3\arcmin$\times$3\arcmin ~in
size.  For each of these regions we average 14\,640 simulated events
(2\,440 per choice of $t_{\rm FWHM}$).

\section{Extinction}
\label{sec:mulresult:extinction}

Microlensing surveys such as MEGA and POINT-AGAPE are motivated, to a
large extent, by the argument that a MACHO population in M31 would
induce a near-far asymmetry in the microlensing event distribution.
In the absence of either extinction or significant intrinsic
asymmetries in the galaxy, the distribution of self-lensing events and
variable stars masquerading as microlensing events would be near-far
symmetric.  The detection of a near-far asymmetry would then provide
compelling evidence in favour of a significant MACHO population.

Recently, \cite{an04} found a near-far asymmetry in the distribution
of variables which they attribute to differential extinction across
the M31 disk. That differential extinction is significant is also
witnessed by several dust features including two prominent dust lanes
on the near side of the disk.

We construct a simple model for differential extinction in M31 and
test it to against the distribution of LPVs.  In the next section, we
incorporate this extinction model into our calculations for the
theoretical event rate.

\begin{figure}[t]
\centering
\includegraphics[width=2.8cm,angle=-90]{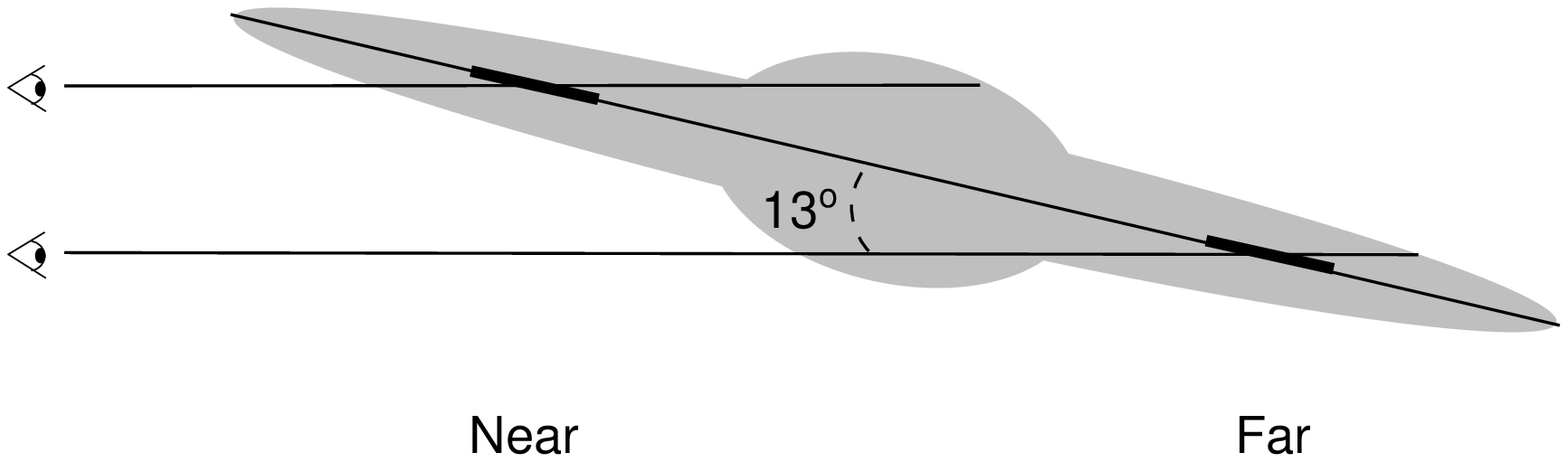}
\caption{Schematic representation of the line-of-sight through the M31
galaxy from an observer on earth. Because of the high inclination of
M31, most of the light observed on the near side of the disk is coming
from behind the dust lanes.}
\label{fig:mulresult:xfraccartoon}
\end{figure}

Following \cite{WK88} we assume that the dust is located in a thin
layer in the mid-plane of the disk.  Along a given line-of-sight, only
light from behind the dust layer is absorbed.  Because of the galaxy's
high inclination, the fraction of stars located behind the dust layer
is higher for lines-of-sight on the near side of the disk than for
those on the far side, as illustrated in Fig.
\ref{fig:mulresult:xfraccartoon}.  Therefore, even if the distribution
of dust is intrinsically symmetric, extinction will have a greater
effect on the near side of the disk.

Based on these assumptions the observed intensity along a particular
line-of-sight is
\begin{equation}
\label{eq:mulresult:extinction1}
\mathcal{I}_{\rm obs} ~=~ \mathcal{I}_{\rm front} ~+~ \mathcal{I}_{\rm
  back} e^{-\tau}
\end{equation}
\noindent where $\mathcal{I}_{\rm front}$ ($\mathcal{I}_{\rm back}$) is the
intensity of light originating from in front of (behind) the dust
layer and $\tau$ is the optical depth.  This equation can be rewritten
in terms of the total intrinsic intensity, $\mathcal{I}_{\rm intr}$, and
the fraction $x$ of light that originates from in front of the dust
layer:
\begin{equation}
\label{eq:mulresult:extinction2}
\mathcal{I}_{\rm obs} ~=~ x \mathcal{I}_{\rm intr} ~+~
(1-x)\mathcal{I}_{\rm intr}e^{-\tau}~.
\end{equation}
The three unknowns in this equation, $\mathcal{I}_{\rm intr}$, $x$, and
$e^{-\tau}$, depend on wavelength.  Rewriting equation
\ref{eq:mulresult:extinction2} for the B-band we have
\begin{equation}
\label{eq:mulresult:extinction3}
e^{-\tau_B} ~=~ \frac{\mathcal{I}_{\rm obs}(B)/
\mathcal{I}_{\rm intr}(B) - x_B}{1-x_B}.
\end{equation}
As a first approximation we assume that $\mathcal{I}_{\rm obs}(I)
=\mathcal{I}_{\rm intr}(I)$ so that
\begin{equation}
e^{-\tau_B} ~=~ \frac{\mathcal{I}_{\rm obs}(B)/
(\mathcal{C}_{BI} \cdot \mathcal{I}_{\rm obs}(I)) ~-~ x}{1-x}
\label{eq:mulresult:taub}
\end{equation}
\noindent where $\mathcal{C}_{BI}\equiv
\mathcal{I}_{\rm intr}(B)/\mathcal{I}_{\rm intr}(I)$ is the intrinsic $I-B$
colour of the stellar population.  An improved estimate of
$\mathcal{I}_{\rm intr}(I)$ is obtained by transforming the extinction
factor from B to I via the standard reddening law
\citep{savagemathis}.  The calculation is repeated several times

We approximate $x_B$ and $x_I$ from a simple model of the galaxy
wherein the intrinsic (i.e., three-dimensional) light distribution
$\eta\left ({\bf x}\right )$ for the disk and bulge are taken to
be double exponentials.  In cylindrical coordinates for M31, we
have

\begin{equation}
\label{eq:mulresult:lightdistribution}
\eta^i\left ({\bf x}\right ) = \eta_0
e^{-r/h_R^i} e^{-z/h_z^i}
\end{equation}
where the superscript $i$ denotes either the disk or bulge, $\eta_0$
is a normalisation constant, and $h_R$ and $h_z$ are the radial and
vertical scale lengths, respectively.  Different scale lengths are
used for B and I because the two bands have different sensitivities to
young and old populations of stars.  Young stars tend to lie closer to
the disk mid-plane than old ones.  Our choices for the parameters are
given in Table \ref{tab:mulresult:xmodels}.  The values of the disk
scale lengths and the bulge-to-disk-ratios are taken from \cite{WK88}.
The scale lengths for bulge are adapted from their de Vaucouleurs fit
while the disk scale heights are based on the distribution of
different stellar populations in the Milky Way disk.  The observables
$\mathcal{I}_{\rm obs}(I)$ and $\mathcal{I}_{\rm obs}(B)$ are from
\cite{guhathakurta} who cover a 1.7$\degr\times$5$\degr$ field
centred on M31.  We derive colour profiles from their mosaics which
are found to be similar to the profiles in \cite{WK88}.  The colour is
approximately constant within 30$\arcsec$ and becomes bluer at larger
radii.

\begin{table}[t]
\centering
\caption{Disk and bulge parameters used to derive $x$, the fraction of
  light originating in front of the mid-plane of M31: the scale length
  and scale height, $h_l$ and $h_z$, for disk and bulge, and the
  fraction of the total light coming from the bulge.
}
\begin{tabular}{c|ccccc}
\hline
\hline
~ & Disk & ~ & Bulge & ~ & $L_{\rm b}/(L_{\rm b} + L_{\rm d})$\\
~ & $h_l$ (kpc) & $h_z$ (kpc) & $h_l$ (kpc) & $h_z$ (kpc) & ~\\
\hline
B & 5.8 & 0.3 & 1.2 & 0.75 & 0.39\\
I & 5.0 & 0.7 & 1.2 & 0.75 & 0.45\\
\hline
\end{tabular}
\begin{small}
\label{tab:mulresult:xmodels}
\end{small}
\end{table}

\begin{figure*}
\centering
\includegraphics[width=12cm,clip=]{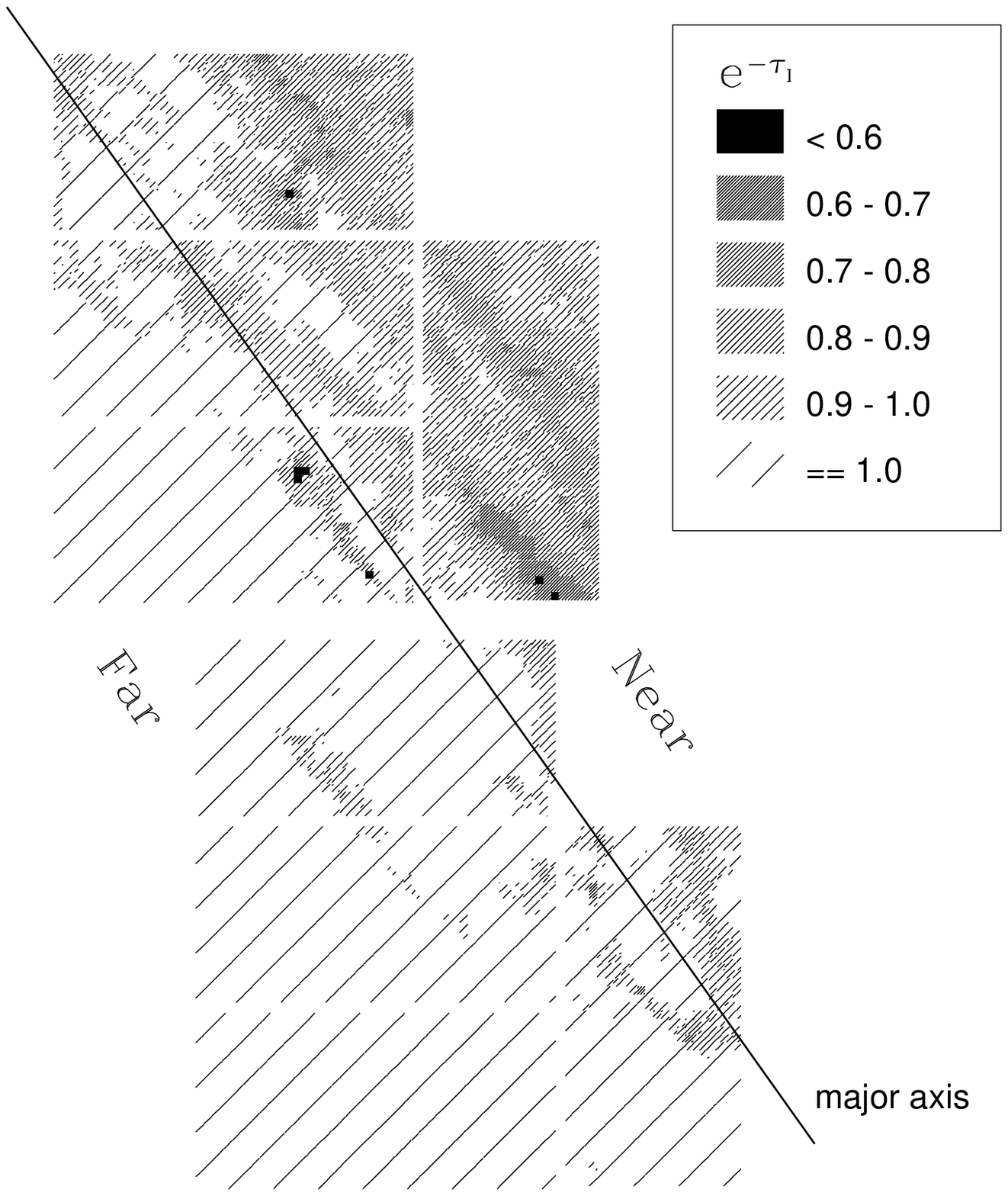}
\caption{
Calculated extinction map in the I-band.  Extinction is
clearly more severe on the near side of the disk.  Note that there
are only a few small patches where the extinction factor rises above
40\%.}
\label{fig:mulresult:extinction}
\end{figure*}

Our I-band extinction map for M31 is shown in Fig.
\ref{fig:mulresult:extinction}.  The major dust lanes are clearly
visible in the northern field and, as expected, the derived extinction
is much larger on the near side of the galaxy than on the far side.
The I-band attenuation is $< 40\%$ and reaches a maximum in the
innermost dust lane and a few smaller complexes.

Our model almost certainly underestimates the effect of extinction
across the M31 disk.  The approximation $\mathcal{I}_{\rm obs}(I)
\simeq \mathcal{I}_{\rm intr}(I)$ is a poor starting point in the
limit of large optical depths.  For $\tau\gg 1$, most of the light in
both B and I from behind the dust layer is absorbed and therefore
$\mathcal{I}_{\rm obs}(B)/ \mathcal{I}_{\rm obs}(I)\simeq
\mathcal{C}_{BI}$.  However substituting this result into equation
\ref{eq:mulresult:extinction3} gives $\exp{\left (-\tau\right )}\simeq
1$, an obvious contradiction.  By the same token, if the dust is
distributed in high-$\tau$ clumps, then $I$ and $B$ wavelengths will
be absorbed by equal amounts given essentially by the geometric cross
section of the clumps.  Moreover, the thin-layer approximation tends
to yield an underestimate of the extinction factor \citep{WK88}.  Finally,
scattering increases the flux observed towards the dust lanes and
therefore also leads one to underestimate the extinction factor.
Some of these problems can be solved by using infrared data in the
construction of the extinction map. In a future paper we plan to use
2MASS data in order to derive a more accurate model for differential
extinction in M31.

\begin{figure}
\centering
\includegraphics[width=8cm,clip=]{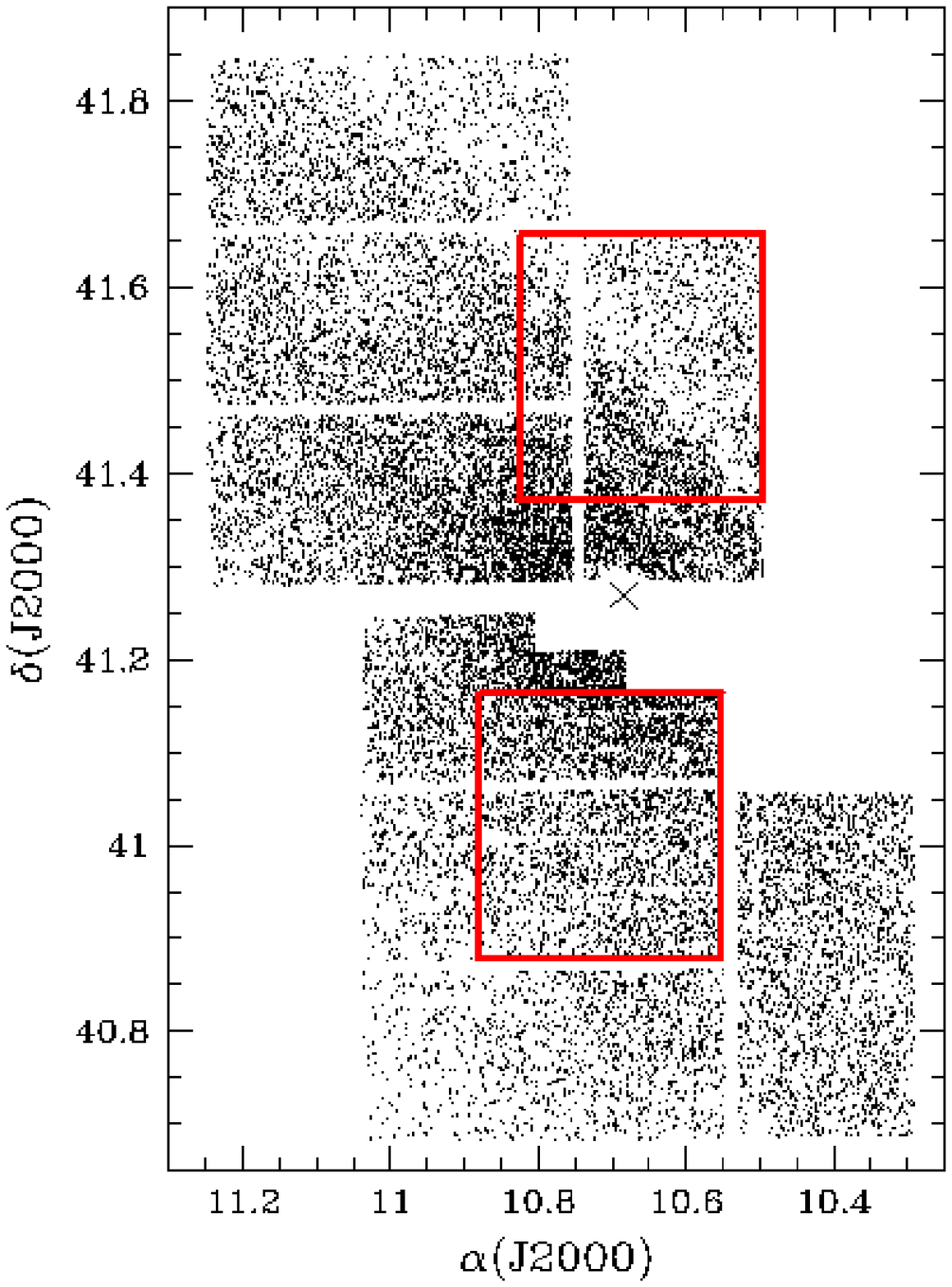}
\caption{The distribution of the LPVs in M31 with the two
  symmetrically placed regions used for the LPV amplitude analysis
  indicated. The northern field is located on the near side and
  contains some of the most heavily extincted parts, the southern
  field is on the far side and hardly affected by extinction. These
  regions are similar to N2 and S2 regions from \cite{an04}, only
  adjusted to avoid the part of the southern INT field that is not
  used in our analysis.  }
\label{fig:mulresult:symregions}
\end{figure}

We can use the distribution of variable stars in our survey to test
and refine the extinction model.  The underlying assumption of this
exercise is that the intrinsic distribution of variables is the same
on the near and far sides of the disk.  We begin by determining the
periods of the variable stars using a multi-harmonic periodogram
\citep{czerny} suitably modified to allow for unevenly sampled data.
A six-term Fourier series is then fit to each lightcurve yielding
additional information such as the amplitude of the flux variations.
Only variables with lightcurves that are well-fit by the Fourier
series are used.

\begin{figure}
\centering \includegraphics[width=8cm]{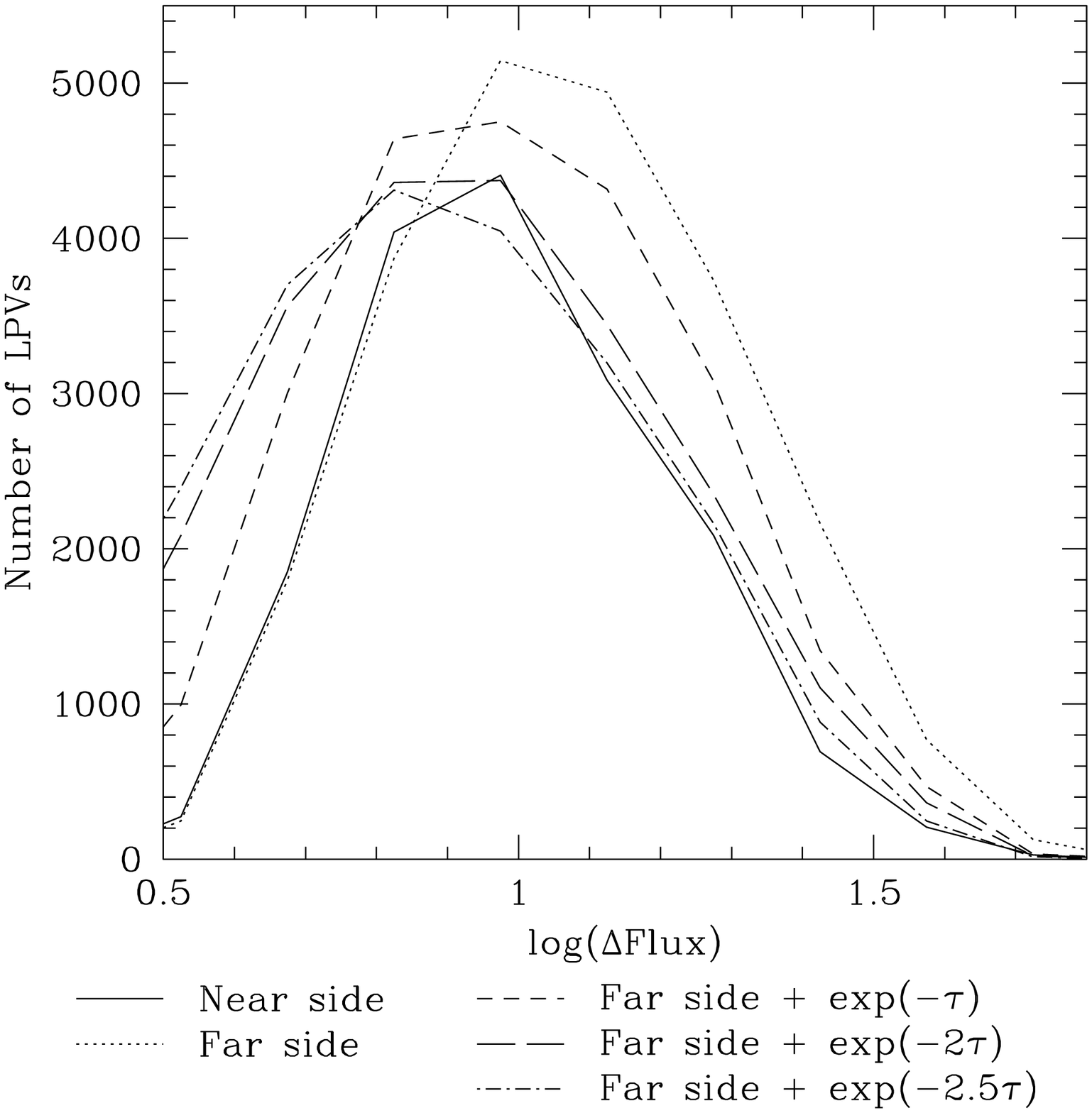}
\caption{Luminosity functions of LPVs in the 2 symmetrically placed
regions. The far side flux distributions were scaled slightly to
correct for small differences in area due to the gaps between the
CCDs. The solid line is for the near side region and the dotted for
the uncorrected far side region. The short-dashed, long-dashed, and
dot-dashed lines are far side distributions corrected for increasing levels of
extinction.  }
\label{fig:mulresult:lpvlf}
\end{figure}

We will use LPVs to test the extinction model because they generally
belong to quite old stellar populations. This is an advantage because
the majority of the microlensing source stars also belong to older
populations which are more smoothly distributed over the galaxy than
younger variables such as Cepheids.  We select LPVs with periods
between 150 and 650 days and focus on two regions of our INT
fields. One of these is located on the near-side of the disk where
extinction is expected to be high while the other is located
symmetrically about the M31 centre on the far side.  Fig.
\ref{fig:mulresult:symregions} shows the spatial distribution of the
LPVs.  Since extinction reduces the amplitude of the flux variations
and the average flux by the same factor we can study extinction by
comparing the distributions in $\Delta F$ for the near and far sides.
These flux variation distributions are shown in
Fig. \ref{fig:mulresult:lpvlf}.  For low $\Delta F$, where the shapes
of the distributions are dominated by the detection efficiency,
results for the near and far side agree.  For high $\Delta F$, where
the detection efficiency for variables approaches 100\%, one finds a
large discrepancy between the near and far-side distributions.

To test whether this discrepancy is indeed due to extinction we
transform the coordinates of LPVs on the far side to their mirror
image on the near side.  The amplitude of the flux variation is then
reduced by the model extinction factor suitably transformed from I to
r$^{\prime}$ \citep{savagemathis}.  The new distribution, shown in
Fig.  \ref{fig:mulresult:lpvlf}, is still significantly above the
near-side distribution at large $\Delta F$ though it does provide a
better match than the original far-side distribution.  The implication
is that our model underestimates extinction.  To explore this point
further we consider models in which $\tau$ is replaced by $c\tau$
where $c>1$.  In Fig. \ref{fig:mulresult:lpvlf}, we show the
distributions of the far side LPVs for $\tau\to 2\tau$ (long-dashed
line) and $\tau\to 2.5\tau$ (dot-dashed line).  Apparently, the bright
end of the (mirror) far-side distribution with $\tau$ increased by a
factor of 2.5 agrees with the bright end of the near-side
distribution.  We therefore conclude that our original model does
indeed underestimate the effects of extinction. In some places this
will be stronger than in others, but over the probed region the model
underestimates extinction effectively by perhaps a factor of 2.5 in
$\tau$.

\section{Theoretical predictions}
\label{sec:mulresult:models}

The detection efficiencies found in Sect.
\ref{sec:mulresult:efficiency} allow us to predict the number and
distribution of events given a specific model for the galaxy.  Though
M31 is one of the best studied galaxies, a number of the parameters
crucial for microlensing calculations, are not well-known.  Chief
among these are the mass-to-light ratios of the disk and bulge, $\mld$
and $\mlb$, respectively.  The light distributions for these
components are constrained by the surface brightness profile while the
mass distributions of the disk, bulge, and halo are constrained by the
rotation curve and line-of-sight velocity dispersion profile.
However, the mass-to-light ratios are poorly constrained primarily
because the shapes of the disk and halo contributions to the rotation
curve are similar \citep[e.g.][]{albada85}.  One can compensate for an
increase in $\mld$ by decreasing the overall density of the halo.
Stellar synthesis models \citep{belldejong}, combined with
observations of the colour profile of M31, can be used to constrain
the mass-to-light ratios though these models come with their own
internal scatter and assumptions.  Another poorly constrained
parameter is the thickness of the disk which affects the disk-disk
self-lensing rate.

In this section we describe theoretical calculations for the expected
number of events in the MEGA-INT survey.  We consider a suite of M31
models which span a wide range of values in $\mld$ and $\mlb$.  The
dependence of the microlensing rate on other parameters is also
explored.

\subsection{Self-consistent models of M31}
\label{subsec:mulresult:models}

The standard practice for modelling disk galaxies is to choose simple
functional forms for the space density of the disk, bulge, and halo
tuned to fit observational data.  For microlensing calculations, velocity
distributions are also required.  Typically, one assumes that the
velocity distribution for each of the components is isotropic,
isothermal, and Maxwellian with a dispersion given by the depth of the
gravitational potential or, in the case of the bulge, the observed
line-of-sight velocity dispersion.  (But see \cite{kerins01} where the
effects of velocity anisotropy are discussed.)  This approach can lead
to a variety of problems.  First, these ``mass models'' do not
necessarily represent equilibrium configurations, that is,
self-consistent solutions to the collisionless Boltzmann and Poisson
equations.  A system initially specified by the model may well relax
to a very different state.  Another issue concerns dynamical
instability.  Self-gravitating rotationally supported disks form
strong bars.  This instability may be weaker or absent altogether if
the disk is supported, at least in part, by the bulge and/or halo.
Therefore, models with very high $\mld$ are the most susceptible to
bar formation and can be ruled out.

In order to overcome these difficulties we use new, multi-component
models for disk galaxies developed by \cite{m31models}.  The models
assume axisymmetry and incorporate an exponential disk, a Hernquist
model bulge \citep{hernquist90}, and an NFW halo \citep{nfw}.  They
represent self-consistent equilibrium solutions to the coupled Poisson
and collisionless Boltzmann equations and are generated using the
approach described in \cite{kdmodels}.

The phase-space distribution functions (DFs) for the disk, bulge, and
halo ($f_{\rm disk},\, f_{\rm bulge},\,$ and $f_{\rm halo}$
respectively) are chosen analytic functions of the integrals of
motion.  For the axisymmetric and time-independent system considered
here, the angular momentum about the symmetry axis, $J_z$, and the
energy, $E$, are integrals of motion.  \cite{m31models} assume that
$f_{\rm halo}$ depends only on the energy while $f_{\rm bulge}$
incorporates a $J_z$-dependence into the Hernquist model DF to allow
for rotation.  For both halo and bulge, the DFs are ``lowered'' as
with the King model \citep{king66} so that the density goes to zero at a
finite ``truncation'' radius.  The disk DF is a function of $E$,
$J_z$, and an approximate third integral of motion, $E_z$, which
corresponds to the energy associated with vertical motions of stars in
the disk \citep{kdmodels}.

Self-consistency requires that the space density, $\rho$, and
gravitational potential, $\psi$, satisfy the following two equations:

\begin{equation}\label{eq:mulresult:densityfromDF}
\rho = \int d^3 v\left (f_{\rm disk} + f_{\rm bulge} + f_{\rm halo}\right )
\end{equation}

\noindent and

\begin{equation}\label{eq:mulresult:potentialfromdensity}
\nabla^2\psi = 4\pi G\rho~.
\end{equation}

\noindent Self-consistency is achieved through an iterative scheme and
spherical harmonic expansion of $\rho$ and $\psi$.  Straightforward
techniques allow one to generate an N-body representation suitable for
pseudo-observations of the type described below.  The N-body
representations also provide very clean initial conditions for
numerical simulations of bar formation and disk warping and heating.

The DFs are described by 15 parameters which can be tuned to fit a
wide range of observations.  In addition, one must specify
mass-to-light ratios if photometric data is used.  Our strategy is to
compare pseudo-observations of M31 with actual observational data to
yield a $\chi^2$-statistic.  Minimisation of $\chi^2$ over the model
parameter space -- performed in \cite{m31models} by the downhill
simplex method \citep[see e.g.][]{numrec} -- leads to a best-fit
model.

Following \cite{m31models} (see, also \citet{WPS} who carried out a
similar exercise with the original \citet{kdmodels} models) we utilise
measurements of the surface brightness profile, rotation curve, and
inner (that is, bulge region) velocity profiles.  We use R-band
surface brightness profiles for the major and minor axes from
\cite{WK88}.  (\cite{m31models} used the global surface brightness
profile from \cite{WK88} which was obtained by averaging the light
distribution in elliptical rings.  The use here of both major and
minor axis profiles should yield a more faithful bulge-disk
decomposition.)  The theoretical profiles are corrected for internal
extinction using the model described in the previous section.  In
addition, a correction for Galactic extinction is included.  We assume
photometric errors of $0.2\,{\rm mag}$.  We use a composite rotation
curve constructed from observations by \cite{kent89} and
\cite{braun91} that run from 2 to 25 kpc in galactocentric radius.
Values and error bars for the circular speed are obtained at intervals
of $10\,{\rm arcmin}\simeq 2.2\,{\rm kpc}$ using kernal smoothing
\citep{WPS}.  Finally, we use kinematic measurements from
\cite{mcelroy83} to constrain the dynamics in the innermost part of
the galaxy.  We smooth his data along the minor axis to give values
for the line-of-sight stellar rotation and velocity dispersion at
$0.5\,{\rm kpc}$ and $1.0\,{\rm kpc}$.  The values at these radii are
insensitive to the effects of a central supermassive object and
reflect the dynamics of the bulge stars with little disk contamination
\citep{mcelroy83}.  An overall $\chi^2$ for the model is calculated by
combining results from the three types of data.  Photometric and
kinematic data are given equal weight; the circular rotation curve
measurements are weighted more heavily than the bulge velocity and
dispersion measurements.  To be precise, we use

\begin{equation}
\label{eq:mulresult:chisquare}
\chi^2 = \frac{1}{\sqrt{2}}
\left (\chi^2_{\rm sbp} + \frac{1}{3}\chi^2_{\rm disp} + 
\frac{2}{3}\chi^2_{\rm rc}\right)
\end{equation}

\noindent where $\chi^2_{\rm sbp}$, $\chi^2_{\rm bulge}$, and
$\chi^2_{\rm rc}$ are the individual $\chi^2$-statistics for the
photometric, bulge kinematics, and rotation curve measurements.

\begin{figure}
\centering
\includegraphics[width=7cm]{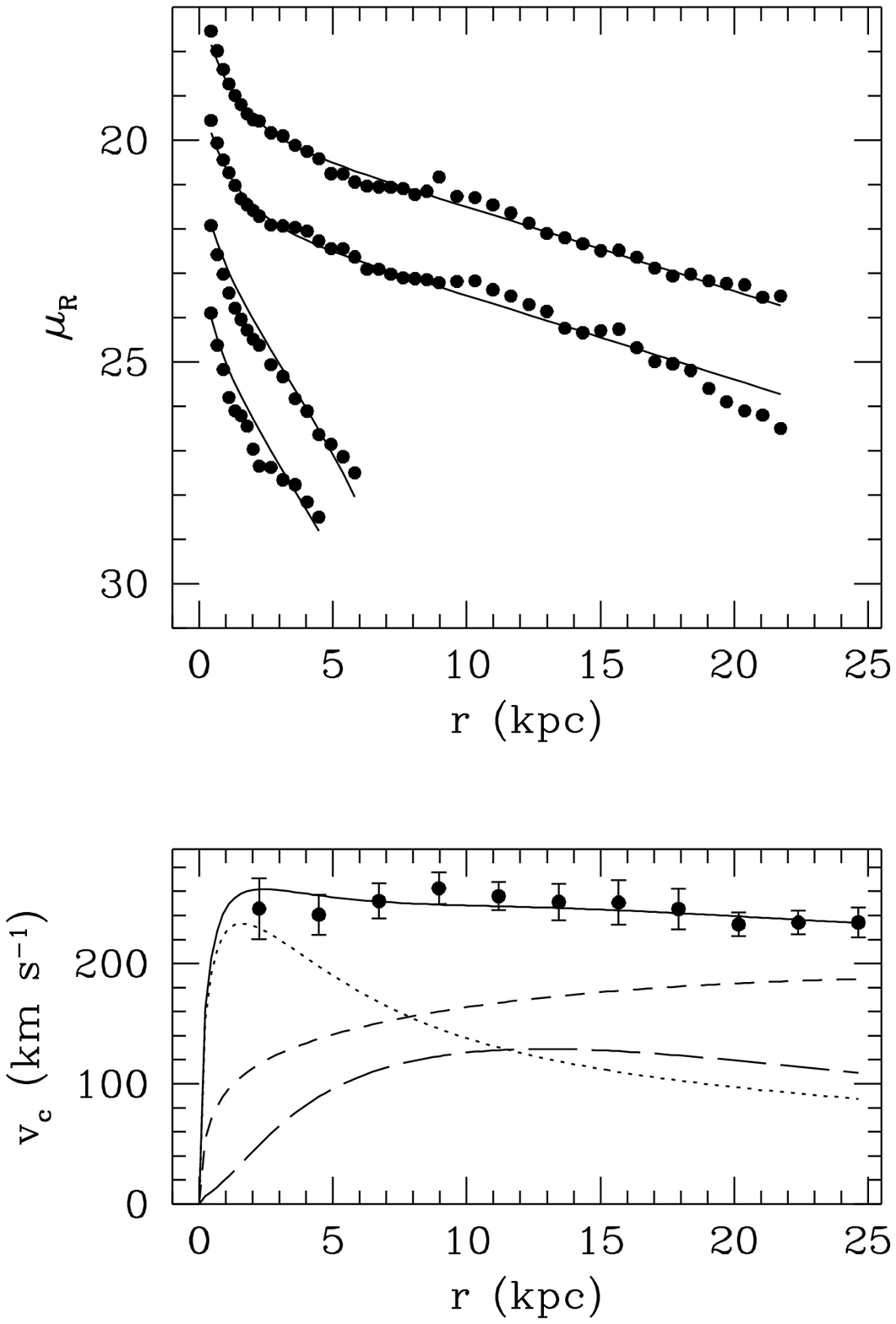}
\caption{Comparison of pseudo-observations of model A1 to real
  observations. {\it Upper panel:} model surface brightness profiles
  (solid lines) along the major and minor axis compared to observations by
  \cite{WK88} (dots). For clarity the profiles are shifted down in steps of 2
  magnitudes. From the top down the profiles correspond to: SW major
  axis, NE major axis, SE minor axis (far side), and NW minor axis
  (near side). {\it Lower panel:} model rotation curve (solid line)
  and combined rotation curve from \cite{kent89} and
  \cite{braun91}. The three lower lines correspond to the
  contributions to the rotation curve of the bulge (dotted), disk
  (long dash) and halo (short dash).
}
\label{fig:mulresult:modela1}
\end{figure}

Our reference model (model A1) is constructed with $\left (M/L\right
)_{\rm d} = 2.4$ and $\left (M/L\right )_{\rm b} = 3.6$. These values
are motivated by the stellar population synthesis models of
\citet{belldejong}.  Along the far side of the
minor axis, where the surface brightness profile is relatively free of
extinction, the $B-R$ colour is $1.8$ in the bulge region and $1.6$ in
the disk region \cite{WK88}.  A correction for Galactic extinction
brings these numbers down by $0.18$.  Substituting into the
appropriate formula from Table 1 of \citet{belldejong} yield the
mass-to-light ratios chosen for this model.
In Fig. \ref{fig:mulresult:modela1} we compare
predictions for model A1 with observations.  Shown are the surface
brightness profiles along major and minor axes and the circular
rotation curve.  Not shown is the excellent agreement between model
and observations for the stellar rotation and dispersion measurements
in the bulge region.  The reduced $\chi^2$ statistic for this model is
$1.06$ (see Table \ref{tab:mulresult:modresults}).

In model A1, the scale height of the disk was fixed to a value of
$1.0\,{\rm kpc}$.  Note that our model uses a ${\rm sech}^2$-law for
the vertical structure of the disk.  A ${\rm sech}^2$-scale height of
$1\,{\rm kpc}$ is roughly equivalent to an exponential scale height of
$0.5-0.7\,{\rm kpc}$.  The observations used in this study do not
provide a tight constraint on the scale height of the disk and so we
appeal to observations of edge-on disk galaxies.  \citet{kregel02}
studied correlations between the (exponential) vertical scale height
and other structural parameters such as the radial scale height and
asymptotic circular speed in a sample of 34 edge-on spirals.  Using
these correlations we arrive at an exponential scale height for $M31$
of $0.6\,{\rm kpc}$ with a fairly large scatter.

We also fix the disk truncation radius for this model to $28\,{\rm
kpc}$ which is at the high end of the range favoured in
\citet{kregel02}.  Lower values appear to be inconsistent with the
measured surface brightness profile.  The remaining parameters for the
disk, bulge, and halo DFs are varied in order to minimise $\chi^2$.

Table \ref{tab:mulresult:modresults} outlines other models considered
in this paper.  Models B1-E1 explore the $\left (M/L\right )_{\rm
b}-\left (M/L\right )_{\rm d}$ plane.  The $\chi^2$ for these models
are generally quite low, a reflection of the model degeneracy
mentioned above.  In these models, disk and bulge ``mass'' are traded
off against halo mass.  Previous investigations \citep{m31models}
suggest that model E1 is unstable to the formation of a strong bar
while the other models are stable against bar formation or perhaps
allow for a weak bar.

The aforementioned models used values for the extinction factor
derived in Sect. \ref{sec:mulresult:extinction}.  As discussed in that
section, there are a number of reasons to expect that this model
underestimates the amount of extinction in M31.  Indeed, our analysis
of the near-far asymmetry in LPVs favours a higher optical depth by a
factor of $2.5$, that is, the substitution $e^{-\tau}\to
e^{-2.5\tau}$.  For this reason, we consider a parallel sequence of
models, A2-F2, with high extinction.  Note that the $\chi^2$ for these
models are as good as if not better than those for the corresponding
low-extinction models.

\subsection{Event rate calculation}
\label{subsec:mulresult:eventrate}

The event rate is calculated by performing integrals over the lens and
source distribution functions.  The rate for lenses to enter the
lensing tube of a single source is
\begin{equation}
d ^5R ~=~ \frac{f_l(l_l,{\bf v}_l)}{\mathcal{M}_l}~ 2 R_E v_{\perp}~ dl_l d{\bf
  v}_l~ d\beta
\label{eq:mulresult:eventrate1}
\end{equation}
where $f_l$ is the DF for the lens population, $l_l$ is the
observer-lens distance ($D_{\rm OL}$ in the notation of equation
\ref{eq:mulresult:r_e}), $v_{\perp}$ is the transverse velocity of the
lens with respect to the observer-source line-of-sight, and
$\mathcal{M}_l$ is the mass of the lens.  In writing this equation, we
assume all lenses have the same mass.

For a distribution of sources described by the DF $f_s$, equation
\ref{eq:mulresult:eventrate1} is replaced by the following expression
for the rate per unit solid angle
\begin{eqnarray}
\frac{dR}{d\Omega} ~=~ \int \frac{f_l(l_l,{\bf v}_l)}{\mathcal{M}_l}
  \frac{f_s(l_s,{\bf v}_s)}{\left (M/L\right )_sL_s}~ 2 R_E v_{\perp}
\nonumber\\~
 \times  dl_l d{\bf v}_l~ l_s^2~ dl_s d{\bf v}_s~ d\beta
\label{eq:mulresult:eventrate2}
\end{eqnarray}
where $l_s$ is the observer-source distance, $\left (M/L\right )_s$ is
the mass-to-light ratio of the source and $L_s$ is the source luminosity.
(For the moment, we treat all sources as being identical.)

We perform the integrals using a Monte Carlo method.  
The DFs are sampled at discrete points: 
\begin{equation}
f_p (l_p,{\bf v}_p) ~=~ 
\frac{\Sigma_p}{N_p}\sum_{i=1}^{N_p} \delta(l_p - l_i)~ 
\delta({\bf v}_p - {\bf v}_i)
\label{eq:mulresult:eventrate3}
\end{equation}
where $p\in \{l,s\}$, $\Sigma_p$ is the surface density of either lens
or source population, and $N_p$ is the number of points used to Monte
Carlo either lens or source populations.  The nine-dimensional
integral in equation \ref{eq:mulresult:eventrate2} is replaced by a
double sum and an integral over $\beta$:
\begin{equation}
\frac{dR}{d\Omega} ~=~ \mathcal{S}_{sl} \sum_{i,j} \int_0^{\beta_u} d\beta \mathcal{R}_{ij}
\label{eq:mulresult:eventrate4}
\end{equation}
where 
\begin{equation}
\label{eq:mulresult:def1}
\mathcal{S}_{sl} = \frac{\Sigma_l\Sigma_s}
{N_l\mathcal{M}_l N_s L_s\left (M/L\right )_s}
\end{equation}
and 
\begin{equation}
\label{eq:mulresult:def2}
\mathcal{R}_{ij} ~\equiv~ (2 R_E v_{\perp})_{ij} l^2_{j}~.
\end{equation}
Note that $\mathcal{S}$ depends on the line of sight densities of the
lens and source distributions along with characteristics of the two
populations.  $\mathcal{R}_{ij}$ depends on the coordinates and
velocities of the lens and source (hence the $ij$ subscripts).  The
sum is restricted to lens-source pairs with $l_l<l_s$.  For each
lens-source pair, the Einstein crossing time, $t_{{\rm E},ij}$ is
easily calculated.  The differential event rate is then

\begin{equation}
\frac{d^2R}{d\Omega dt_E} ~=~ \mathcal{S}_{sl}
\sum_{i,j} \int_0^{\beta_{u}} d\beta \mathcal{R}_{ij}
\delta(t_{{\rm E},ij}-t_{\rm E})~.
\label{eq:mulresult:eventrate5}
\end{equation}

\subsection{Stellar and MACHO populations}

The formulae in the previous section apply to the six lens-source
combinations in our model: disk-disk, disk-bulge, bulge-disk,
bulge-bulge, halo-disk, and halo-bulge.  As written the formulae
assume homogeneous populations.  For the disk and bulge populations,
we modify equation \ref{eq:mulresult:eventrate5} to include integrals
over the mass and luminosity functions as appropriate.  We write the
luminosity function (LF) as
\begin{equation}
\frac{dN}{dM_R} ~=~ A g(M_R)~
\label{eq:mulresult:eventrate7}
\end{equation}
and the mass function as
\begin{equation}
\frac{dN}{d\mathcal{M}} ~=~ B h(\mathcal{M},\mathcal{M}_0)
\label{eq:mulresult:eventrate8}
\end{equation}
where $A$ and $B$ are normalisation constants and $\mathcal{M}_0$ is
the lower bound for the mass function (MF).  We take the function $g$
from \cite{mamonsoneira} and the function $h$ from
\citet{binneymerrifield} (their equation 5.16) with the power-law form
$dN/d\mathcal{M}\propto \mathcal{M}^{-1.8}$ extended to
$\mathcal{M}_0$.  $A$ and $B$ are evaluated separately for the disk
and bulge populations.  In the case of the disk, we assume that $30\%$
of the mass is in the form of gas.  The LF is normalised to give
$\overline{L}=L_\odot$ with the proviso that $L_s$ in equation
\ref{eq:mulresult:eventrate2} is given in solar units.  To determine
the normalisation constant $B$ of the mass function, we
write
\begin{equation}
B h(\mathcal{M}_\odot,\mathcal{M}_0) ~=~ \left .
\left (\frac{dN}{dM_V}\frac{dM_V}{d\mathcal{M}}\right )
\right |_{\mathcal{M}=\mathcal{M}_\odot}
\label{eq:mulresult:eventrate9}
\end{equation}
where the V-band LF is again from \cite{mamonsoneira} and
$dM_V/d\mathcal{M}$ is from \cite{kroupa93}.  Equation
\ref{eq:mulresult:eventrate9} is evaluated at solar values for
convenience.  The relation
\begin{equation}
\left(\frac{M}{L} \right)_R ~=~ \frac{\int B h(\mathcal{M},\mathcal{M}_0)
\mathcal{M} d\mathcal{M}}
{\int A g(M_R) L(M_R) dM_R}
\label{eq:mulresult:eventrate10}
\end{equation}
can then be solved for $\mathcal{M}_0$.  Thus, a disk with high $M/L$
contains more low-mass stars than a disk with low $M/L$.

For simplicity, and because we lack a model for what MACHOs actually
are, we assume all MACHOs have the same mass, $\mathcal{M}_{\rm M}$ that is

\begin{equation}
\label{eq:mulresult:machomass}
\frac{dN}{d\mathcal{M}} = \delta \left
(\mathcal{M}-\mathcal{M}_M\right)~.
\end{equation}

The value of $\mathcal{M}_{\rm M}$ will directly determine the
  number density of MACHOs for a given halo mass density. Since the
  MACHOs only provide lenses and no sources for microlensing, a higher
  value of $\mathcal{M}_{\rm M}$ and thus a lower number density,
  will result in a lower number of microlensing events. A given value
  of $\mathcal{M}_{\rm M}$ can be considered as the average mass of a
  more elaborate MACHO mass function.

\subsection{Theoretical prediction for the number of events}

Recall that the efficiency $\epsilon$ is written as a function of
$t_{\rm FWHM}$ and $\Delta F_{\rm max}$.  (The efficiency also depends
on the line of sight.)  These quantities are explicit functions of
$\beta$, $F_r$, and $t_{\rm E}$.  Thus, the expected number of events per
unit solid angle is

\begin{eqnarray}
\label{eq:mulresult:expected}
\frac{d\mathcal{E}}{d\Omega} = E\,A\,B\mathcal{S}_{ls}
\sum_{i,j}\int_0^{\beta_u} d\beta\int dM_R
g\left (M_R\right )\nonumber \\
\times \int d\mathcal{M}_l h\left (\mathcal{M},\mathcal{M}_0\right )
\mathcal{R}_{ij}\epsilon\left (t_{\rm FWHM},\,\Delta F\right )
\end{eqnarray}
where $E$ is the overall duration of the experiment.  Our survey
covers four half-year seasons and so, with our choice of units for 
$\epsilon$ and $dR/d\Omega$, we have $E=2$.

The number of events expected in each of the 250 bins
used for the extinction calculation and labelled by ``k'' is
\begin{equation}
\label{eq:mulresult:expected2}
\mathcal{E}_k = \Delta\Omega\left (\frac{dE}{d\Omega}\right )_k
\end{equation}
where $\Delta\Omega = 9\,{\rm arcmin}^2$ is the angular area of a bin.
$\mathcal{E}_k$ carries an additional label (suppressed for notational
simplicity) which denotes the lens-source combination.  The total
number of events is $\mathcal{E} = \sum\mathcal{E}_k$.

\subsection{Binary lenses}
\label{subsec:mulresult:binary}

Our microlensing selection criteria are based on the assumption that
the lenses are single point-mass objects.  However, at least half of
all stars are members of multiple star systems.  Microlensing
lightcurves for a lens composed of two or more point masses can
deviate significantly from the standard lightcurve \citep{schneider86}
and may therefore escape detection.  The deviations are strongest when
the source crosses or comes close to the so-called caustics, positions
in the source plane where the magnification factor is formally
infinite.  (The actual magnification factor is finite due to the
finite size of the source.)  The size of the caustic region is largest
when the separation of the components of the lens is comparable to the
Einstein radius corresponding to the total mass (equation
\ref{eq:mulresult:r_e}).  \cite{mao91} estimated that $\sim$10\% of
microlensing events towards the bulge of the Milky Way (mainly
self-lensing events) should show strong binary characteristics such as
caustic crossings.  Since the Einstein radius for bulge-bulge
self-lensing toward the Milky Way and M31 are comparable, we can
expect a similar 10\% effect in our survey. 
\cite{baltzgondolo} perform a similar analysis for
pixel-lensing surveys and estimate that in the order of 6\% of
self-lensing events from normal stellar populations will exhibit
caustic crossings. Since the majority of detected events will have low
signal-to-noise, we can assume that deviations other than caustic
crossings in most cases will not strongly affect our detection
efficiency.
Therefore, to account for binary lenses, the calculated theoretical
predictions for self-lensing are revised downward by $10\%$.

\subsection{Results}
\label{sec:mulresult:modelresults}

Table 5 presents the theoretical predictions for the total number of
events expected in the MEGA-INT four-year survey.  The results are
given for both self-lensing ($\mathcal{E}_{\rm self}$) and halo
lensing ($\mathcal{E}_{\rm halo}$).  The values quoted for
$\mathcal{E}_{\rm halo}$ assume 100\% of the halo is in the form of
MACHOs.  In other words, these values should be multiplied by the
MACHO halo fraction in order to get the expected number of events for
a MACHO component. We note that lensing by the Milky Way halo is not
included in these results. This possible contribution is expected to
be small, since the number of microlensing events from a 100\% MW halo
is a few times lower than for a 100\% M31 halo
\citep{gyukcrotts00,baillon93} for MACHO masses around
0.5$\mbox{M$_{\scriptsize \odot}$}$.

\begin{figure}
\centering
\includegraphics[width=8cm]{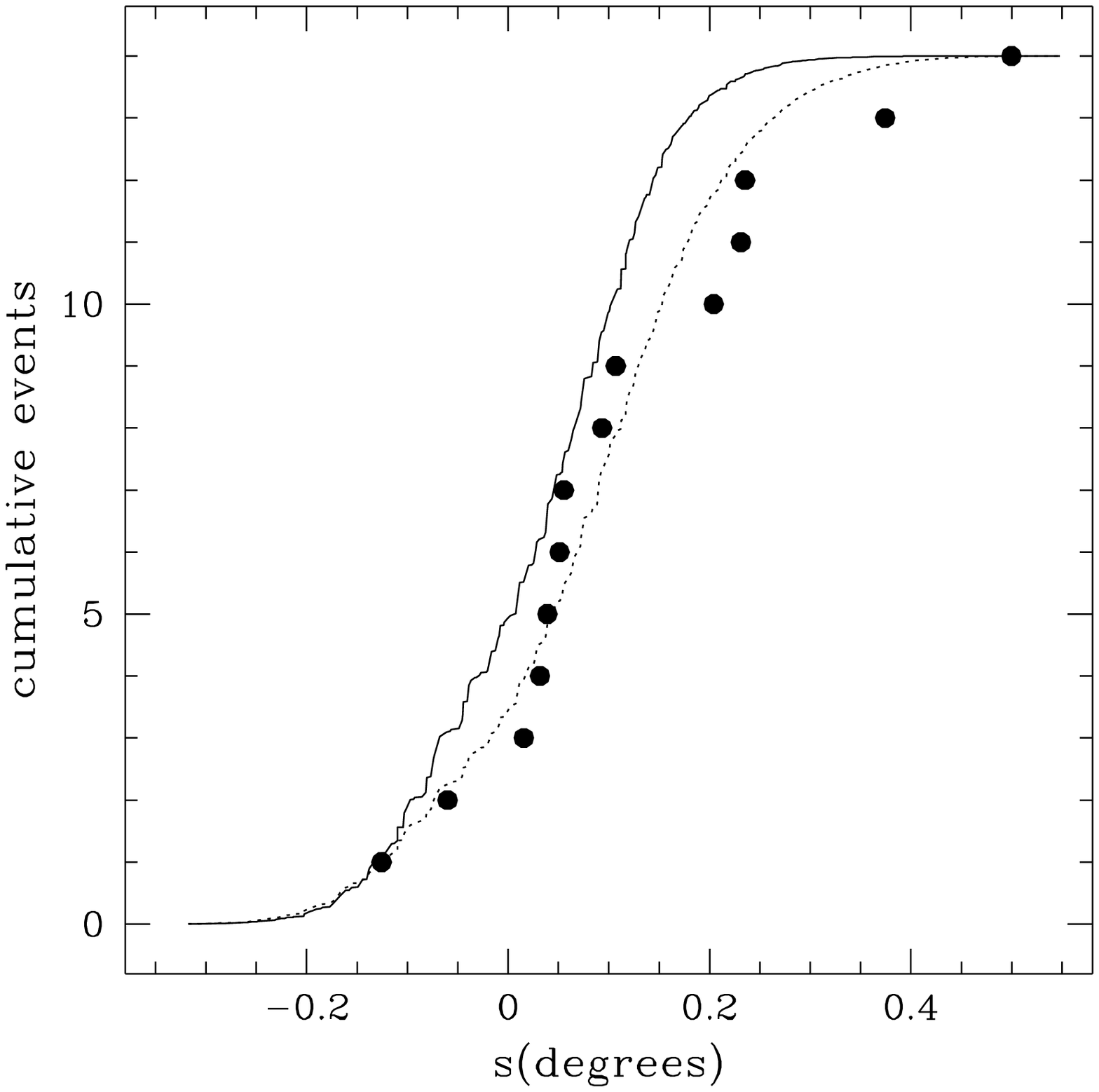}
\caption{Cumulative event distribution as a function of distance
  from the major axis (in degrees).  Shown are the 
  data (dots), self-lensing distribution (solid line), and halo-lensing
  distribution (dotted line). Both self- and halo-lensing lines are
  scaled to give a total of 14 events.}
\label{fig:mulresult:asymmetryplot}
\end{figure}

We also consider the near-far asymmetry for self and halo lensing.  In
Fig. \ref{fig:mulresult:asymmetryplot}, we show the cumulative
distribution of events for self and halo lensing as a function of the
distance from the major axis, $s$.  We take $s$ to be positive on the
far side of the disk.  For this plot, we choose model A1 but since the
distributions are normalised to give 14 total events, the difference
between the models is rather inconsequential.  We see that both self
and halo lensing models do a good job of describing the event
distribution in the inner $0.2^\circ$.  The halo distribution
does a somewhat better job of modelling the three events between
$s=0.2^\circ$ and $s=0.3^\circ$.  Neither halo nor self lensing models
predict anywhere near two events for $s>0.35^\circ$.

To further explore the distribution, we define the asymmetry parameter
$\mathcal{A}$:
\begin{equation}
\mathcal{A} ~=~ \frac{\sum \mathcal{E}_k \cdot s_k}{\mathcal{E}}~.
\end{equation}
In Table 5 we give values for $\mathcal{A}_{\rm self}$ and
$\mathcal{A}_{\rm halo}$.  We also provide an average
$\mathcal{A}_{\rm ave}$ which assumes that MACHOs make up the shortfall
between the expected number of events and the observed value of 14.
In cases where the expected number of events is greater than 14, we
set $\mathcal{A}_{\rm ave} = \mathcal{A}_{\rm self}$.  The asymmetry
parameter for the 14 candidate events is $\mathcal{A}_{\rm data} =
0.125$.

\begin{table*}
\centering
\caption{Results of the microlensing modelling using self-consistent
M31 models. In the first columns some model parameters and the
combined $\chi^2$ are listed. The remaining columns contain the
predicted number of events due to self-lensing ($\mathcal{E}_{self}$),
due to halo-lensing ($\mathcal{E}_{halo}$), the asymmetry of the
self-lensing ($\mathcal{A}_{self}$), of the halo-lensing
($\mathcal{A}_{halo}$), and of the combination of both
($\mathcal{A}_{ave}$). The number of self-lensing events
$\mathcal{E}_{self}$ has been corrected for the fact that $\sim$10\%
of the events will show strong binary effects and therefore be
selected against.  The microlensing event rate due to the halo
$\mathcal{E}_{halo}$ is for a 100\% MACHO halo, i.e. all of the halo
mass is assumed to be in the MACHOs. For calculating the combined
self- and halo-lensing asymmetry parameter $\mathcal{A}_{ave}$ a
smaller fraction of the halo mass is assumed to be in MACHOs, namely
the amount necessary to make up the difference, if any, between
$\mathcal{E}_{self}$ and the observed number of 14 candidate events.
The disk scale heights $h_z$ are $sech^2$ scale heights. The upper,
low extinction part of the table contains models with internal
extinction values as derived in section \ref{sec:mulresult:extinction},
while the lower, high extinction part contains models with increased 
extinction, as motivated by our analysis of the LPV amplitudes.}
\begin{tabular}{l|ccc|cc|ccc}
\multicolumn{9}{c}{{\bf Low extinction}}\\
\hline
\hline
\multicolumn{9}{l}{Models with $m_\textsc{macho}$=0.5 $\mbox{M$_{\scriptsize \odot}$}$ and
  $h_z$=1.0 kpc}\\
\hline
~ & $(M/L)_{\rm d}$ & $(M/L)_{\rm b}$ & $\chi^2$ & $\mathcal{E}_{\rm self}$ & 
$\mathcal{E}_{\rm halo}$ &  $\mathcal{A}_{\rm self}$ & $\mathcal{A}_{\rm halo}$ &
$\mathcal{A}_{\rm ave}$ \\
\hline
A1 & 2.4 & 3.6 & 1.06 & 14.2 & 30.9 & 0.037 & 0.086 & 0.037 \\
\hline
B1 & 2.4 & 2.9 & 1.17 & 13.4 & 31.5 & 0.031 & 0.085 & 0.033 \\
C1 & 2.4 & 4.3 & 1.02 & 13.1 & 29.6 & 0.039 & 0.092 & 0.043 \\
\hline
D1 & 1.8 & 2.4 & 1.34 & 11.3 & 35.5 & 0.031 & 0.082 & 0.041 \\
E1 & 3.6 & 4.4 & 1.03 & 15.8 & 24.6 & 0.030 & 0.091 & 0.030 \\
\hline
\multicolumn{9}{l}{~}\\
\multicolumn{9}{l}{Models with $(M/L)_{\rm d}$=2.4 and $(M/L)_{\rm b}$=3.6}\\
\hline
~ & $h_z$ & $\mathcal{M}_{\rm M}$ & $\chi^2$ & $\mathcal{E}_{\rm self}$ &
$\mathcal{E}_{\rm halo}$ &  $\mathcal{A}_{\rm self}$ & $\mathcal{A}_{\rm halo}$ &
$\mathcal{A}_{\rm ave}$ \\
\hline
F1 & 0.5 & 0.5 & 1.10 & 12.5 & 30.7 & 0.037 & 0.084 & 0.042 \\
\hline
G1 & 1.0 & 0.1 & 1.06 & 14.2 & 43.1 & 0.037 & 0.088 & 0.037 \\
H1 & 1.0 & 1.0 & 1.06 & 14.2 & 25.9 & 0.037 & 0.085 & 0.037 \\
\hline
\multicolumn{9}{l}{~}\\
\multicolumn{9}{l}{~}\\
\multicolumn{9}{c}{{\bf High extinction}}\\
\hline
\hline
\multicolumn{9}{l}{Models with $\mathcal{M}_{\rm M}=0.5 M_{\odot}$ and
  $h_z$=1.0 kpc}\\
\hline
~ & $(M/L)_{\rm d}$ & $(M/L)_{\rm b}$ & $\chi^2$ & $\mathcal{E}_{\rm self}$ &
$\mathcal{E}_{\rm halo}$ &  $\mathcal{A}_{\rm self}$ & $\mathcal{A}_{\rm halo}$ &
$\mathcal{A}_{\rm ave}$ \\
\hline
A2 & 2.4 & 3.6 & 0.99 & 12.4 & 28.6 & 0.052 & 0.095 & 0.057 \\
\hline
B2 & 2.4 & 2.9 & 1.08 & 12.2 & 32.6 & 0.046 & 0.094 & 0.052 \\
C2 & 2.4 & 4.3 & 0.99 & 14.5 & 29.6 & 0.056 & 0.098 & 0.056 \\
\hline
D2 & 1.8 & 2.4 & 1.23 & 10.3 & 34.5 & 0.045 & 0.095 & 0.058 \\
E2 & 3.6 & 4.4 & 1.04 & 14.2 & 22.8 & 0.046 & 0.105 & 0.046 \\
\hline
\multicolumn{9}{l}{~}\\
\multicolumn{9}{l}{Models with $(M/L)_{\rm d}$=2.4 and $(M/L)_{\rm b}$=3.6}\\
\hline
~ & $h_z$ & $\mathcal{M}_{\rm M}$ & $\chi^2$ & $\mathcal{E}_{\rm self}$ &
$\mathcal{E}_{\rm halo}$ &  $\mathcal{A}_{\rm self}$ & $\mathcal{A}_{\rm halo}$ &
$\mathcal{A}_{\rm ave}$ \\
\hline
F2 & 0.5 & 0.5 & 1.06 & 11.2 & 30.5 & 0.052 & 0.095 & 0.061 \\
\hline
G2 & 1.0 & 0.1 & 0.99 & 12.4 & 39.1 & 0.052 & 0.098 & 0.057 \\
H2 & 1.0 & 1.0 & 0.99 & 12.4 & 23.8 & 0.052 & 0.093 & 0.057 \\
\hline
\end{tabular}
\label{tab:mulresult:modresults}
\end{table*}

The general trend, in terms of total expected number of events, is
that as the mass-to-light ratios are increased, $\mathcal{E}_{\rm
self}$ increases and $\mathcal{E}_{\rm halo}$ decreases.  There are
counter examples.  In model C1, the $\left (M/L\right )_{\rm b}$ (as
compared with model A1) leads to a less massive disk and lower
$\mathcal{E}_{\rm self}$.  Recall that for each choice of
mass-to-light ratios, the remaining parameters are adjusted to
minimise $\chi^2$.  The process can lead to rather complicated
interdependencies between the model parameters.  The self-lensing rate
decreases with decreasing $h_z$ as illustrated with model F1.  The
self-lensing rate is generally reduced in the high extinction models
relative to the low extinction ones.  Finally we see that the halo
event rate decreases with increasing MACHO mass.  Models G and H
illustrate this point and span the range in $\mathcal{M}_M$ identified
by \cite{macho5.7} as the most probable mass range for Milky Way
MACHOs.

\begin{figure}
\centering
\includegraphics[width=8cm]{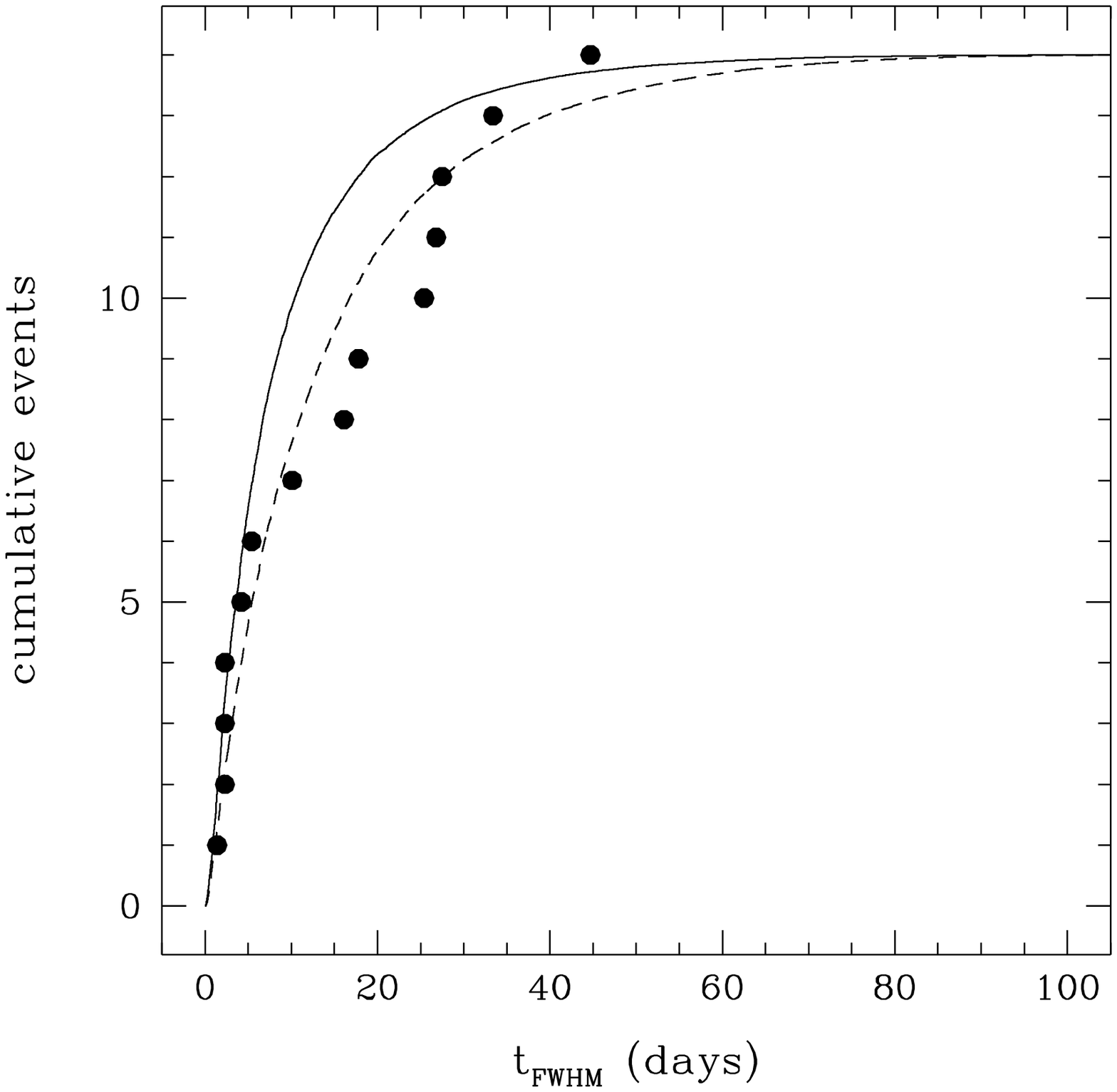}
\caption{Cumulative microlensing event distribution as a function of timescale.
The line and point-types are the same as in Fig. \ref{fig:mulresult:asymmetryplot}.}
\label{fig:mulresult:timescales}
\end{figure}

The timescale distribution is easily calculated using the method
outlined in the previous section.  Essentially, one calculates $t_{\rm
FWHM}$ for each lens-source pair in the Monte Carlo sum.  In Fig.
\ref{fig:mulresult:timescales} we show the cumulative timescale
distribution of our candidate microlensing event sample and model A1.
In constructing the curves for self and halo lensing, we have scaled the
distributions to give a total of 14 events.

\section{Discussion}
\label{sec:mulresult:discussion}

The numbers expected for events due to self-lensing across the models
probed in Table \ref{tab:mulresult:modresults} fall within the narrow
range of 10-16.  The relative insensitivity of $\mathcal{E}_{\rm
self}$ to changes in the mass-to-light ratios is a result of our
approach to constructing models; changes in $\left (M/L\right )_{\rm
b}$ and $\left (M/L\right )_{\rm d}$ are compensated by changes in the
structural parameters of the disk, bulge, and halo so as to minimise
$\chi^2$ for the fit to the rotation curve and surface brightness
data.  Consider models D1 and E1.  The mass-to-light ratios differ by
a factor of $\sim 2$ while $\mathcal{E}_{\rm self}$ differs by only a
factor of 1.4; with the low $M/L$ values in model D1, the rotation curve
data drive up the disk and bulge luminosity distributions at the
expense of a poorer fit for the photometric data.  A balance is struck
and the net result is that the change in $\mathcal{E}_{\rm self}$ is
significantly smaller than what one might expect.

The consistency of the number of candidate events with the number of
predicted self-lensing events is contrary to the results of the
analysis of the first three seasons of INT data by the POINT-AGAPE
collaboration. \cite{calchi05} present six high quality, short
duration microlensing candidates with one of these events attributed
to M32-M31 lensing. They also model the detection
efficiency and calculate number of expected self- and halo-lensing
events for a variety of M31 models.  In all of their models, the
number of events for self-lensing is predicted to be less than $\sim
1.5$.  Since this number is significantly less than the observed
number, they conclude that some of the events are due to MACHOs and
estimate that the MACHO halo fraction is at least 20\%.

\cite{calchi05} use the model from \cite{kerins01} which features
a bulge following \cite{kent89}, an exponential $sech^2$ disk and a
spherical, nearly isothermal halo. They use the same structural
parameters for the three components as \cite{kerins01} but take
$(M/L_B)_b=3$ and $(M/L)_d=4$.  This model for the stellar mass
distribution in M31 predicts an inner rotation curve that is
significantly lower than the observed one, and so an extra `dark
bulge' component is required as well as the isothermal
halo. \cite{calchi05} do not consider microlensing by this dark bulge
in their model, but instead attribute all surplus microlensing to the
halo. In our model the stellar bulge is more massive, with $M/L$ that
is sufficient to reproduce the inner rotation curve, and there is no
non-lensing dark bulge component. It appears to provide sufficient
microlensing events to explain the observations.

Furthermore, the choice of $0.3\,{\rm kpc}$ for the $sech^2$ scale
height is small by perhaps a factor of 3 if M31 is a typical spiral
galaxy as represented in the survey by \cite{kregel02}.  Thickening
the disk increases the disk-disk self-lensing rate.

For our models, the number of events due to self-lensing is consistent
with the total number of events observed but not inconsistent with a
significant MACHO fraction for the halo of M31.  We can make this
statement more quantitative by treating halo events as a Poisson
process with background due to self-lensing and employing the approach
of \cite{feldmancousins}.  We let $n$ be the number of observed
events consisting of MACHO events with mean $f\mathcal{E}_{\rm halo}$,
where $f$ is the MACHO fraction, and a background due to self-lensing
with known mean $\mathcal{E}_{\rm self}$.  For this analysis, we ignore
the background due to variables and background supernovae.  The
probability distribution function is
\begin{equation}
\label{eq:mulresult:probability}
P\left (n|f\right ) = \left (f\mathcal{E}_{\rm halo}+\mathcal{E}_{\rm self}
\right )^n\exp\left [-\left (f\mathcal{E}_{\rm halo}+\mathcal{E}_{\rm self}
\right )\right ]/n!~.
\end{equation}

To obtain confidence intervals for $f$:
\begin{enumerate}
\item Calculate $P\left (n|f\right )$ for $N$ values of $f\in\{0,1\}$
  and sort from high to low. The maximum of $P$ defines the most
  probable value of $f$. The values of $P$ are normalised so
  that the sum of all sampled values of $P$ is 1.
\item Accept values of $f$ starting from the highest value of $P$
  until the sum of $P$ exceeds the desired confidence level.  The
  largest and smallest values of accepted $f$ define the confidence
  interval.
\end{enumerate}

In Table \ref{tab:mulresult:confidenceintervals} we provide most
probable values of $f$ and 95\% confidence intervals for all of the
models in Table \ref{tab:mulresult:modresults}. We provide these
values both for the case of the full sample of 14 observed candidate
events ($n$=14), as well as for the case of 11 observed events
($n$=11), for reasons discussed below.

We next turn to the distribution of events across the M31 disk as
represented by the asymmetry parameters.  From Table
\ref{tab:mulresult:modresults} we see that $\mathcal{A}_{\rm
self}<\mathcal{A}_{\rm halo}<\mathcal{A}_{\rm data}$.  The (weak)
asymmetry in the self-lensing distribution is due to extinction.  Note
that the values are significantly below $\mathcal{A}_{\rm data}$ even
for the high extinction models.

The asymmetry parameter for the halo is significantly higher than that
for self-lensing events and close to, though still below,
$\mathcal{A}_{\rm data}$.  However, the asymmetry parameter for
combinations of self and halo lensing are well below $\mathcal{A}_{\rm
data}$.  Evidently, the distribution of candidate events is difficult
to explain with any reasonable combination of self and halo lensing.

The large asymmetry in the data is due, for the most part, to events
11, 13, and 14 (see Table \ref{tab:mulresult:dataasym}).  It is
therefore worth considering alternative explanations for these events.
As argued in \cite{paulin02}, the lens for event 11 likely resides in
M32 and since we have not included M32 in our model, this event should
be removed from the analysis.  Doing so leads to a modest reduction in
$\mathcal{A}_{\rm data}$.

Events 13 and 14 may be more difficult to explain.  For model A1, the
predicted number of self-lensing events with $s>s\left ({\rm
event}\,18\right )$ is $0.005$ while the predicted number of MACHO
events in the same range in $s$ is $0.14f$.  Thus, the probability of
having two events either from self or halo lensing is exceedingly
small, unless the halo fraction is very large.  However, since some
contamination by variable stars of our sample can not be excluded, one
or both of these events may be a variable star.  We note, for example,
that event 13 has the lowest S/N in our sample.  The probability of
having one event for MACHO lensing with $f=0.20$ is $\sim 3\%$, small,
but not vanishingly so.

A closer inspection of the model is also warranted.  Recall that our
models assume axisymmetry whereas M31 exhibits a variety of
non-axisymmetric features such as disk warping.  This point is
illustrated in the isophotal map by \cite{hodgekenn}.  From the map,
one finds that event 13 lies on the B=24 (R=22.6) contour while model
A1 predicts R=23.5.  Thus, the model may in fact underestimate the
surface brightness of the disk by a factor of 2, and hence the disk-disk
self-lensing rate by a factor of 4.  (The reason for the
discrepancy is not completely clear.  The contours on the far side do
appear to be ``boxier'' than those predicted by the model.)

It is interesting to note that events 13 and 14 are coincident with
the location of the giant stellar stream discovered by \cite{ibata01}.
This stream runs across the southern INT field, approximately
perpendicular to the major axis and over M32.  Indeed, M32 may be the
progenitor of the stream \citep{merrett03}.  The average V-band
surface brightness of the stream is $\Sigma_V \approx 30\pm0.5$ mag
arcsec$^{-2}$ \citep{ibata01} but this is measured far from the
projected positions of events 13 and 14.  The surface brightness of
the stream might be significantly higher near the position of M32.
Perhaps the most conservative statement one can make about the stream
is that it is not bright enough to distort the contours near events 13
and 14, that is, it cannot be brighter than the disk at these radii.
The microlensing event rate due to stars in the stream is of course
enhanced relative to the rate for self-lensing by the ratio of the
distance from the stream to the disk and the thickness of the disk,
that is, by a factor of $\sim 20$.  The stream-disk lensing rate might
be further enhanced if the stars in the stream have a large proper
motion relative to the disk.  These arguments suggest that the number
of stream-disk events in the vicinity of M32 might be $0.03-0.1$;
perhaps high enough to explain one event.

\begin{table}
\centering
\caption{Most probable value and 95\% confidence limits for the MACHO
halo fraction $f$ from the \cite{feldmancousins} analysis, for the
full sample and the case without candidate events 11, 13, and 14.}
\begin{tabular}{lcccc}
\hline
\hline
~ & \multicolumn{2}{c}{14 events} & \multicolumn{2}{c}{11 events}\\
model & $f_{\rm best}$ & conf. interval & $f_{\rm best}$ & conf. interval\\
\hline
A1 & 0. & [0,0.28] & 0. & [0.,0.21] \\
B1 & 0.02 & [0,0.29] & 0. & [0.,0.22] \\
C1 & 0.03 & [0,0.32] & 0. & [0.,0.24] \\
D1 & 0.08 & [0,0.30] & 0. & [0.,0.22] \\
E1 & 0. & [0,0.32] & 0. & [0.,0.25] \\
F1 & 0.05 & [0,0.32] & 0. & [0.,0.24] \\
G1 & 0. & [0,0.20] & 0. & [0.,0.15] \\
H1 & 0. & [0,0.34] & 0. & [0.,0.25] \\
\hline
A2 & 0.06 & [0.,0.35] & 0. & [0.,0.25] \\
B2 & 0.06 & [0.,0.31] & 0. & [0.,0.23] \\
C2 & 0. & [0.,0.29] & 0. & [0.,0.22] \\
D2 & 0.11 & [0.,0.33] & 0.02 & [0.,0.24] \\
E2 & 0. & [0.,0.39] & 0. & [0.,0.29] \\
F2 & 0.09 & [0.,0.35] & 0. & [0.,0.26] \\
G2 & 0.04 & [0.,0.25] & 0. & [0.,0.18] \\
H2 & 0.07 & [0.,0.42] & 0. & [0.,0.31] \\
\hline
\end{tabular}
\begin{small}
\label{tab:mulresult:confidenceintervals}
\end{small}
\end{table}

\begin{table}
\centering
\caption{Observed number of events and the asymmetry of their spatial
  distribution, shown for the full sample of 14 events and for cases
  where the probable M32 event (11) and candidate events 13
  and 14 are ignored. The quoted errors are 1$\sigma$ errors,
  determined with the bootstrap method. Also shown is the asymmetry
  for the long-period variable stars (LPVs).
}
\begin{tabular}{lcc}
\hline
\hline
Events used & $\mathcal{E}_{data}$ & $\mathcal{A}_{data}$ \\
\hline
Full sample & 14 & 0.125 $\pm$ 0.046 \\
without 11 & 13 & 0.120 $\pm$ 0.049 \\
without 13, 14 & 12 & 0.076 $\pm$ 0.034 \\
without 11, 13, 14 & 11 & 0.066 $\pm$ 0.034 \\
\hline
LPVs & 20,864 & 0.071 $\pm$ 0.001 \\
\hline
\end{tabular}
\begin{small}
\label{tab:mulresult:dataasym}
\end{small}
\end{table}

\begin{figure}
\centering
\includegraphics[width=8cm]{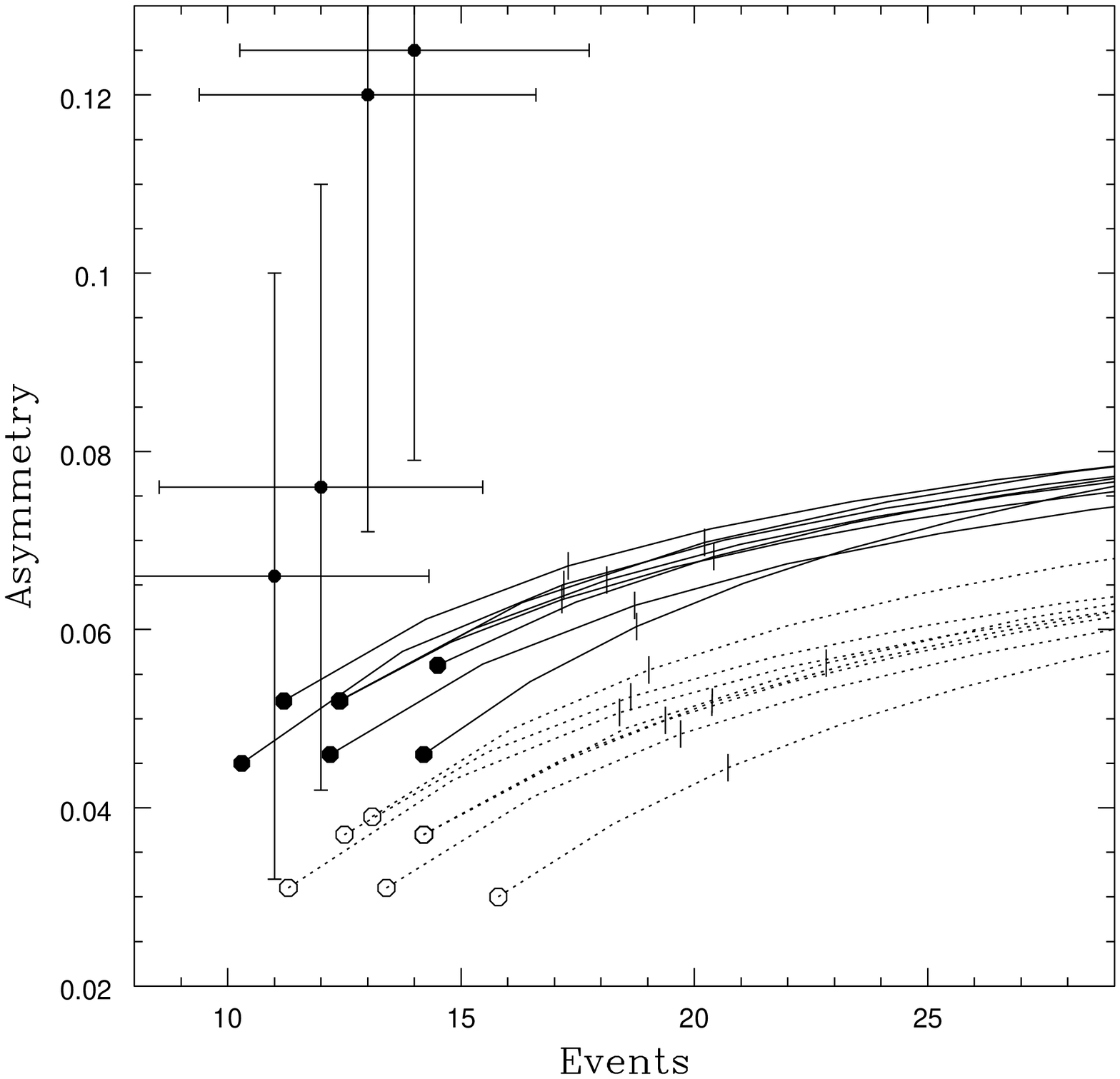}
\caption{Asymmetries and event numbers for data and models. The points
  with error bars on the left show $\mathcal{E}$ and $\mathcal{A}$ for
  the sample of candidate events and the subsamples listed in table
  \ref{tab:mulresult:dataasym}. The solid lines correspond to the high
  and the dotted lines to the low extinction models from table
  \ref{tab:mulresult:modresults}. The dots are the pure self-lensing
  points, with the MACHO mass fraction increasing along the
  line. MACHO fractions of 20\% are indicated with vertical lines.
}
\label{fig:mulresult:resultplot}
\end{figure}

Fig. \ref{fig:mulresult:resultplot} provides a summary of our results
with respect to the expected number of events and the asymmetry
parameter.  The points with error bars represent the data for the 4
cases considered in Table \ref{tab:mulresult:dataasym}.  The solid
circles and lines correspond to the high extinction case; the open
circles and dotted lines correspond to the low extinction case.  The
circles assume pure self lensing while the lines trace out the values
for increasing MACHO fraction with the tick-mark indicating the
position of $f=0.2$.  Once again, we see that the asymmetry parameter
for the data is higher than that for any of the models.  Removing
events 13 and 14 does improve the situation as does increasing the
optical depth $\tau$; the asymmetry remains a little higher but
consistent with the models.

\section{Conclusions}
\label{sec:mulresult:conclusions}

This paper presents the analysis of four
seasons of M31 observations at the INT, a subset of the MEGA
survey of M31.  The observations were carried
out to search for MACHOs in the halo of M31.  Our fully automated
search algorithm identified 14 candidate microlensing events from over
$10^5$ variable sources.  Three of the candidates were previously
unpublished. The spatial and timescale distributions are consistent
with microlensing.

The core of this paper is the comparison of this candidate event
sample with a calculation of the expected number of events from self
and halo lensing.  This calculation breaks into three parts: a model
for the extinction across the M31 disk; a model for the detection
efficiency; and a suite of self-consistent disk-bulge-halo models for
M31.

The results with regard to the fundamental question of whether there
is a significant MACHO fraction in the halo are inconclusive. Based
on the total number of events, we find that the most probable MACHO
halo fraction $f$ varies between $0$ and $0.1$ depending on the model.
Our event rate analysis is consistent with a total absence of MACHOs as the
confidence intervals for all of our models include $f=0$. On the other
hand we can not exclude some MACHO component, since the confidence
intervals extend typically up to $f=0.25$ and even up to $f=0.4$ for a
few models.

The spatial distribution of the candidate events is highly asymmetric
and does seem to favour a MACHO component.  However, for different
reasons it is questionable whether the 3 candidate events that largely
determine the asymmetry signal should be used in this analysis.  Thus,
we conclude that both from the observed number of events, and from
their spatial distribution we find no compelling evidence for the
presence of MACHOs in the halo of M31.

\begin{acknowledgements}
We would like to thank Raja Guhathakurta and Phil Choi for making
their M31 map available to us and Stephane Courteau for useful
conversations.  We also thank all observers who have performed the
observations for this survey and the ING staff. Support is
acknowledged from STScI (GO 10273) and NSF (grants 0406970 and
0070882). JdJ and KK thank the LKBF for travel support.
\end{acknowledgements}

\bibliographystyle{aa}
\bibliography{biblio_thesis.bib}

\begin{thebibliography}{53}
\expandafter\ifx\csname natexlab\endcsname\relax\def\natexlab#1{#1}\fi

\bibitem[{{Afonso} {et~al.}(2003){Afonso}, {Albert}, {Andersen}, {Ansari},
  {Aubourg}, {Bareyre}, {Beaulieu}, {Blanc}, {Charlot}, {Couchot}, {Coutures},
  {Ferlet}, {Fouqu{\' e}}, {Glicenstein}, {Goldman}, {Gould}, {Graff}, {Gros},
  {Haissinski}, {Hamadache}, {de Kat}, {Lasserre}, {Le Guillou}, {Lesquoy},
  {Loup}, {Magneville}, {Marquette}, {Maurice}, {Maury}, {Milsztajn}, {Moniez},
  {Palanque-Delabrouille}, {Perdereau}, {Pr{\' e}vot}, {Rahal}, {Rich},
  {Spiro}, {Tisserand}, {Vidal-Madjar}, {Vigroux}, \& {Zylberajch}}]{afonso03}
{Afonso}, C., {Albert}, J.~N., {Andersen}, J., {et~al.} 2003, A\&A, 400, 951

\bibitem[{{Alcock} {et~al.}(2000){Alcock}, {Allsman}, {Alves}, {Axelrod},
  {Becker}, {Bennett}, {Cook}, {Dalal}, {Drake}, {Freeman}, {Geha}, {Griest},
  {Lehner}, {Marshall}, {Minniti}, {Nelson}, {Peterson}, {Popowski}, {Pratt},
  {Quinn}, {Stubbs}, {Sutherland}, {Tomaney}, {Vandehei}, \&
  {Welch}}]{macho5.7}
{Alcock}, C., {Allsman}, R.~A., {Alves}, D.~R., {et~al.} 2000, ApJ, 542, 281

\bibitem[{{An} {et~al.}(2004{\natexlab{a}}){An}, {Evans}, {Hewett}, {Baillon},
  {Calchi Novati}, {Carr}, {Cr{\' e}z{\' e}}, {Giraud-H{\' e}raud}, {Gould},
  {Jetzer}, {Kaplan}, {Kerins}, {Paulin-Henriksson}, {Smartt}, {Stalin}, \&
  {Tsapras}}]{an04}
{An}, J.~H., {Evans}, N.~W., {Hewett}, P., {et~al.} 2004{\natexlab{a}}, MNRAS,
  351, 1071

\bibitem[{{An} {et~al.}(2004{\natexlab{b}}){An}, {Evans}, {Kerins}, {Baillon},
  {Calchi Novati}, {Carr}, {Cr{\' e}z{\' e}}, {Giraud-H{\' e}raud}, {Gould},
  {Hewett}, {Jetzer}, {Kaplan}, {Paulin-Henriksson}, {Smartt}, {Tsapras}, \&
  {Valls-Gabaud}}]{pa99n2}
{An}, J.~H., {Evans}, N.~W., {Kerins}, E., {et~al.} 2004{\natexlab{b}}, ApJ,
  601, 845

\bibitem[{{An} {et~al.}(2004{\natexlab{c}}){An}, {Evans}, {Kerins}, {Baillon},
  {Calchi Novati}, {Carr}, {Cr{\' e}z{\' e}}, {Giraud-H{\' e}raud}, {Gould},
  {Hewett}, {Jetzer}, {Kaplan}, {Paulin-Henriksson}, {Smartt}, {Tsapras}, \&
  {Valls-Gabaud}}]{event7anomalies}
{An}, J.~H., {Evans}, N.~W., {Kerins}, E., {et~al.} 2004{\natexlab{c}}, ApJ,
  601, 845

\bibitem[{{Baillon} {et~al.}(1993){Baillon}, {Bouquet}, {Giraud-Heraud}, \&
  {Kaplan}}]{baillon93}
{Baillon}, P., {Bouquet}, A., {Giraud-Heraud}, Y., \& {Kaplan}, J. 1993, A\&A,
  277, 1

\bibitem[{{Baltz} \& {Gondolo}(2001)}]{baltzgondolo}
{Baltz}, E.~A. \& {Gondolo}, P. 2001, ApJ, 559, 41

\bibitem[{{Baltz} {et~al.}(2003){Baltz}, {Gyuk}, \& {Crotts}}]{bgc}
{Baltz}, E.~A., {Gyuk}, G., \& {Crotts}, A. 2003, ApJ, 582, 30

\bibitem[{{Baltz} \& {Silk}(2000)}]{baltzsilk00}
{Baltz}, E.~A. \& {Silk}, J. 2000, ApJ, 530, 578

\bibitem[{{Bell} \& {de Jong}(2001)}]{belldejong}
{Bell}, E.~F. \& {de Jong}, R.~S. 2001, ApJ, 550, 212

\bibitem[{{Belokurov} {et~al.}(2005){Belokurov}, {An}, {Evans}, {Hewett},
  {Baillon}, {Novati}, {Carr}, {Cr{\'e}z{\'e}}, {Giraud-H{\'e}raud}, {Gould},
  {Jetzer}, {Kaplan}, {Kerins}, {Paulin-Henriksson}, {Smartt}, {Stalin},
  {Tsapras}, \& {Weston}}]{belokurov04}
{Belokurov}, V., {An}, J., {Evans}, N.~W., {et~al.} 2005, MNRAS, 357, 17

\bibitem[{{Bertin} \& {Arnouts}(1996)}]{sextractor}
{Bertin}, E. \& {Arnouts}, S. 1996, A\&AS, 117, 393

\bibitem[{{Binney} \& {Merrifield}(1998)}]{binneymerrifield}
{Binney}, J. \& {Merrifield}, M. 1998, {Galactic astronomy} (Princeton, NJ :
  Princeton University Press (Princeton series in astrophysics))

\bibitem[{{Braun}(1991)}]{braun91}
{Braun}, R. 1991, ApJ, 372, 54

\bibitem[{{Calchi Novati} {et~al.}(2003){Calchi Novati}, {Jetzer}, {Scarpetta},
  {Giraud-H{\' e}raud}, {Kaplan}, {Paulin-Henriksson}, \& {Gould}}]{calchi03}
{Calchi Novati}, S., {Jetzer}, P., {Scarpetta}, G., {et~al.} 2003, A\&A, 405,
  851

\bibitem[{{Calchi Novati} {et~al.}(2005){Calchi Novati}, {Paulin-Henriksson},
  {An}, {Baillon}, {Belokurov}, {Carr}, {Cr\'ez\'e}, {Evans}, {Giraud-H{\'
  e}raud}, {Gould}, {Jetzer}, {Kaplan}, {Kerins}, {Hewett}, {Smartt}, {Stalin},
  {Tsapras}, \& {Weston}}]{calchi05}
{Calchi Novati}, S., {Paulin-Henriksson}, S., {An}, J., {et~al.} 2005, {subm.},
  {astro-ph/0504188}

\bibitem[{{Crotts}(1992)}]{crotts92}
{Crotts}, A.~P.~S. 1992, ApJL, 399, L43

\bibitem[{{Darnley} {et~al.}(2004){Darnley}, {Bode}, {Kerins}, {Newsam}, {An},
  {Baillon}, {Novati}, {Carr}, {Cr{\' e}z{\' e}}, {Evans}, {Giraud-H{\'
  e}raud}, {Gould}, {Hewett}, {Jetzer}, {Kaplan}, {Paulin-Henriksson},
  {Smartt}, {Stalin}, \& {Tsapras}}]{darnley04}
{Darnley}, M.~J., {Bode}, M.~F., {Kerins}, E., {et~al.} 2004, MNRAS, 353, 571

\bibitem[{{de Jong} {et~al.}(2004){de Jong}, {Kuijken}, {Crotts}, {Sackett},
  {Sutherland}, {Uglesich}, {Baltz}, {Cseresnjes}, {Gyuk}, {Widrow}, \& {The
  MEGA collaboration}}]{dejong04}
{de Jong}, J.~T.~A., {Kuijken}, K., {Crotts}, A.~P.~S., {et~al.} 2004, A\&A,
  417, 461

\bibitem[{{Feldman} \& {Cousins}(1998)}]{feldmancousins}
{Feldman}, G.~J. \& {Cousins}, R.~D. 1998, {Phys. Rev. D}, 57, 3873

\bibitem[{{Gondolo}(1999)}]{gondolo99}
{Gondolo}, P. 1999, ApJL, 510, L29

\bibitem[{{Gould}(1996)}]{gould96}
{Gould}, A. 1996, ApJ, 470, 201

\bibitem[{{Guhathakurta} {et~al.}(2005){Guhathakurta}, {Choi}, \&
  {Raychaudbury}}]{guhathakurta}
{Guhathakurta}, P., {Choi}, P.~I., \& {Raychaudbury}, S. 2005, in prep

\bibitem[{{Gyuk} \& {Crotts}(2000)}]{gyukcrotts00}
{Gyuk}, G. \& {Crotts}, A. 2000, ApJ, 535, 621

\bibitem[{{Hernquist}(1990)}]{hernquist90}
{Hernquist}, L. 1990, ApJ, 356, 359

\bibitem[{{Hodge} \& {Kennicutt}(1982)}]{hodgekenn}
{Hodge}, P.~W. \& {Kennicutt}, R.~C. 1982, AJ, 87, 264

\bibitem[{{Ibata} {et~al.}(2001){Ibata}, {Irwin}, {Lewis}, {Ferguson}, \&
  {Tanvir}}]{ibata01}
{Ibata}, R., {Irwin}, M., {Lewis}, G., {Ferguson}, A.~M.~N., \& {Tanvir}, N.
  2001, Nature, 412, 49

\bibitem[{{Kent}(1989)}]{kent89}
{Kent}, S.~M. 1989, AJ, 97, 1614

\bibitem[{{Kerins} {et~al.}(2001){Kerins}, {Carr}, {Evans}, {Hewett},
  {Lastennet}, {Le Du}, {Melchior}, {Smartt}, \& {Valls-Gabaud}}]{kerins01}
{Kerins}, E., {Carr}, B.~J., {Evans}, N.~W., {et~al.} 2001, MNRAS, 323, 13

\bibitem[{{King}(1966)}]{king66}
{King}, I.~R. 1966, AJ, 71, 64

\bibitem[{{Kregel} {et~al.}(2002){Kregel}, {van der Kruit}, \& {de
  Grijs}}]{kregel02}
{Kregel}, M., {van der Kruit}, P.~C., \& {de Grijs}, R. 2002, MNRAS, 334, 646

\bibitem[{{Kroupa} {et~al.}(1993){Kroupa}, {Tout}, \& {Gilmore}}]{kroupa93}
{Kroupa}, P., {Tout}, C.~A., \& {Gilmore}, G. 1993, MNRAS, 262, 545

\bibitem[{{Kuijken} \& {Dubinski}(1995)}]{kdmodels}
{Kuijken}, K. \& {Dubinski}, J. 1995, MNRAS, 277, 1341

\bibitem[{{Lasserre} {et~al.}(2000){Lasserre}, {Afonso}, {Albert}, {Andersen},
  {Ansari}, {Aubourg}, {Bareyre}, {Bauer}, {Beaulieu}, {Blanc}, {Bouquet},
  {Char}, {Charlot}, {Couchot}, {Coutures}, {Derue}, {Ferlet}, {Glicenstein},
  {Goldman}, {Gould}, {Graff}, {Gros}, {Ha{\i}ssinski}, {Hamilton}, {Hardin},
  {de Kat}, {Kim}, {Lesquoy}, {Loup}, {Magneville}, {Mansoux}, {Marquette},
  {Maurice}, {Milsztajn}, {Moniez}, {Palanque-Delabrouille}, {Perdereau},
  {Pr{\' e}vot}, {Regnault}, {Rich}, {Spiro}, {Vidal-Madjar}, {Vigroux},
  {Zylberajch}, \& {The EROS collaboration}}]{lasserre00}
{Lasserre}, T., {Afonso}, C., {Albert}, J.~N., {et~al.} 2000, A\&A, 355, L39

\bibitem[{{Mamon} \& {Soneira}(1982)}]{mamonsoneira}
{Mamon}, G.~A. \& {Soneira}, R.~M. 1982, ApJ, 255, 181

\bibitem[{{Mao} \& {Paczynski}(1991)}]{mao91}
{Mao}, S. \& {Paczynski}, B. 1991, ApJL, 374, L37

\bibitem[{{McElroy}(1983)}]{mcelroy83}
{McElroy}, D.~B. 1983, ApJ, 270, 485

\bibitem[{{Merrett} {et~al.}(2003){Merrett}, {Kuijken}, {Merrifield},
  {Romanowsky}, {Douglas}, {Napolitano}, {Arnaboldi}, {Capaccioli}, {Freeman},
  {Gerhard}, {Evans}, {Wilkinson}, {Halliday}, {Bridges}, \&
  {Carter}}]{merrett03}
{Merrett}, H.~R., {Kuijken}, K., {Merrifield}, M.~R., {et~al.} 2003, MNRAS,
  346, L62

\bibitem[{{Navarro} {et~al.}(1996){Navarro}, {Frenk}, \& {White}}]{nfw}
{Navarro}, J.~F., {Frenk}, C.~S., \& {White}, S.~D.~M. 1996, ApJ, 462, 563

\bibitem[{{Paczy\'nski}(1986)}]{paczynski86}
{Paczy\'nski}, B. 1986, ApJ, 304, 1

\bibitem[{{Paulin-Henriksson} {et~al.}(2002){Paulin-Henriksson}, {Baillon},
  {Bouquet}, {Carr}, {Cr{\' e}z{\' e}}, {Evans}, {Giraud-H{\' e}raud}, {Gould},
  {Hewett}, {Kaplan}, {Kerins}, {Lastennet}, {Le Du}, {Melchior}, {Smartt}, \&
  {Valls-Gabaud}}]{paulin02}
{Paulin-Henriksson}, S., {Baillon}, P., {Bouquet}, A., {et~al.} 2002, ApJL,
  576, L121

\bibitem[{{Paulin-Henriksson} {et~al.}(2003){Paulin-Henriksson}, {Baillon},
  {Bouquet}, {Carr}, {Cr{\' e}z{\' e}}, {Evans}, {Giraud-H{\' e}raud}, {Gould},
  {Hewett}, {Kaplan}, {Kerins}, {Le Du}, {Melchior}, {Smartt}, {Valls-Gabaud},
  \& {The POINT-AGAPE Collaboration}}]{paulin03}
{Paulin-Henriksson}, S., {Baillon}, P., {Bouquet}, A., {et~al.} 2003, A\&A,
  405, 15

\bibitem[{{Press} {et~al.}(1992){Press}, {Teukolsky}, {Vetterling}, \&
  {Flannery}}]{numrec}
{Press}, W.~H., {Teukolsky}, S.~A., {Vetterling}, W.~T., \& {Flannery}, B.~P.
  1992, {Numerical recipes in C. The art of scientific computing} (Cambridge:
  University Press, |c1992, 2nd ed.)

\bibitem[{{Riffeser} {et~al.}(2003){Riffeser}, {Fliri}, {Bender}, {Seitz}, \&
  {G{\" o}ssl}}]{riffeser03}
{Riffeser}, A., {Fliri}, J., {Bender}, R., {Seitz}, S., \& {G{\" o}ssl}, C.~A.
  2003, ApJL, 599, L17

\bibitem[{{Savage} \& {Mathis}(1979)}]{savagemathis}
{Savage}, B.~D. \& {Mathis}, J.~S. 1979, ARA\&A, 17, 73

\bibitem[{{Schneider} \& {Weiss}(1986)}]{schneider86}
{Schneider}, P. \& {Weiss}, A. 1986, A\&A, 164, 237

\bibitem[{{Schwarzenberg-Czerny}(1996)}]{czerny}
{Schwarzenberg-Czerny}, A. 1996, ApJL, 460, L107+

\bibitem[{{Tomaney} \& {Crotts}(1996)}]{tc96}
{Tomaney}, A.~B. \& {Crotts}, A.~P.~S. 1996, AJ, 112, 2872

\bibitem[{{Uglesich} {et~al.}(2004){Uglesich}, {Crotts}, {Baltz}, {de Jong},
  {Boyle}, \& {Corbally}}]{colvatt}
{Uglesich}, R.~R., {Crotts}, A.~P.~S., {Baltz}, E.~A., {et~al.} 2004, ApJ, 612,
  877

\bibitem[{{van Albada} {et~al.}(1985){van Albada}, {Bahcall}, {Begeman}, \&
  {Sancisi}}]{albada85}
{van Albada}, T.~S., {Bahcall}, J.~N., {Begeman}, K., \& {Sancisi}, R. 1985,
  ApJ, 295, 305

\bibitem[{{Walterbos} \& {Kennicutt}(1988)}]{WK88}
{Walterbos}, R.~A.~M. \& {Kennicutt}, R.~C. 1988, A\&A, 198, 61

\bibitem[{{Widrow} \& {Dubinski}(2005)}]{m31models}
{Widrow}, L.~M. \& {Dubinski}, J. 2005, {ApJ., in press}

\bibitem[{{Widrow} {et~al.}(2003){Widrow}, {Perrett}, \& {Suyu}}]{WPS}
{Widrow}, L.~M., {Perrett}, K.~M., \& {Suyu}, S.~H. 2003, ApJ, 588, 311

\end{thebibliography}


\appendix

\section{Candidate event lightcurves}
\label{ap:mulresult:lightcurves}

On the following pages, for each of the 14 candidate microlensing
events in our sample, the r$^{\prime}$ and i$^{\prime}$ lightcurves
and thumbnails taken from the difference centred on the event
positions are shown, together with a short discussion. Apart from the
INT r$^{\prime}$ and i$^{\prime}$ data, KP4m R and I data points are
also plotted in the lightcurves. The fits shown are however the fits
done to only the INT data.

\clearpage

\begin{figure*}[h!]
\vspace{0.1\hsize}{\tt figapp01a.gif}\vspace{0.1\hsize}
\caption{{\bf(a)} Event 1: lightcurves. The two upper panels show the full r$^{\prime}$
  and i$^{\prime}$ lightcurves of the microlensing event. In the lower
  left corner are zooms on the peak region. In the lower right
  corner the r$^{\prime}$ flux is plotted versus the i$^{\prime}$
  flux; if the colour is constant, the points should lie on a straight
  line. Also drawn is the best fit microlensing model. The solid
  circles are points from the INT data, the open circles are from the
  KP4m data. The start of the INT survey, August 1st 1999, is used as
  the zeropoint for the timescale.}
\label{fig:mulresult:event1}
\end{figure*}

\addtocounter{figure}{-1}
\begin{figure*}[h!]
\vspace{0.1\hsize}{\tt 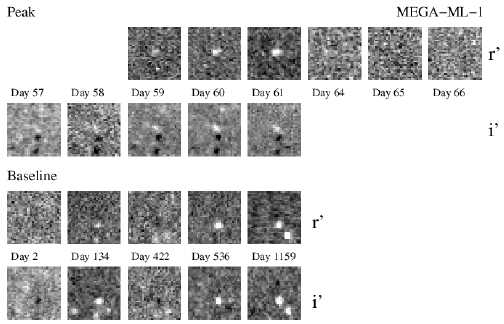}\vspace{0.1\hsize}
\caption{{\bf(b)} Event 1: thumbnails. The two upper rows of
  thumbnails show are taken from r$^{\prime}$ and i$^{\prime}$
  difference images during the peak of the candidate event. Selected
  thumbnails from the baseline are also shown in the two bottom
  rows. Each thumbnail is 30$\times$30 pixels or 10$\times$10\arcsec~
  in size.}
\label{fig:mulresult:event1tn}
\end{figure*}

\subsection*{MEGA-ML-1}
Located close to the centre of M31, this event has a rather noisy
baseline. Apart from the background of very faint variables there are
some variable sources clearly visible in the difference images. As can
be seen in the thumbnails in figure \ref{fig:mulresult:event1tn}(b) a
bright variable is located just a few pixels from the position of the
candidate event. Another, fainter variable is seen at a similar
distance above and to the left. The other variable sources are further
away and should have no influence on the photometry.


\begin{figure*}
\vspace{0.1\hsize}{\tt figapp02a.gif}\vspace{0.1\hsize}
\caption{{\bf(a)} Event 2: lightcurves. See caption of \ref{fig:mulresult:event1}(a).}
\label{fig:mulresult:event2}
\end{figure*}

\addtocounter{figure}{-1}
\begin{figure*}[t]
\vspace{0.1\hsize}{\tt 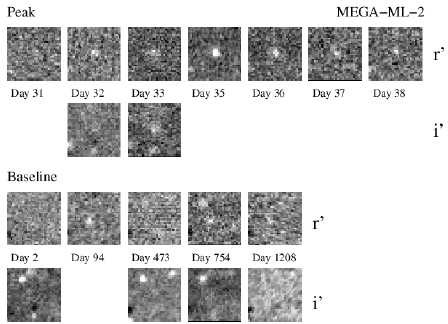}\vspace{0.1\hsize}
\caption{{\bf(b)} Event 2: thumbnails. See caption of \ref{fig:mulresult:event1tn}(b).}
\label{fig:mulresult:event2tn}
\end{figure*}

\subsection*{MEGA-ML-2}

This candidate event is located very close to MEGA-ML-1 and therefore
has the same problems connected to being close the centre of M31. In
the thumbnails of days 94, 754, and 1208 we see a variable source a
few pixels to the left of the event position. This variable is brighter
in r$^{\prime}$ than in i$^{\prime}$, which causes the r$^{\prime}$
baseline to be the most noisy.


\begin{figure*}
\vspace{0.1\hsize}{\tt figapp03a.gif}\vspace{0.1\hsize}
\caption{{\bf(a)} Event 3: lightcurves. See caption of \ref{fig:mulresult:event1}(a).}
\label{fig:mulresult:event3}
\end{figure*}

\addtocounter{figure}{-1}
\begin{figure*}[t]
\vspace{0.1\hsize}{\tt 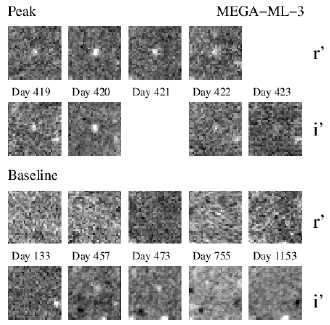}\vspace{0.1\hsize}
\caption{{\bf(b)} Event 3: thumbnails. See caption of \ref{fig:mulresult:event1tn}(b).}
\label{fig:mulresult:event3tn}
\end{figure*}

\subsection*{MEGA-ML-3}
This candidate event is also located close to the M31 centre.
In figure \ref{fig:mulresult:event3_kp97} we already demonstrated that
a very faint variable source is positioned $\sim$0.25\arcsec away from
this candidate event. In the i$^{\prime}$ thumbnails another variable
is visible just above and to the right of the event. This variable has
a bright episode between days 440 and 480, causing the bump in the
baseline in the i$^{\prime}$ lightcurve.


\begin{figure*}
\vspace{0.1\hsize}{\tt figapp04a.gif}\vspace{0.1\hsize}
\caption{{\bf(a)} Event 7: lightcurves. See caption of \ref{fig:mulresult:event1}(a).}
\label{fig:mulresult:event7}
\end{figure*}

\addtocounter{figure}{-1}
\begin{figure*}[t]
\vspace{0.1\hsize}{\tt 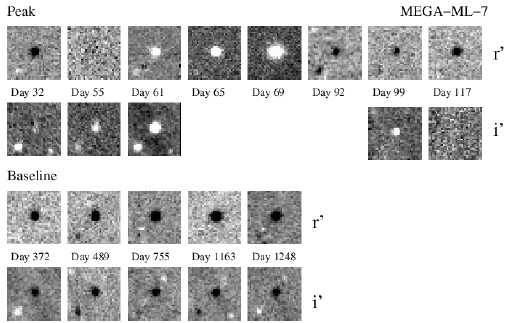}\vspace{0.1\hsize}
\caption{{\bf(b)} Event 7: thumbnails. See caption of \ref{fig:mulresult:event1tn}(b).}
\label{fig:mulresult:event7tn}
\end{figure*}

\subsection*{MEGA-ML-7}
By far the brightest event in our sample, the thumbnails of MEGA-ML-7
show a very bright residual close to the peak centre. Since the peak
occurs during the first season, some of the exposures used for
creating the reference image contained a significant amount of the
magnified flux, so that the baseline lies at a negative
difference flux. There are some variables nearby, but none of them are
close or bright enough to significantly influence the photometry. The
distance to the centre of M31 is also quite large ($\sim$22\arcmin),
reducing the background of faint variable sources. As pointed out by
\cite{paulin03}, there are some systematic deviations from the best fit
microlensing model. \cite{event7anomalies} find that this anomaly can
be explained by a binary lens.


\begin{figure*}
\vspace{0.1\hsize}{\tt figapp05a.gif}\vspace{0.1\hsize}
\caption{{\bf(a)} Event 8: lightcurves. See caption of \ref{fig:mulresult:event1}(a).}
\label{fig:mulresult:event8}
\end{figure*}

\addtocounter{figure}{-1}
\begin{figure*}[t]
\vspace{0.1\hsize}{\tt 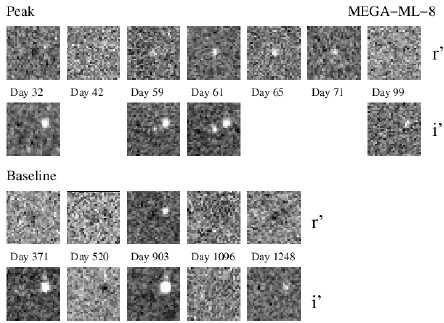}\vspace{0.1\hsize}
\caption{{\bf(b)} Event 8: thumbnails. See caption of \ref{fig:mulresult:event1tn}(b).}
\label{fig:mulresult:event8tn}
\end{figure*}

\subsection*{MEGA-ML-8}
This near side event is located $\sim$23\arcmin~ from the centre of
M31. A variable that is particularly bright in i$^{\prime}$ is
situated about 2.4\arcsec~ NW of the candidate event, but should not
have much of an effect on the photometry. The baselines of the
lightcurves indeed look stable and well-behaved.


\begin{figure*}
\vspace{0.1\hsize}{\tt figapp06a.gif}\vspace{0.1\hsize}
\caption{{\bf(a)} Event 9: lightcurves. See caption of \ref{fig:mulresult:event1}(a).}
\label{fig:mulresult:event9}
\end{figure*}

\addtocounter{figure}{-1}
\begin{figure*}[t]
\vspace{0.1\hsize}{\tt 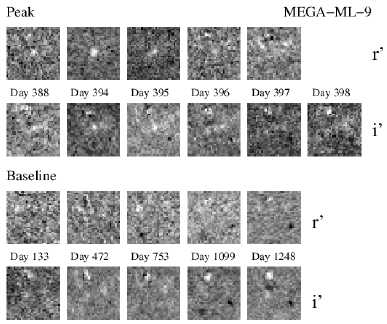}\vspace{0.1\hsize}
\caption{{\bf(b)} Event 9: thumbnails. See caption of \ref{fig:mulresult:event1tn}(b).}
\label{fig:mulresult:event9tn}
\end{figure*}

\subsection*{MEGA-ML-9}
Peak coverage is poor for this candidate event, but the baselines are
stable. The thumbnails show quite a lot of faint variables, two of
which are located very close, approximately 1\arcsec~ to the left of
the event position, accounting for the noise in the i$^{\prime}$
baseline that is higher than in the r$^{\prime}$ lightcurve.


\begin{figure*}
\vspace{0.1\hsize}{\tt figapp07a.gif}\vspace{0.1\hsize}
\caption{{\bf(a)} Event 10: lightcurves. See caption of \ref{fig:mulresult:event1}(a).}
\label{fig:mulresult:event10}
\end{figure*}

\addtocounter{figure}{-1}
\begin{figure*}[t]
\vspace{0.1\hsize}{\tt 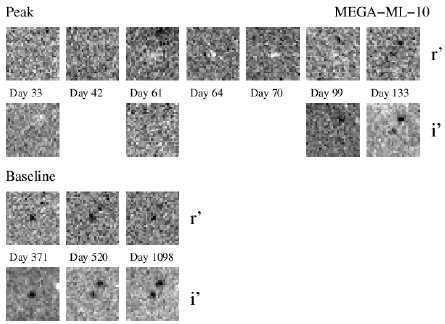}\vspace{0.1\hsize}
\caption{{\bf(b)} Event 10: thumbnails. See caption of \ref{fig:mulresult:event1tn}(b).}
\label{fig:mulresult:event10tn}
\end{figure*}

\subsection*{MEGA-ML-10}
This event is a beautiful example of a combined lightcurve with KP4m
and INT data. Peak coverage in INT i$^{\prime}$ is poor, but the KP4m
I data points follow the fit (derived only from INT data) very well. A
fairly bright variable is situated slightly above and to the right of
the event position and there is a hint of a very faint variable about
1\arcsec~ to the left. Although the INT baseline in i$^{\prime}$ is
noisy, the r$^{\prime}$ and both KP4m R and I lightcurves show an very
stable and well-behaved baseline.

\clearpage

\begin{figure*}
\vspace{0.1\hsize}{\tt figapp08a.gif}\vspace{0.1\hsize}
\caption{{\bf(a)} Event 11: lightcurves. See caption of \ref{fig:mulresult:event1}(a).}
\label{fig:mulresult:event11}
\end{figure*}

\addtocounter{figure}{-1}
\begin{figure*}[t]
\vspace{0.1\hsize}{\tt 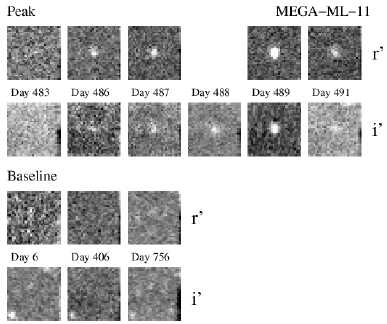}\vspace{0.1\hsize}
\caption{{\bf(b)} Event 11: thumbnails. See caption of \ref{fig:mulresult:event1tn}(b).}
\label{fig:mulresult:event11tn}
\end{figure*}

\subsection*{MEGA-ML-11}
A high signal-to-noise event with a good fit and stable
baseline. There is some noise in the i$^{\prime}$ baseline, caused by
the variable source that is visible in the thumbnails of days 6 and
756 at $\sim$1.3\arcsec~ above the event position. During the fourth
observing season a few bad columns were lying exactly on top of the
event position, so that there is only 1 INT data point
available. However, the KP4m data show that the baseline remains flat
everywhere.


\begin{figure*}
\vspace{0.1\hsize}{\tt figapp09a.gif}\vspace{0.1\hsize}
\caption{{\bf(a)} Event 13: lightcurves. See caption of \ref{fig:mulresult:event1}(a).}
\label{fig:mulresult:event13}
\end{figure*}

\addtocounter{figure}{-1}
\begin{figure*}[t]
\vspace{0.1\hsize}{\tt 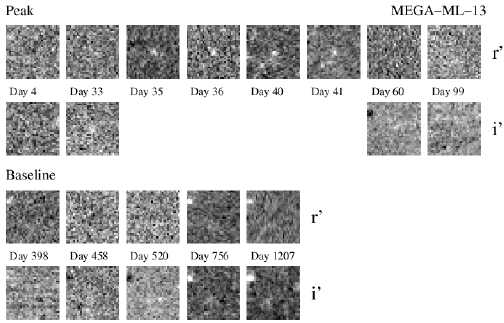}\vspace{0.1\hsize}
\caption{{\bf(b)} Event 13: thumbnails. See caption of \ref{fig:mulresult:event1tn}(b).}
\label{fig:mulresult:event13tn}
\end{figure*}

\subsection*{MEGA-ML-13}
This candidate event has the lowest signal-to-noise of our sample. It
is situated far out in the far side of the disk at $\sim$31\arcmin~
from the centre of the galaxy and the relatively low galaxy background
makes it possible to detect these kind of faint events. Due
to the y-axis scale the i$^{\prime}$ the baseline looks quite noisy,
but it is in fact not significantly more so than for other candidate
events. The thumbnails of days 398 and 520 show that the closest
variable source is located $\sim$1.4\arcsec~ below and to the left of
the event, which explains the scatter in the i$^{\prime}$ baseline.


\begin{figure*}
\vspace{0.1\hsize}{\tt figapp10a.gif}\vspace{0.1\hsize}
\caption{{\bf(a)} Event 14: lightcurves. See caption of \ref{fig:mulresult:event1}(a).}
\label{fig:mulresult:event14}
\end{figure*}

\addtocounter{figure}{-1}
\begin{figure*}[t]
\vspace{0.1\hsize}{\tt 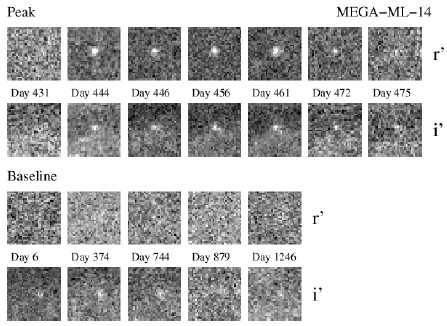}\vspace{0.1\hsize}
\caption{{\bf(b)} Event 14: thumbnails. See caption of \ref{fig:mulresult:event1tn}(b).}
\label{fig:mulresult:event14tn}
\end{figure*}

\subsection*{MEGA-ML-14}
At $\sim$35.5\arcmin~ from the M31 centre, this candidate event is the
most far out in the disk of all events in our sample.  The
i$^{\prime}$ photometry of this candidate event is compromised by the
variable source at $\sim$1.3\arcsec. From the i$^{\prime}$ thumbnails
one can also see that the event lies at the edge of a fringe, making
the background in the lower half of the thumbnails brighter than in
the upper half. This can also cause some extra scatter in the
photometry. Overall, however, the microlensing fit is very good and
both INT and KP4m lightcurves show a stable baseline.


\begin{figure*}
\vspace{0.1\hsize}{\tt figapp11a.gif}\vspace{0.1\hsize}
\caption{{\bf(a)} Event 15: lightcurves. See caption of \ref{fig:mulresult:event1}(a).}
\label{fig:mulresult:event15}
\end{figure*}

\addtocounter{figure}{-1}
\begin{figure*}[t]
\vspace{0.1\hsize}{\tt 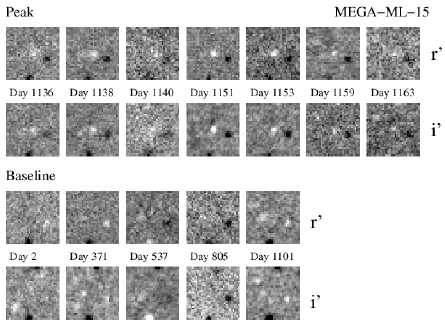}\vspace{0.1\hsize}
\caption{{\bf(b)} Event 15: thumbnails. See caption of \ref{fig:mulresult:event1tn}(b).}
\label{fig:mulresult:event15tn}
\end{figure*}

\subsection*{MEGA-ML-15}
This event is again located close to the centre of M31 and presumably
has a strong background of faint variable sources. In the thumbnails
also several variables are visible very close to the event position,
both in r$^{\prime}$ and in i$^{\prime}$. The lightcurve baselines are
rather noisy because of this, but show no coherent secondary bumps and
the KP4m baselines are very stable.


\begin{figure*}
\vspace{0.1\hsize}{\tt figapp12a.gif}\vspace{0.1\hsize}
\caption{{\bf(a)} Event 16: lightcurves. See caption of \ref{fig:mulresult:event1}(a).}
\label{fig:mulresult:event16}
\end{figure*}

\addtocounter{figure}{-1}
\begin{figure*}[t]
\vspace{0.1\hsize}{\tt 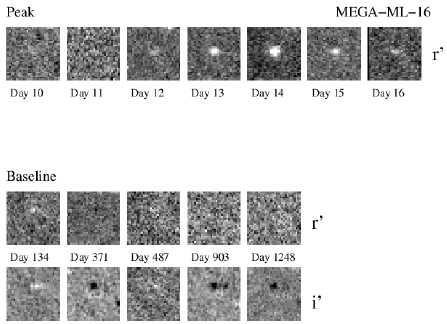}\vspace{0.1\hsize}
\caption{{\bf(b)} Event 16: thumbnails. See caption of \ref{fig:mulresult:event1tn}(b).}
\label{fig:mulresult:event16tn}
\end{figure*}

\subsection*{MEGA-ML-16}
Not selected in our first analysis of the first two seasons of INT
data \citep{dejong04} due to baseline variability, the i$^{\prime}$
lightcurve of this event is strongly influenced by a bright variable
situated just 1.1\arcsec~ to the north. Using a smaller extraction
aperture for the photometry in the present analysis, the i$^{\prime}$
baseline is still very noisy and the same is true for the KP4m I-band
data. The INT r$^{\prime}$ and KP4m R data are much better behaved and
the r$^{\prime}$ peak is fit very well by the microlensing fit.


\begin{figure*}
\vspace{0.1\hsize}{\tt figapp13a.gif}\vspace{0.1\hsize}
\caption{{\bf(a)} Event 17: lightcurves. See caption of \ref{fig:mulresult:event1}(a).}
\label{fig:mulresult:event17}
\end{figure*}

\addtocounter{figure}{-1}
\begin{figure*}[t]
\vspace{0.1\hsize}{\tt 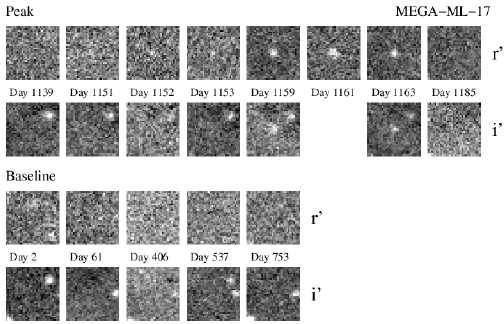}\vspace{0.1\hsize}
\caption{{\bf(b)} Event 17: thumbnails. See caption of \ref{fig:mulresult:event1tn}(b).}
\label{fig:mulresult:event17tn}
\end{figure*}

\subsection*{MEGA-ML-17}
The i$^{\prime}$ baseline is slightly noisy, but the r$^{\prime}$ and
both KP4m lightcurves are well-behaved. In the thumbnails no very
close variables are visible.


\begin{figure*}
\vspace{0.1\hsize}{\tt figapp18a.gif}\vspace{0.1\hsize}
\caption{{\bf(a)} Event 18: lightcurves. See caption of \ref{fig:mulresult:event1}(a).}
\label{fig:mulresult:event18}
\end{figure*}

\addtocounter{figure}{-1}
\begin{figure*}[t]
\vspace{0.1\hsize}{\tt figapp18b.jpg}\vspace{0.1\hsize}
\caption{{\bf(b)} Event 18: thumbnails. See caption of \ref{fig:mulresult:event1tn}(b).}
\label{fig:mulresult:event18tn}
\end{figure*}

\subsection*{MEGA-ML-18}
This candidate event shows quite large scatter in the baseline and
also in the peak. Faint variables might be the culprits, although the
event is not located very close to the galaxy centre
($\sim$15.1\arcmin). The thumbnails show no variable sources very
close to the event position, however they do show that this event is
situated on the edge of a fringe running diagonally across the
thumbnails. This fringe and the fact that it can change position
slightly between frames is the most probable cause for the noisy
i$^{\prime}$ photometry.

\end{document}